\documentclass[12pt, a4paper]{article}
\usepackage[a4paper, left=3cm,right=2cm]{geometry}
\usepackage{hyperref}
\usepackage{amsfonts} 
\usepackage{amsmath}
\usepackage{amssymb}
\usepackage{setspace}
\usepackage{color}
\usepackage{verbatim}
\usepackage{pdflscape}
\usepackage{empheq,fancybox}
\usepackage[utf8]{inputenc}
\usepackage{tensor}
\usepackage{cite}
\usepackage{tikz}
\usepackage{graphicx,subfigure}
\usepackage[font=small, font+=it, format=hang, margin={1cm, 1cm}]{caption}
\graphicspath{{figure/}}
\usepackage{amsthm}
\usepackage{array,multirow}
\usepackage{dcolumn}
\usepackage{bm}
\usepackage{physics}
\usepackage{enumitem}

\numberwithin{equation}{section}

\newcommand{\bea}{\begin{eqnarray}}
\newcommand{\eea}{\end{eqnarray}}
\newcommand{\be}{\begin{equation}}
\newcommand{\ee}{\end{equation}}

\def\p{\partial}
\def\eps{\epsilon}

\newcommand{\sign}[1]{\,\text{sign}(#1)}

\renewcommand{\d}{\textrm{d}}
\newcommand{\sint}{\rotatebox[origin=c]{-90}{$\backslash$}\hspace{-14pt}\int}
\newcommand{\sintline}{\scriptsize\rotatebox[origin=c]{-90}{$\backslash$}\hspace{-10pt}\int}

\newcommand{\that}{{\hat{t}}}
\newcommand{\rhat}{{\hat{r}}}
\newcommand{\phihat}{{\hat{\varphi}}}

\newcommand{\sn}{\,\text{sn}\,}

\newcommand{\dn}{\,\text{dn}\,}

\newcommand{\arctanh}{\text{arctanh}}
\newcommand{\ehat}{{\hat{E}}}

\newtheorem{prop}{Proposition}

\begin{document}

\setcounter{tocdepth}{2}

\begin{titlepage}

\begin{flushright}\vspace{-3cm}
{\small
\today }\end{flushright}
\vspace{0.5cm}

\begin{center}

{{ \LARGE{\bf{Near-horizon geodesics \vspace{10pt}\\ 
of high-spin black holes}}}}
\vspace{5mm}

\bigskip
\bigskip

\centerline{\large{\bf{Geoffrey Comp\`{e}re and Adrien Druart}}}

\vspace{2mm}
\normalsize
\bigskip\medskip
\textit{Universit\'{e} Libre de Bruxelles, Gravitational Wave Centre, \\
 International Solvay Institutes, CP 231, B-1050 Brussels, Belgium}\\\vspace{2mm}

\bigskip

\vspace{15mm}

\begin{abstract}
\noindent

We provide an exhaustive and illustrated classification of timelike and null geodesics in the near-horizon region of near-extremal Kerr black holes. The classification of polar motion extends to Kerr black holes of arbitrary spin. The classification of radial motion leads to a simple parametrization of the separatrix between bound and unbound motion. Furthermore, we prove that each timelike or null geodesic is related via conformal transformations and discrete symmetries to spherical orbits and we provide the explicit mappings. We detail the high-spin behavior of both the innermost stable and the innermost bound spherical orbits.

\end{abstract}


\end{center}

\end{titlepage}

\tableofcontents

\newpage

\section{Introduction and summary}

The study of timelike and null geodesics of the Kerr metric has a long history which is still ongoing \cite{PhysRev.174.1559,Wilkins:1972rs,Bardeen:1973aa,Chandrasekhar:1983aa,Rauch:1994aa,Neill:1995aa,Schmidt:2002qk,Mino:2003yg,Vazquez:2003zm,Kraniotis:2005zm,Dexter:2009fg,Fujita:2009bp,Kraniotis:2010gx,Hackmann:2015ewa,Hackmann:2015vla,Porfyriadis:2016gwb,Compere:2017hsi,Kapec:2019hro,Gralla:2019ceu,Rana:2019bsn,Stein:2019buj}. There are at least two motivations for studying Kerr geodesics. First, null geodesics (together with the modeling of light sources) underpin the field of black hole imaging \cite{Bardeen:1973aa,Luminet:1979nyg,Falcke:1999pj,Vazquez:2003zm,James:2015yla,Luminet:2019hfx,Gralla:2019xty,Gralla:2019drh}, which has recently become an observational science \cite{Akiyama:2019cqa}. Second, timelike geodesics provide the zeroth-order motion of binary systems in the perturbative small-mass-ratio expansion, which leads in the adiabatic approximation to the leading-order gravitational waveforms of extreme-mass-ratio inspirals (EMRIs) \cite{1973ApJ...185..635T,Sasaki:1981sx,Ryan:1995zm,Finn:2000sy}. 

In the extremely high-spin limit, conformal $SL(2,\mathbb R)$ symmetry appears in the near-horizon limit \cite{Bardeen:1999px,Amsel:2009ev,Dias:2009ex,Bredberg:2009pv}. Previous studies of geodesic motion in the high-spin near-horizon limit were restricted either to equatorial orbits \cite{Porfyriadis:2014fja,Hadar:2014dpa,Hadar:2015xpa,Gralla:2015rpa,Hadar:2016vmk,Compere:2017hsi,Hod:2017uof}, to specific orbits \cite{AlZahrani:2010qb}, or to parametrically generic geodesics \cite{Kapec:2019hro} that discard relevant measure-zero sets in parameter space such as the separatrix between bound and unbound motion. The first objective of this paper is to derive the complete classification of geodesics in the high-spin near-horizon Kerr region. While our treatment of polar motion applies to arbitrary geodesics in Kerr, the radial motion is only studied in the high-spin limit and in the near-horizon region. For the ease of readability, we provide extended tables and figures summarizing our classification. In their range of applicability, all the results of Kapec and Lupsasca \cite{Kapec:2019hro} agree with ours.  We will be attentive to defining an intuitive nomenclature for each geodesic class. 

It was previously shown in Refs. \cite{AlZahrani:2010qb,Porfyriadis:2016gwb} that any null orbit that enters or leaves the near-horizon region has a polar motion bounded by the minimal angle $\cos^2\theta_{\text{\text{min}}} = 2 \sqrt{3}-3$ ($47^\circ \lessapprox \theta \lessapprox 133^\circ$), which corresponds to the polar inclination of the velocity-of-light surface in the near-horizon and high-spin limit. As an outcome of our analysis, we confirm this property and further prove that it holds for any timelike geodesic as well. The polar motion is more restricted for the innermost bound spherical orbits (IBSOs): $\cos^2\theta_{\text{\text{min}}} = 1/3$ ($55^\circ \lessapprox \theta \lessapprox 125^\circ$), as already shown in Refs. \cite{Hod:2017uof,Stein:2019buj} and even more restricted for the innermost stable spherical orbits (ISSOs): $\cos^2\theta_{\text{\text{min}}} = 3-2 \sqrt{2}$ ($65^\circ \lessapprox \theta \lessapprox 115^\circ$), as independently shown in Ref. \cite{Stein:2019buj} at the time of finalizing this draft. We also prove that the separatrix between bound and unbound motion consists of all orbits with an angular momentum equal to that of the ISSO.

Conformal symmetry in the near-horizon high-spin Kerr geometry leads to potentially observable signatures if such high-spin black holes are realized in nature. The behavior of null geodesics on the image of an extremely spinning Kerr black hole leads to the NHEKline \cite{Bardeen:1973aa,Gralla:2017ufe} and to specific polarization whorls \cite{Gates:2018hub}. Gravitational waveforms on adiabatic inspirals lead to exponentially decaying tails at fixed oscillation frequencies with amplitudes suppressed as $(1-J^2/M^4)^{1/6}$ \cite{Porfyriadis:2014fja,Gralla:2016qfw}, while plunging trajectories lead to impact-dependent polynomial quasinormal ringing with a power ranging from inverse time to square root of inverse time \cite{Compere:2017hsi}. With the aim of deriving further consequences of conformal symmetry, we provide in this paper the action of conformal symmetry on arbitrary timelike and null geodesics in near-horizon high-spin Kerr. It was shown in Ref. \cite{Compere:2017hsi} that conformal symmetry together with a discrete symmetry leads to equivalence classes of equatorial timelike geodesics with circular orbits as distinguished representatives. This allows us to simplify the computation of Teukolsky waveforms by applying conformal transformations to the seed circular waveform \cite{Hadar:2014dpa,Compere:2017hsi}. In this paper, we show that in whole generality conformal symmetry and discrete symmetries lead to equivalence classes which each admit spherical orbits as distinguished representatives. For orbits with an angular momentum lower than the ISSO, which we denote as subcritical orbits, no spherical orbit exists but a ``complex spherical orbit'' exists that generates the equivalence class. Such a complex spherical orbit can be used as a seed and complexified conformal transformations allow to reach all real subcritical geodesics. 

We provide as ancillary files the following four Mathematica sheets: the explicit polar integrals describing geodesics in Kerr and in (near-)NHEK, the explicit radial integrals describing geodesics in (near-)NHEK, the list of conformal maps between (near-) NHEK radial geodesics and finally the code allowing to draw the Penrose diagram of NHEK geodesics.

The rest of the paper is organized as follows. In Sec. \ref{sec:polar} we provide the complete classification of polar geodesic motion in Kerr. In Sec. \ref{sec:nhg}, we present the complete classification of timelike and null geodesics in the near-horizon region of a high-spin Kerr black hole. We discuss the properties of spherical geodesics in Sec. \ref{sec:spherical_properties} and we derive the conformal mappings between near-horizon geodesics in high-spin Kerr in Sec. \ref{sec:classes}. Several appendixes contain the explicit form of all prograde near-horizon geodesics as well as relevant integrals. A comparison with the radial geodesic motion in high-spin Kerr obtained in \cite{Kapec:2019hro} is performed in Appendix \ref{app:comparison}.

\section{Polar geodesic motion in Kerr}\label{sec:polar}

After a short summary of some essential features of Kerr geodesic motion, we will provide a complete classification of  polar geodesic motion in Kerr. Our results complete the inspiring work of Ref. \cite{Kapec:2019hro} by treating all marginal cases (i.e., measure-zero sets) that were discarded such as $Q=0$ and $\ell = 0$ (for the notation, see below). We will also provide extensive tables and figures that summarize the classification and allow us to visually grasp all main features. We will be attentive to defining an intuitive nomenclature for each geodesic class. We will also derive manifestly real and positive explicit analytical forms for each polar motion with $\ell \neq 0$. We do not provide the exhaustive real analytical forms for $\ell = 0$ by lack of astrophysical interest. 

In Boyer-Lindquist coordinates $\hat x^\mu = (\that,\rhat,\theta,\phihat)$, the Kerr metric is
\begin{equation}
\dd s^2=-\frac{\Delta}{\Sigma}\qty(\dd \that-a\sin^2\theta\dd\phihat)^2+\Sigma\qty(\frac{\dd\rhat^2}{\Delta}+\dd\theta^2)+\frac{\sin^2\theta}{\Sigma}\qty((\rhat^2+a^2)\dd\phihat-a\dd\that)^2\label{eq:kerr_bl}
\end{equation}
with
\begin{align}
\Delta(\rhat)&\triangleq\rhat^2-2M\rhat+a^2,\qquad \Sigma(\rhat,\theta)\triangleq\rhat^2+a^2\cos^2\theta.
\end{align}
Assuming the validity of the cosmic censorship conjecture, the angular momentum per unit mass $a$ of the hole is bounded by its mass $M$: $\abs{a}\leq M$.

The Kerr metric admits a Killing-Yano tensor $J_{\mu\nu}=J_{[\mu\nu]}$ \cite{10.2307/1969287,1952AnMat..55..328Y,1973NYASA.224..125P,Floyd:1973aa} and a derived Killing-St\"ackel tensor $K_{\mu\nu}=K_{(\mu\nu)}$ \cite{Walker:1970un}, where (see also Ref. \cite{Kapec:2019hro})
\bea \label{Kmunu}
\begin{split}
K_{\mu\nu} &= -J_\mu^{\;\; \lambda}J_{\lambda \nu} , \\ 
\frac{1}{2}J_{\mu\nu} \dd\hat x^\mu \wedge \dd \hat x^\nu &= a \cos\theta \dd\hat r \wedge (\dd\hat t - a \sin^2\theta \dd \hat \phi)+\hat  r \sin\theta \dd \theta \wedge \left( (\hat r^2+a^2) \dd \hat \phi - a \dd \hat t\right) .
\end{split}
\eea
It obeys $K^{\mu\nu}=2 \Sigma \ell^{(\mu} n^{\nu)}+\hat r^2 g^{\mu\nu}$, where the two principal null directions of Kerr are given by $ \ell^\mu \p_\mu =\Delta^{-1}( (\hat r^2+a^2)\p_{\hat t}+a \p_{\hat \phi})$ and $ n^\mu \p_\mu=(2 \Sigma)^{-1}( (\hat r^2+a^2)\p_{\hat t}-\Delta \p_{\hat r} +a \p_{\hat \phi} )$. 

\subsection{Geodesic equations}

As anyone can check, geodesic motion of a test pointwise particle in the Kerr geometry can be described by the following set of four first-order differential equations:
\begin{align}
\Sigma\,\dv{\that}{\tau}&=a \ell -a^2\ehat\sin^2\theta+(\rhat^2+a^2)\frac{P(\rhat)}{\Delta(\rhat)},\label{eq:kerr_t}\\
\Sigma\,\dv{\rhat}{\tau}&=\pm_{\hat r}\sqrt{\hat R(\rhat)},\label{eq:kerr_r}\\
\Sigma\,\dv{\cos \theta}{\tau}&=\pm_\theta \sqrt{\Theta(\cos^2 \theta)},\label{eq:kerr_theta}\\
\Sigma\,\dv{\phihat}{\tau}&=-a\ehat+ \ell \csc^2\theta+a\frac{P(\rhat)}{\Delta(\rhat)}\label{eq:kerr_phi}
\end{align}
where $\mu$ is the mass, $\tau$ is either $\mu$ times the proper time for timelike geodesics or an affine parameter for massless particles (i.e. $\mu = 0$). The momentum is denoted as $p^\mu = \frac{d\hat x^\mu}{d\tau}$. The geodesic is characterized by its energy $\hat E = -p_{\hat t}$, its angular momentum $\ell =p_{\hat \phi}$ and its non-negative Carter constant $k= K_{\mu\nu}p^\mu p^\nu = p_\theta^2 + a^2 \mu^2 \cos^2\theta + (p_{\hat \phi} \csc \theta + p_{\hat t} a \sin\theta)^2 \geq 0$ or, equivalently, its Carter constant $Q = k- ( \ell - a \hat E)^2 $ \cite{Carter:1968ks,Carter:1968rr}. We define
\begin{align}
P(\rhat)&\triangleq\ehat(\rhat^2+a^2)-a  \ell,\\
\hat R(\rhat)&\triangleq P^2(\rhat)-\Delta(\rhat)\qty(\mu^2\rhat^2+k),\label{eq:kerr_vr}\\
\Theta(\cos^2\theta)&\triangleq Q \sin^2 \theta+ a^2(\ehat^2-\mu^2)\cos^2\theta \sin^2 \theta-  \ell^2 \cos^2\theta.\label{eq:kerr_vtheta}
\end{align}
In a given Kerr geometry $(M,a)$, a geodesic is fully characterized by the quadruplet of parameters $(\mu \geq 0,\ehat \geq 0, \ell \in \mathbb R,Q \geq -(\ell - a \hat E)^2$), its initial spacetime position and two signs, $s_r^i \equiv \pm_{\hat r}|_{\tau = \tau_i}$, $s_\theta^i \equiv \pm_{\theta}|_{\tau = \tau_i}$, that correspond to the signs of the radial and polar velocity at the initial time $\tau_i$. 

The motion can be integrated using Mino time defined as $\d\lambda = d\tau/\Sigma$ thanks to the property
\bea
\dd\lambda = \frac{\dd\hat r}{\pm_{\hat r}  \sqrt{\hat R(\rhat)} } = \frac{\dd\cos \theta}{\pm_\theta \sqrt{\Theta(\cos^2\theta)}}. 
\eea
We consider a geodesic path linking the initial event $(\that_i,\rhat_i,\theta_i,\phihat_i)$ at  Mino time $\lambda_i$ and the final event  $(\that_f,\rhat_f,\theta_f,\phihat_f)$ at Mino time $\lambda_f$. The equations of motion can be formally integrated as 
\bea
\hat t(\lambda_f)-\hat{t}(\lambda_i) &=& a(\ell - a \hat E)(\lambda_f-\lambda_i)  +a^2 \hat E\,  \qty(\hat T_{\theta}(\lambda_f)-\hat T_{\theta}(\lambda_i) )+\hat T_{\hat r}(\lambda_f)-\hat T_{\hat r}(\lambda_i) ,\nonumber \\
\phihat(\lambda_f) -\phihat(\lambda_i) &=& (\ell- a \hat E) (\lambda_f-\lambda_i) + \ell \qty(\hat \Phi_\theta (\lambda_f)-\hat \Phi_\theta (\lambda_i)) + a \qty(\hat\Phi_{\hat r}(\lambda_f)-\hat \Phi_{\hat r}(\lambda_i)) \nonumber
\eea
where
\bea
\lambda &=& \sint \frac{\dd\hat r}{\pm_{\hat r}  \sqrt{\hat R(\rhat)} } = \sint \frac{\dd\cos\theta}{\pm_\theta \sqrt{\Theta(\cos^2\theta)}},\label{eqn:lambda}\\
\hat T_\theta(\lambda) &\triangleq& \sint \frac{\cos^2 \theta \,\dd\cos \theta}{\pm_\theta \sqrt{\Theta (\cos^2 \theta)}}, \qquad \hat \Phi_\theta(\lambda) \triangleq  \sint \frac{ \dd\cos\theta}{\pm_\theta \sqrt{\Theta (\cos^2\theta)}}\qty(\csc^2 \theta-1), \label{eqn:polarIntegrals}\\
\hat T_{\hat r}(\lambda) &\triangleq & \sint \frac{(\hat r^2 + a^2)P(\hat r) \,\dd\hat r}{\pm_{\hat r}  \sqrt{\hat R(\rhat)} \Delta(\rhat)},\qquad \hat \Phi_{\hat r}(\lambda) \triangleq \sint \frac{P(\hat r) \,\dd\hat r}{\pm_{\hat r}  \sqrt{\hat R(\rhat)} \Delta(\rhat)}. 
\eea
The notation $\sintline$ indicates that the signs $\pm_r$ and $\pm_\theta$ are flipped each time a zero of $\hat R$ and $\Theta$, respectively, is encountered. Since the signs $\pm_r$, $\pm_\theta$ are identical to the signs of $\dd r$ and $\dd\theta$, respectively, the integral \eqref{eqn:lambda} is monotonic around each turning point, as it should be in order to define an increasing Mino time $\lambda$ along the geodesic. It is the same notation used in Ref. \cite{Kapec:2019hro}. Note that $\hat{T}_\theta,~\hat\Phi_\theta$ are normalized to be vanishing for equatorial motion. The initial signs $s_r^i \equiv \pm_{\hat r}|_{\lambda = \lambda_i}$, $s_\theta^i \equiv \pm_{\theta}|_{\lambda = \lambda_i}$, as well as the initial spacetime position, are fixed as a part of the specification of the orbit. If we denote by $w(\lambda),m(\lambda)$ the number of turning points in the radial and polar motion, respectively, at Mino time $\lambda$, then as the velocity changes sign at each turning point,
\bea
\pm_r = s_r^i (-1)^w, \qquad \pm_\theta = s_\theta^i (-1)^m.\label{signs}
\eea

\subsection{Taxonomy of polar geodesic motion in Kerr}\label{sec:polarKerr}

\begin{figure}[!ph]\vspace{0pt}
    \centering\vspace{-70pt}
    \includegraphics[width=11cm]{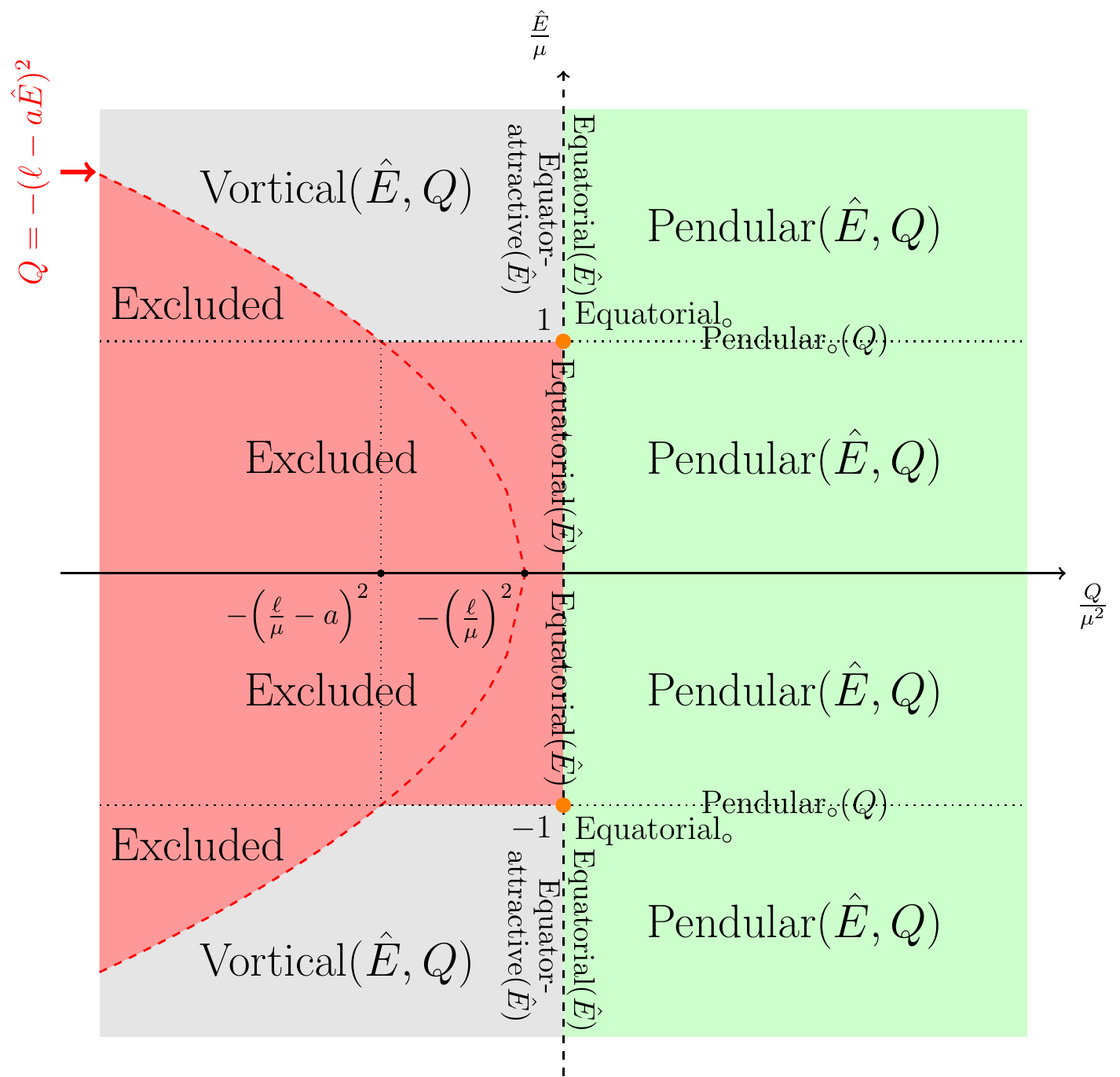}
    \caption{Polar taxonomy of $\ell\neq 0$ Kerr geodesics. Equator-attractive$(\hat{E})$ orbits become Equatorial$(\hat{E})$ orbits when the initial angle is at the equator. }\vspace{0pt}
    \label{fig:angularKerrTaxonomy}  
    \centering
    \includegraphics[width=11cm]{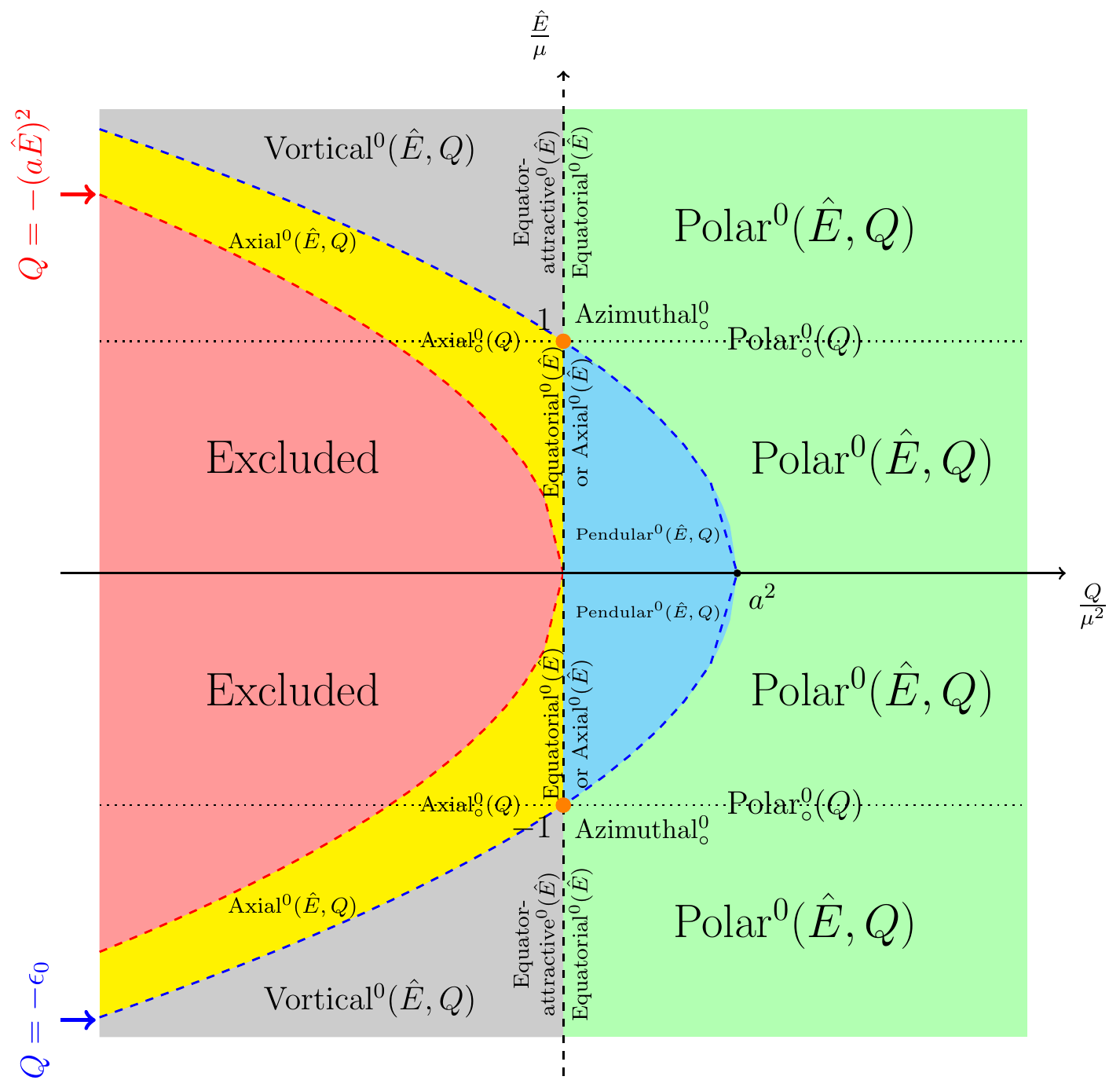}
    \caption{Polar taxonomy of $\ell= 0$ Kerr geodesics. In addition to the possible motions depicted in the figure, an axial motion is always possible for any value of $\epsilon_0$ and $Q\geq-(a\ehat)^2$.}
    \label{fig:angularNullKerrTaxonomy}\vspace{0pt}
\end{figure}

This section aims to describe in full detail the classification of the polar geodesic motion around a Kerr black hole. We will complete the classification given in the recent analysis \cite{Kapec:2019hro} to all geodesics including in particular $\ell = 0$\footnote{Note the conversion of notations $Q_{\text{here}}=Q_{\text{there}}$, $\ehat_{\text{here}} = \omega_{\text{there}}$, $G_\theta^{(\text{there})}=\lambda^{\text{here}}$, $G_\phi^{(\text{there})}=\hat \Phi^{(\text{here})}_\theta$, $G_t^{(\text{there})}=\hat T^{(\text{here})}_\theta$. In addition, we name Type A as Pendular and Type B as Vortical.}. Our taxonomy is depicted in Figs. \ref{fig:angularKerrTaxonomy} and \ref{fig:angularNullKerrTaxonomy}. The phenomenology of the polar behavior is principally governed by the sign of $Q$. The details of the classification are summarized in Tables \ref{table:taxPolarKerr1} and \ref{table:taxPolarKerrNull}. 

Due to mirror symmetry between the two hemispheres, the natural variable to describe the polar motion is $z\triangleq\cos^2\theta$. Let us define
\begin{equation}
    \eps_0(\hat E,\mu)  \triangleq a^2 (\hat E^2 - \mu^2), \label{defeps0}
\end{equation}
and
\begin{equation}\label{defzpm}
    z_\pm\triangleq\Delta_\theta\pm\sign {\epsilon_0} \sqrt{\Delta_\theta^2+\frac{Q}{\epsilon_0}},\qquad\Delta_\theta\triangleq\frac{1}{2}\qty(1-\frac{Q+\ell^2}{\epsilon_0}),\qquad z_0\triangleq\frac{Q}{Q+\ell^2}. 
\end{equation}

The classification is determined by the roots of the polar potential \eqref{eq:kerr_vtheta}. Assuming $a \neq 0$, we can rewrite it as
\begin{equation}
    \Theta(z)=-\ell^2 z+(Q+\eps_0 z)(1-z)= \left\lbrace\begin{array}{ll}
        \epsilon_0(z_+-z)(z-z_-), & \epsilon_0\neq 0 ;\\
        (Q+\ell^2)(z_0-z), & \epsilon_0=0.
    \end{array}\right.
\end{equation}

Our definition of the roots $z_\pm$ implies the ordering $z_-<z_+$ (and respectively $z_+<z_-$) for $\eps_0>0$ (respectively $\eps_0<0$). This is a convenient convention because in both cases the maximal angle will be related to $z_+$. The positivity of the polar potential implies that the poles $z=1$ ($\theta=0,\pi$) can only be reached if $\ell=0$. Note that when a geodesic crosses a pole, its $\hat\varphi$ coordinates discontinuously jump by $\pi$. The invariance of the polar geodesic equation under $(\hat{E},\ell)\to(-\hat{E},-\ell)$ allows us to reduce the analysis to prograde $\ell\geq 0$ orbits. We distinguish the orbits with angular momentum $\ell \neq 0$ and without, $\ell = 0$:

\vspace{4pt}
\noindent
\textit{\underline{I. Nonvanishing angular momentum $\ell\neq0$.}} We must consider the following cases:
\begin{enumerate}
    \item \underline{$-(\ell-a\hat{E})^2\leq Q<0$} can only occur if $\eps_0>0$, otherwise leading to $\Theta<0$. For $\eps_0>0$, the motion is \textit{vortical}; \textit{i.e.}, it takes place only in one of the two hemispheres without crossing the equatorial plane and is bounded by
    \begin{equation}
        0<z_-\leq z\leq z_+<1.
    \end{equation}
This vortical motion can only occur provided $\ell^2\leq\qty(\sqrt{\eps_0}-\sqrt{-Q})^2$.

    \item \underline{$Q>0$} leads to motion crossing the equator and symmetric with respect to it, bounded by
    \begin{align}
       0 &\leq z \leq z_+<1\qquad(\eps_0\neq0),\\
        0 & \leq z \leq z_0<1\qquad(\eps_0=0).
    \end{align}
    We will refer to such a motion as \textit{pendular}; 
    
    \item \underline{$Q=0$} allows us to write 
    \begin{equation}
        \Theta(z)=\eps_0\, z\, (1-\frac{\ell^2}{\eps_0}-z).
    \end{equation}
    If $\eps_0\leq 0$, the positivity of the polar potential enforces the motion to be equatorial. For $\eps_0\geq 0 $, equatorial motion exists at $z=0$. For $\eps_0\geq 0 $  and $\ell^2 \leq \eps_0$, another motion exists bounded by
    \begin{equation}
        0 < z \leq 1-\frac{\ell^2}{\eps_0}<1,
    \end{equation}
    which is a marginal case separating the pendular and vortical regimes; the motion then admits only one turning point and asymptotes to the equator both at future and at past times. Since we could not find a terminology for such a motion in the literature, we propose to call it \emph{equator-attractive}\footnote{This neologism accurately reflects the fact that the motion is polar and that the equator is an attractor. The terminology ``homoclinic'' is already used in the literature to refer to radial motion.}. In the special case where $z=0$ at the initial time, the motion remains $z=0$ at all times: it is \emph{equatorial}. 
\end{enumerate}

\vspace{4pt}
\noindent
\textit{\underline{II. Vanishing angular momentum $\ell=0$.}} The polar potential reduces to
\begin{equation}
\Theta(z)=\left\lbrace\begin{array}{ll}
\epsilon_0\qty(\frac{Q}{\epsilon_0}+z)\qty(1-z), & \epsilon_0\neq 0\\
Q\qty(1-z), & \epsilon_0=0. 
\end{array}\right.
\end{equation}
We distinguish the following cases:
\begin{enumerate}
    \item \underline{$\eps_0=0$} leads to motion over the whole polar range $0\leq z \leq 1$ for $Q>0$; we called it \textit{polar} motion. The only turning point is located at $z=1$. For $Q=0$, the potential vanishes identically and the polar angle remains constant; we call it \textit{azimuthal} motion; for $Q<0$ the potential is positive only if the motion takes place along the black hole axis $z=1$; we call it \textit{axial} motion.
    \item \underline{$\eps_0>0$} leads to a \textit{polar} motion $0\leq z\leq1$ for {$Q>0$}. For $Q=0$ and $z=0$, the motion is \textit{equatorial}. For $Q=0$ and $z\neq0$, $z=0$ is an asymptotic attractor of the motion which only takes place in one of the hemispheres. It is therefore a special case of \textit{equator-attractive} motion where the turning point is at the pole $z=1$. For {$Q<0$}, the motion is either \textit{vortical} ({$0<-\frac{Q}{\epsilon_0}\leq z \leq 1$}) for {$-\epsilon_0<Q<0$} or \textit{axial} with $z=1$ for {$Q\leq-\epsilon_0<0$}.
    \item \underline{$\eps_0<0$} leads to a \textit{polar} motion $0\leq z\leq1$ for {$Q\geq-\epsilon_0>0$} and to a \textit{pendular} one ({$0\leq z \leq -\frac{Q}{\epsilon_0}<1$}) for {$0<Q<-\epsilon_0$}. For $Q=0$, the motion is either \textit{equatorial} or \textit{axial} for the potential to be positive. For {$Q<0$}, the motion also has to take place along the axis.
\end{enumerate}
Let us finally notice that, for any value of $\epsilon_0$ and $Q\geq -(a\ehat)^2$, an axial motion is always possible.

 \begin{table}[!hbt]
    \centering
    \begin{tabular}{|c|c|c|c|}
    \hline
        \textbf{Energy} & \textbf{Carter constant} & \textbf{Polar range} & \textbf{Denomination}\\\hline
        $\eps_0<0$ ($|\hat{E}|<\mu$) & $-(\ell-a\hat{E})^2\leq Q<0$ & $\emptyset$ &\rule{0pt}{13pt} $\emptyset$\\\cline{2-4}
         & $Q=0$ & $z=0$ &\rule{0pt}{13pt} Equatorial$(\hat{E})$\\\cline{2-4}
          & $Q>0$ & $0\leq z\leq z_+<1$ & \rule{0pt}{13pt}Pendular$(\hat{E},Q)$\\\hline
           $\eps_0=0$ ($|\hat{E}|=\mu$) & $-(\ell-a\hat{E})^2\leq Q<0$ & $\emptyset$ & \rule{0pt}{13pt}$\emptyset$\\\cline{2-4}
         & $Q=0$ & $z=0$ &\rule{0pt}{13pt} Equatorial${}_\circ$\\\cline{2-4}
          & $Q>0$ & $0\leq z\leq z_0<1$ &\rule{0pt}{13pt} Pendular${}_\circ(Q)$\\\hline
           $\eps_0>0$ ($|\hat{E}|>\mu$) & $-(\ell-a\hat{E})^2\leq Q<0$ & $0<z_-\leq z \leq z_+<1$ &\rule{0pt}{13pt} Vortical$(\hat{E},Q)$\\\cline{2-4}
         & $Q=0$ and $\eps_0\geq\ell^2$ & \begin{tabular}{c} \rule{0pt}{13pt}$0 < z \leq 1-\frac{\ell^2}{\eps_0} <1$\\ ($\sign{\cos\theta}$ fixed)\end{tabular} &\begin{tabular}{c}\rule{0pt}{13pt} Equator-\\attractive$(\hat{E})$\end{tabular}\\\cline{3-4}
        & &$z=0$&Equatorial$(\hat{E})$  \\\cline{2-4}
               & $Q>0$ & $0\leq z\leq z_+<1$ &\rule{0pt}{13pt} Pendular$(\hat{E},Q)$\\\hline
         \end{tabular}
       \caption{Polar taxonomy of Kerr geodesics  with $\ell\geq 0$. The orbits with $\ell<0$ are obtained from $\ell>0$ by flipping the signs of both $\hat{E}$ and $\ell$.\\ }
    \label{table:taxPolarKerr1}
\end{table}

\vspace{-0pt}
\begin{table}[!h]
    \centering
        \begin{tabular}{|c|c|c|c|}
\hline
        \textbf{Energy} &\textbf{ Carter constant} & \textbf{Polar range} & \textbf{Denomination}\\\hline
        $\eps_0<0$ ($|\hat{E}|<\mu$) & $ -(a \hat E)^2 \leq Q<0$ & $z=1$ &\rule{0pt}{13pt} Axial${}^0(\hat{E},Q)$\\\cline{2-4}
        & $Q=0$ & $z=0,1$ &\rule{0pt}{13pt} \begin{tabular}{c}\rule{0pt}{13pt}Equatorial${}^0(\hat{E})$\\ Axial${}^0(\hat{E})$\end{tabular}\\\cline{2-4}
         & $0<Q<-\eps_0$ & $0\leq z\leq- \frac{Q}{\eps_0}<1$ &\rule{0pt}{13pt} Pendular${}^0(\hat{E},Q)$\\\cline{2-4}
          & $0<-\epsilon_0\leq Q$ & $0\leq z\leq 1$ & \rule{0pt}{13pt}Polar${}^0(\hat{E},Q)$\\\hline
           $\eps_0=0$ ($|\hat{E}|=\mu$) & $-(a\hat{E})^2\leq Q<0$ & $z=1$ & \rule{0pt}{13pt}Axial${}^0_\circ(Q)$\\\cline{2-4}
         & $Q=0$ & $z=constant$ &\rule{0pt}{13pt} Azimuthal${}^0_\circ$\\\cline{2-4}
          & $Q>0$ & $0\leq z\leq 1$ &\rule{0pt}{13pt} Polar${}^0_\circ(Q)$\\\hline
           $\eps_0>0$ ($|\hat{E}|>\mu$) & $ -(a \hat E)^2 \leq Q\leq -\eps_0<0$ & $z=1$ &\rule{0pt}{13pt} Axial${}^0(\hat{E},Q)$\\\cline{2-4}
           & $-\eps_0<Q<0$ & $0<-\frac{Q}{\eps_0}\leq z\leq 1$ &\rule{0pt}{13pt} Vortical${}^0(\hat{E},Q)$\\\cline{2-4}
         & $Q=0$  & $0 < z \leq  1$ ($\sign{\cos\theta}$ fixed)&\rule{0pt}{13pt} \begin{tabular}{c}Equator-\\attractive${}^0(\hat{E})$\end{tabular}\\\cline{3-4}
         &   & $z=0$ & \rule{0pt}{13pt}Equatorial${}^0(\hat{E})$ \\\cline{2-4}
                        
           & $Q>0$ & $0\leq z\leq1$ & \rule{0pt}{13pt}Polar${}^0(\hat{E},Q)$\\\hline
\rule{0 pt}{13 pt}$\eps_0\in\mathbb{R}$ & $Q\geq - (a \hat E)^2$ & $z=1$ & Axial$^0(\ehat,Q)$\\\hline
\end{tabular}
    \caption{Polar taxonomy of Kerr  geodesics  with $\ell =  0$.\vspace{20pt}}
    \label{table:taxPolarKerrNull}
\end{table}

\clearpage

\subsection{Solution to the polar integrals} 

After having classified the different types of motion allowed, we will provide manifestly real and positive explicit solutions in terms of elliptic integrals for each type of polar motion with $\ell \neq 0$ in line with the recent analysis \cite{Kapec:2019hro}. All such integrals will turn out to agree with Ref. \cite{Kapec:2019hro}, but our presentation will be slightly simpler.

The solution to the polar integrals \eqref{eqn:lambda} and \eqref{eqn:polarIntegrals} can be organized in terms of the categories of polar motion with $\ell \neq 0$: 
\begin{center}
    \begin{tabular}{|c|c|c|c|}
    \cline{2-4}
        \multicolumn{1}{c|}{ } & \textbf{Vortical} & \textbf{Equator-attractive} & \textbf{Pendular} \\\hline
       \rule{0pt}{13pt} $\mathbf{\eps_0<0}$ & $\emptyset$ & $\emptyset$ & Pendular$(\hat E,Q)$\\\hline
       \rule{0pt}{13pt} $\mathbf{\eps_0=0}$ & $\emptyset$ & $\emptyset$ & Pendular${}_*(Q)$\\\hline
       \rule{0pt}{13pt} $\mathbf{\eps_0>0}$ & Vortical$(\hat E,Q)$ & Equator-attractive$(\hat E)$ & Pendular$(\hat E,Q)$\\\hline
    \end{tabular}
\end{center}

Each type of motion yields to a specific decomposition of the line integrals $\sintline$ in terms of basic integrals. In order to simplify the notations, we drop the ``$f$" indices labeling the final event and define $h \equiv \sign{ \cos\theta}$, $\theta_a \triangleq \arccos \sqrt{z_a}$ ($a=+,-,0$), as well as the initial and final signs $\eta_i$, $\eta$:
\bea
\eta_i \triangleq   - s_\theta^i \sign{\cos\theta_i}  ,\qquad \eta \triangleq  - (-1)^m s_\theta^i \sign{\cos\theta }.
\eea
We are now ready to perform the explicit decomposition:
\begin{enumerate}
    \item \underline{Pendular motion.}  We have $0< z_+ \leq 1$, and $\theta$ therefore belongs to the interval $\theta_+ \leq \theta \leq \pi - \theta_+$. The polar integral can be written (see Ref. \cite{Kapec:2019hro}) 
\begin{align}
\sint_{\cos \theta_i}^{\cos \theta} &= 2 m \left| \int_{0}^{\cos\theta_+} \right|  -\eta \left| \int_0^{\cos\theta}\right|+\eta_i \left| \int_0^{\cos \theta_i} \right|,\qquad\eps_0\neq 0,\label{int1}\\
\sint_{\cos \theta_i}^{\cos \theta} &= 2 m \left| \int_{0}^{\cos\theta_0} \right|  -\eta \left| \int_0^{\cos\theta}\right|+\eta_i \left| \int_0^{\cos \theta_i} \right|,\qquad\eps_0= 0.
\end{align}
It is useful to note that our definitions of the roots imply
\begin{equation}
    \eps_0\,z_-<0,\qquad\eps_0(z-z_-)>0,\qquad\frac{z_+}{z_-}\leq 1.
\end{equation}

    \item \underline{Vortical motion.} We have $\eps_0>0$ and $0 < z_-\leq \cos^2\theta \leq z_+ < 1$. The motion therefore never reaches the equator. The sign of $\cos\theta$ is constant and determines whether the motion takes place in the northern or the southern hemisphere. Without loss of generality, let us focus on the northern hemisphere: $0 \leq \theta_+\leq \theta \leq \theta_- < \frac{\pi}{2}$; we denote again as $m$ the number of turning points at Mino time $\lambda$. The polar integral can be written (see Ref. \cite{Kapec:2019hro} and Appendix A of Ref. \cite{Gralla:2019ceu}):
\bea
\sint_{\cos \theta_i}^{\cos \theta} = \left( m -\eta_i \frac{1-(-1)^m}{2} \right) \left| \int_{\cos\theta_-}^{\cos\theta_+} \right|  -\eta  \left| \int_{\cos\theta_-}^{\cos\theta} \right|+ \eta_i \left| \int_{\cos\theta_-}^{\cos \theta_i} \right| .\label{int2}
\eea
    
    \item \underline{Equator-attractive motion.} This is a limit case of the vortical motion reached in the limit $z_-\to 0$, $z_+\to 2\Delta_\theta$. As detailed in Ref. \cite{Kapec:2019hro}, the turning point $z_-=0$ corresponds to a nonintegrable singularity of the polar integrals and the motion exhibits consequently at most one turning point at $z_+=2\Delta_\theta$, leading to the line-integral decomposition
    \begin{equation}
    \sint_{\cos \theta_i}^{\cos \theta} = \eta\abs{\int_{\cos\theta_+}^{\cos\theta}}-\eta_i\abs{\int_{\cos\theta_+}^{\cos\theta_i}}.
    \end{equation}
\end{enumerate}
In all cases but the equator-attractive case, the polar motion is periodic. Denoting by $\Lambda_\theta$ its period, one can easily give an explicit formula for the number of turning points $m$ as a function of the Mino time:
\begin{equation}
    m(\lambda)=\left\lbrace\begin{array}{ll}
    \rule{0pt}{6pt}\left\lfloor \frac{2}{\Lambda_\theta}(\lambda-\lambda_i^\theta)+\frac{1}{2}\right\rfloor, & Q>0\\
    \rule{0pt}{16pt}\left\lfloor \frac{2}{\Lambda_\theta}(\lambda-\lambda_i^\theta)\right\rfloor + \left\lfloor \frac{2}{\Lambda_\theta}(\lambda_i^\theta-\lambda_i)\right\rfloor + \frac{3-s^i_\theta}{2}, & Q<0
    \end{array}\right. 
\end{equation}
with $\lambda_i^\theta\triangleq\lambda_i-s^i_\theta\int_{0}^{\cos\theta_i}\frac{\dd \cos \theta}{\sqrt{\Theta(\cos^2 \theta)}}$ and where the floor function is defined as $\left\lfloor x \right\rfloor\triangleq\max\qty{n\in\mathbb{Z}|n\leq x}$. For the equator-attractive case, one has simply $m(\lambda)=\theta(\lambda-\lambda_i^\theta)$ where $\theta$ is here the Heaviside step function.

\begin{figure}[!htb]
    \centering
    \begin{tabular}{cc}
      \includegraphics[width=4cm]{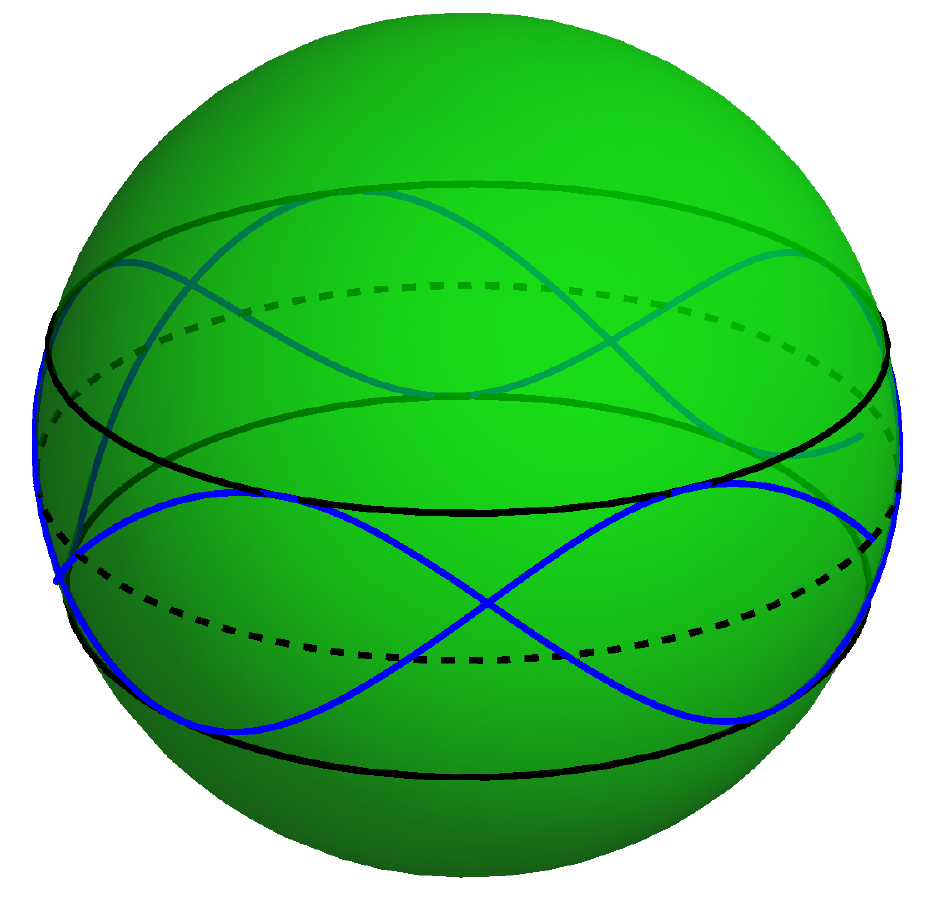} & \includegraphics[width=4cm]{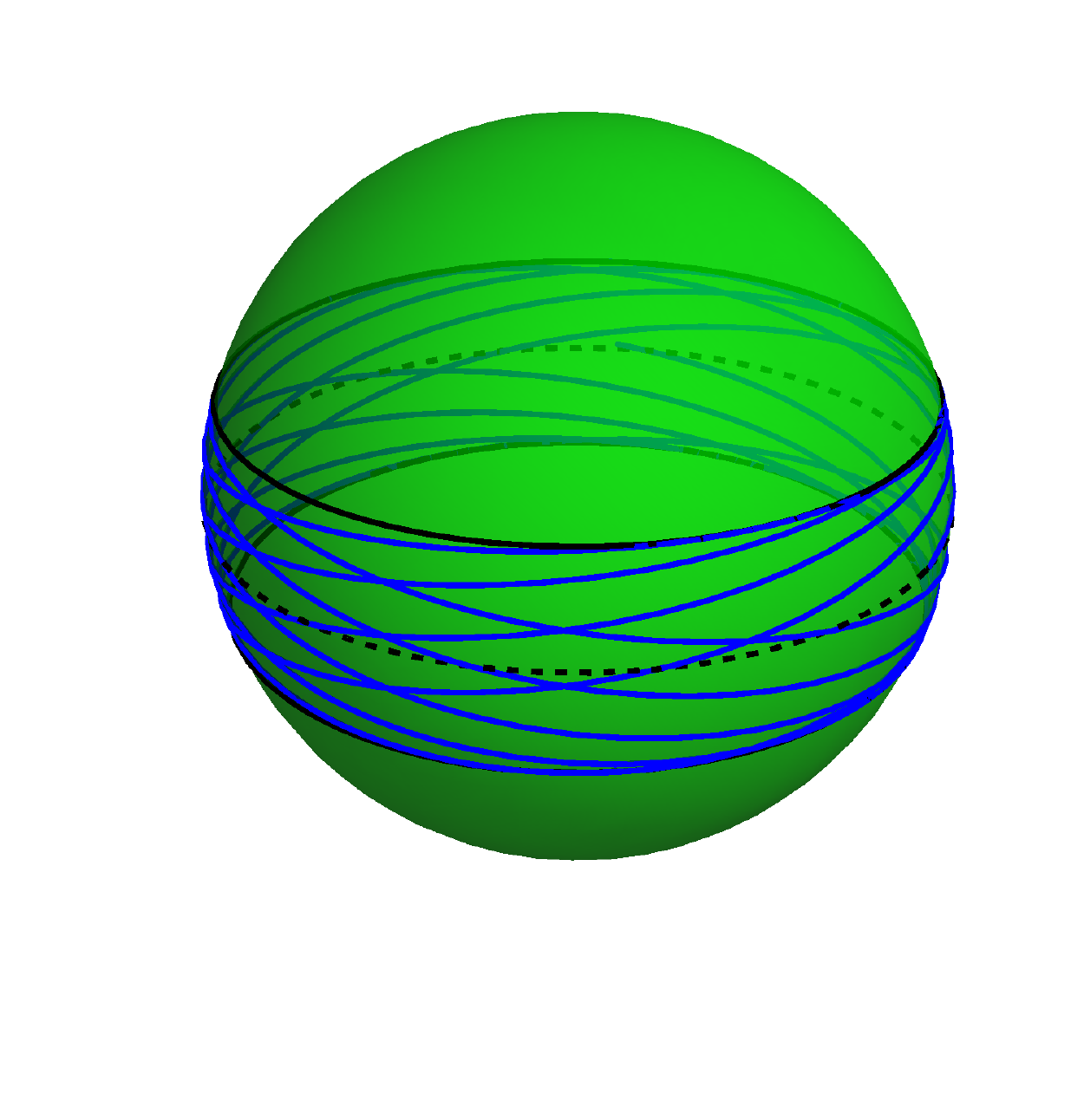} \\ (a) Pendular$(\hat E, Q)$ & (b) Pendular${}_\circ(Q)$\vspace{0.6cm}\\
        \includegraphics[width=4cm]{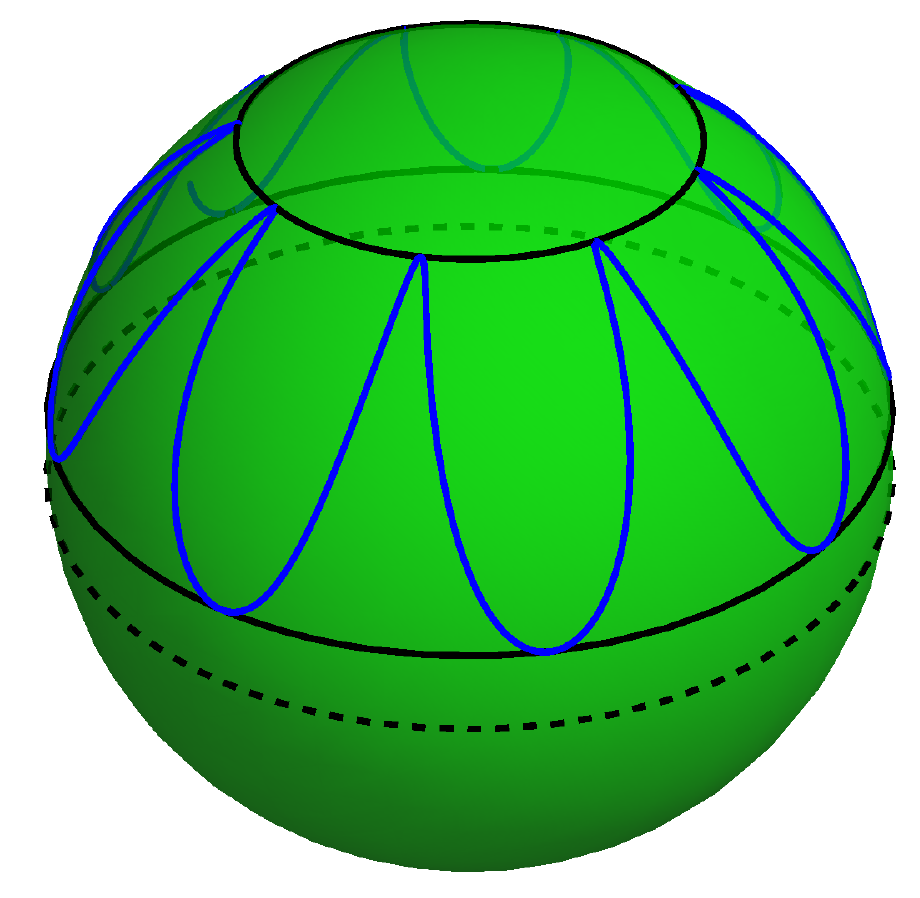} & \includegraphics[width=4cm]{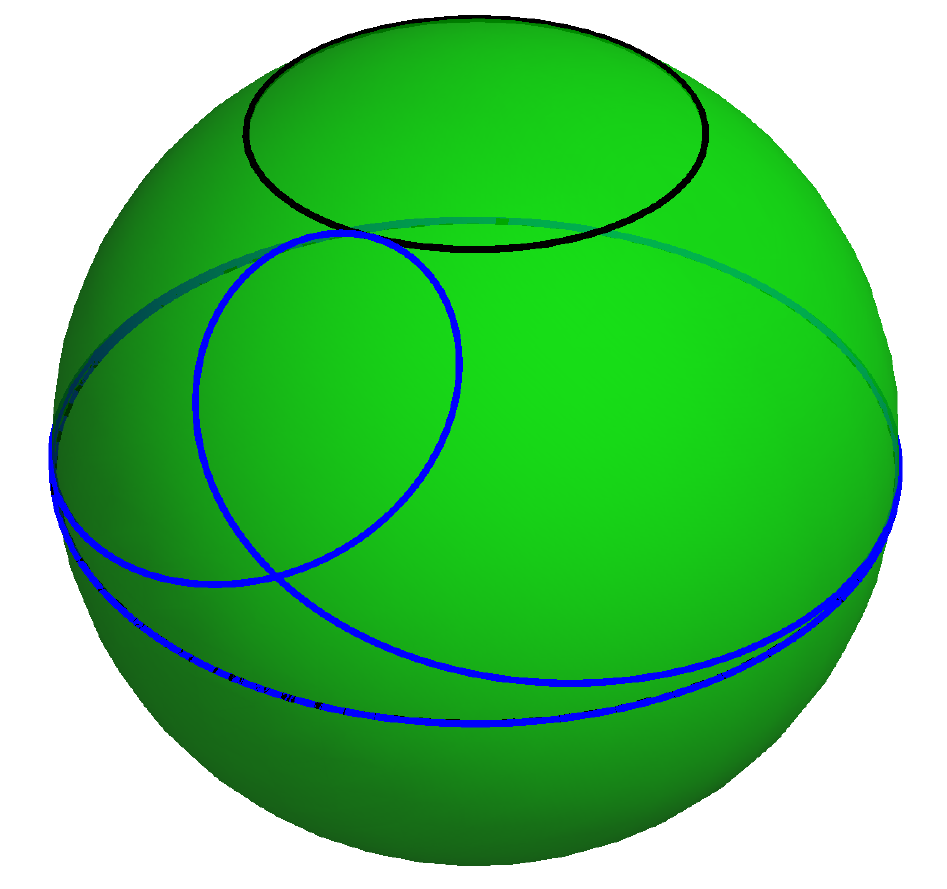} \\ (c) Vortical$(\hat E,Q)$ & (d) Equator-attractive$(\hat E)$
    \end{tabular}
    \caption{Angular taxonomy of $\ell\neq 0$ Kerr geodesics. The angular behavior is depicted in spherical coordinates on the unit sphere: the polar angle is $\theta(\lambda)$, and the azimuthal angle is the purely angular part of the Kerr azimuthal angle $(\ell- a \hat E) (\lambda-\lambda_i) + \ell \hat \Phi_\theta (\lambda)$.}
    \label{fig:polarClassesFigure}
\end{figure}

The integrals introduced above are solved explicitly in Appendix \ref{app:basicIntegrals}. For each case, the corresponding solutions are detailed below and schematically depicted in Fig. \ref{fig:polarClassesFigure}.

\paragraph{Pendular$(\hat E,Q)$ motion.} The motion exhibits a positive Carter constant $Q$ and can occur for any $\eps_0\neq 0$; our definition of the roots $z_\pm$ allows us to treat simultaneously the two cases $\eps_0<0$ and $\eps_0>0$, which is a simplification with respect to the analysis carried out in Ref. \cite{Kapec:2019hro}. The period of the polar motion (comprising two turning points) in Mino time is given by 
\begin{equation}
\Lambda_\theta = 4  \int_0^{\cos\theta_+} \frac{\text{d} \cos\theta}{\sqrt{\Theta(\cos^2\theta)}}\triangleq 4\hat I^{(0)}(\sqrt{z_+}) =\frac{4}{\sqrt{-\epsilon_0z_-}}K\qty(\frac{z_+}{z_-}).
\end{equation}
Using the basic integrals of Appendix \ref{app:ellipticFunctions}, one can write \eqref{eqn:lambda} as
\begin{align}
    \lambda-\lambda_i&=\frac{1}{\sqrt{-\eps_0z_-}}\left[2mK\qty(\frac{z_+}{z_-})+s^i_\theta(-1)^mF\qty(\Psi^+(\cos\theta),\frac{z_+}{z_-})\right.\nonumber\\
    &~\left.-s_\theta^iF\qty(\Psi^+(\cos\theta_i),\frac{z_+}{z_-})\right]\label{eqn:pendularLambda}
\end{align}
where we define $\Psi^+(x)\triangleq\arcsin\qty(\frac{x}{\sqrt{z_+}})$. 
Using \eqref{eqn:inversionEpsNeq0}, one can invert \eqref{eqn:pendularLambda} as
\begin{equation}
    \cos\theta=s^i_\theta(-1)^m\sqrt{z_+}\sn \qty(\sqrt{-\eps_0 z_-}\qty(\lambda-\lambda_i^\theta)-2mK\qty(\frac{z_+}{z_-}),\frac{z_+}{z_-})
\end{equation}
where we introduce
\begin{align}
        \lambda_i^\theta&\triangleq\lambda_i-\frac{s^i_\theta}{\sqrt{-\eps_0z_-}}F\qty(\Psi^+(\cos\theta_i),\frac{z_+}{z_-}).
\end{align}
This expression matches with Eq.(38) of Ref. \cite{Fujita:2009bp}. Using the periodicity property \eqref{per} of the elliptic sine, we can further simplify it to 
\begin{equation}
\cos\theta(\lambda)=s^i_\theta \sqrt{z_+}\sn\qty(  \sqrt{-\epsilon_0 z_-}(\lambda-\lambda_i^\theta), \frac{z_+}{z_-}).\label{eq:costh}
\end{equation}
It consistently obeys $\cos\theta(\lambda_i)=\cos\theta_i$ and $\sign{\cos\theta'(\lambda_i)}=s_\theta^i$. This formula agrees with (53) of Ref. \cite{Kapec:2019hro} but it is written in a simpler form. We also obtain
\begin{align}
\hat T_\theta &=\frac{-2z_+}{\sqrt{-\eps_0 z_-}} \left[ 2 m E' \qty(z_+,\frac{z_+}{z_-}) +( \pm_\theta) E'\qty(z_+,\Psi^+(\cos\theta),\frac{z_+}{z_-})\right.\nonumber\\
&\left.~- s_\theta^i  E'\qty(z_+, \Psi^+(\cos\theta_i),\frac{z_+}{z_-} )\right], 
\end{align}
\begin{align}
\hat\Phi_\theta &=  \frac{1}{\sqrt{-\eps_0 z_-}} \left[ 2 m \Pi \qty(z_+,\frac{z_+}{z_-}) +( \pm_\theta) \Pi\qty(z_+,\Psi^+(\cos\theta),\frac{z_+}{z_-})\right.\nonumber\\
&~\left.- s_\theta^i  \Pi\qty(z_+,\Psi^+(\cos\theta_i) ,\frac{z_+}{z_-})\right]-(\lambda-\lambda_i).
\end{align}
where $\lambda-\lambda_i$ is given by \eqref{eqn:pendularLambda}. All quantities involved are manifestly real. These final expressions agree with Ref. \cite{Kapec:2019hro}.

\paragraph{Pendular${}_\circ(Q)$ motion.} We now consider the critical case $|\hat E| = \mu$. The period of the polar motion is
\begin{equation}
    \Lambda_\theta=4\hat I^{(0)}(\sqrt{z_0})=2\pi\sqrt{\frac{z_0}{Q}}.
\end{equation}
In this critical case, \eqref{eqn:lambda} leads to 
\begin{equation}
    \lambda-\lambda_i=\sqrt{\frac{z_0}{Q}}\qty[m\pi+s^i_\theta(-1)^m\arcsin{\frac{\cos\theta}{\sqrt{z_0}}}-s^i_\theta\arcsin{\frac{\cos\theta_i}{\sqrt{z_0}}}],
\end{equation}
which can be simply inverted as
\begin{equation}
    \cos\theta=s_\theta^i\sqrt{z_0}\,\sin\qty(\sqrt{\frac{Q}{z_0}}(\lambda-\lambda_i^\theta)),\qquad\lambda_i^\theta\triangleq\lambda_i-\sqrt{\frac{z_0}{Q}}\arcsin{\frac{\cos\theta_i}{\sqrt{z_0}}}.
\end{equation}
The other polar integrals are
\begin{align}
    \hat T_\theta&=\frac{1}{2}\qty{z_0(\lambda-\lambda_i)-\sqrt{\frac{z_0}{Q}}\qty[(\pm_\theta)\cos\theta\sqrt{z_0-\cos^2\theta}-s^i_\theta\cos\theta_i\sqrt{z_0-\cos^2\theta_i}]},\\
    \hat\Phi_\theta&=\sqrt{\frac{z_0}{Q(1-z_0)}}\left[m\pi+(\pm_\theta)\arcsin\qty(\sqrt{\frac{1-z_0}{z_0}}\cot\theta)-s^i_\theta\arcsin\qty(\sqrt{\frac{1-z_0}{z_0}}\cot\theta_i)\right]\nonumber\\
    &~-(\lambda-\lambda_i).
\end{align}
 
\paragraph{Vortical$(\hat E,Q)$ motion.} 
The period in Mino time is given by 
\begin{equation}
\Lambda_\theta = 2 \left| \int_{\cos\theta_-}^{\cos\theta_+} \frac{\text{d} \cos\theta}{\sqrt{\Theta(\cos^2\theta)}} \right|=\frac{2}{\sqrt{\eps_0 z_+}} K\qty(1-\frac{z_-}{z_+}).
\end{equation}
Using the basic integrals of Appendix \ref{app:basicIntegrals}, one has
\begin{align}
    \lambda-\lambda_i&=\frac{1}{\sqrt{\eps_0z_+}}\left[\qty(m-h s^i_\theta\frac{1-(-1)^m}{2})K(\tilde m)-s^i_\theta(-1)^mF\qty(\Psi^-(\cos\theta),\tilde{m})\right.\nonumber\\
    &~\left.+s^i_\theta F\qty(\Psi^-(\cos\theta_i),\tilde m)\right]
\end{align}
where
\begin{equation}
    \tilde{m}\triangleq 1-\frac{z_-}{  z_+},\qquad\Psi^-(x)=\arcsin{\sqrt{\frac{z_+-x^2}{z_+-z_-}}}.
\end{equation}
Using the inversion formula \eqref{eqn:inversionFormula} and the periodicity property \eqref{eqn:perDn}, we obtain
\begin{equation}
    \cos\theta=h\sqrt{z_+}\dn\qty(\sqrt{\eps_0z_+}(\lambda-\lambda_\theta^i),\tilde m)
\end{equation}
with
\begin{equation}
    \lambda_i^\theta\triangleq\lambda_i+\frac{s^i_\theta h}{\sqrt{\eps_0z_+}}F\qty(\Psi^-(\cos\theta_i),\tilde m).
\end{equation}
Again, one has $\cos\theta(\lambda_i)=\cos\theta_i$ and $\sign{\cos\theta'(\lambda_i)}=s_\theta^i$. The two other polar integrals are
\begin{align}
    \hat T_\theta&=\sqrt{\frac{z_+}{\epsilon_0}}\left[\qty(m-h s^i_\theta\frac{1-(-1)^m}{2})E(\tilde m)-(\pm_\theta)E\qty(\Psi^-(\cos\theta),\tilde{m})\right.\nonumber\\
    &~\left.+s^i_\theta E\qty(\Psi^-(\cos\theta_i),\tilde m)\right],\\
    \hat \Phi_\theta&=\frac{1}{(1-z_+)\sqrt{\eps_0z_+}}\left[\qty(m-h s^i_\theta\frac{1-(-1)^m}{2})\Pi\qty(\frac{z_--z_+}{1-z_+},\tilde m)\right.\nonumber\\&~\left.-(\pm_\theta)\Pi\qty(\frac{z_--z_+}{1-z_+},\Psi^-(\cos\theta),\tilde{m})+s^i_\theta \Pi\qty(\frac{z_--z_+}{1-z_+},\Psi^-(\cos\theta_i),\tilde m)\right]\nonumber\\& ~-(\lambda-\lambda_i)
\end{align}
in agreement with the results of Ref. \cite{Kapec:2019hro}.

\paragraph{Equator-attractive$(\hat E)$ motion.}
This is the only polar motion which is not periodic. One has
\begin{equation}
    \lambda-\lambda_i=\frac{h}{\sqrt{\eps_0z_+}}\qty[-(\pm_\theta)\,\arctanh\sqrt{1-\frac{\cos^2\theta}{z_+}}+s^i_\theta\,\arctanh\sqrt{1-\frac{\cos^2\theta_i}{z_+}}]
\end{equation}
leading to
\begin{equation}
    \cos\theta=h\sqrt{z_+}\,\text{sech}\qty(\sqrt{\eps_0z_+}(\lambda-\lambda_i^\theta)),\qquad\lambda_i^\theta\triangleq\lambda_i+\frac{s^i_\theta h}{\sqrt{\eps_0z_+}}\arctanh\sqrt{1-\frac{\cos^2\theta_i}{z_+}}.
\end{equation}
The polar integrals are
\begin{align}
    \hat T_\theta&=\frac{h}{\sqrt{\eps_0}}\qty[-(\pm_\theta)\sqrt{z_+-\cos^2\theta}+s^i_\theta\sqrt{z_+-\cos^2\theta_i}],\\
    \hat\Phi_\theta&=\frac{h}{\sqrt{\eps_0(1-z_+)}}\qty[-(\pm_\theta)\arctan\sqrt{\frac{z_+-\cos^2\theta}{1-z_+}}+s^i_\theta\arctan\sqrt{\frac{z_+-\cos^2\theta_i}{1-z_+}}].
\end{align}
This agrees with the results of Ref. \cite{Kapec:2019hro}.

\section{Near-horizon geodesics in the high-spin limit}
\label{sec:nhg}

In this section, we derive a complete classification of timelike and null geodesic trajectories lying in the near-horizon region of a quasiextremal Kerr black hole, the so-called \textit{near-horizon extremal Kerr} (NHEK) region. We will provide explicit manifestly real analytic expressions for all geodesic trajectories. We will present the classification in terms of the geodesic energy, angular momentum, and Carter constant $Q$. We will also illustrate each radial motion in NHEK with a Penrose diagram. 

Partial classifications were performed in Refs. \cite{Compere:2017hsi} and \cite{Kapec:2019hro}. In Ref. \cite{Compere:2017hsi}, equatorial timelike prograde incoming (i.e. that originate from the Kerr exterior geometry) geodesics were classified. Such geodesics reach the spatial boundary of the near-horizon region at infinite past proper time and therefore physically reach the asymptotically flat Kerr region once the near-horizon is glued back to the exterior Kerr region. It turns out that bounded geodesics in the near-horizon Kerr region also arise in the study of gravitational waves since they correspond to the end point of the transition motion \cite{Compere:2019cqe}. Timelike outgoing geodesics originating from the white hole horizon and reaching the near-horizon boundary are also relevant for particle emission within the near-horizon region \cite{Gralla:2017ufe}. In addition, null outgoing  geodesics are relevant for black hole imaging around high-spin black holes \cite{Gralla:2019drh}. 

The generic nonequatorial geodesics were obtained in Ref. \cite{Kapec:2019hro}. In particular, real forms were obtained for each angular integral involved in geodesic motion. However, zero-measure sets of parameters were discarded. These zero-measure sets include in particular the separatrix between bounded and unbounded radial motion which plays a key role in EMRIs.  

 In the following, we do not make any assumption on the geodesic parameters. We will treat both timelike and null geodesics, prograde or retrograde, and with any boundary conditions. Without loss of generality, we will consider future-directed orbits. Past-directed geodesics can be obtained from future-directed geodesics using the $\mathbb Z_2$ map: 
\bea
T \rightarrow -T,\quad \Phi \rightarrow -\Phi,\quad  E\rightarrow -E,\quad  \ell \rightarrow -\ell,\label{PTflip}
\eea
which will play an important role in Sec. \ref{sec:classes}. We will denote it as the $\uparrow\!\downarrow$-flip. We will compare our classification with \cite{Kapec:2019hro} in Appendix \ref{app:comparison}.

\subsection{Near-horizon extremal Kerr (NHEK)}
\label{sec:NHEK}

We now consider the near-extremal limit 
\begin{equation}
\lambda\triangleq\sqrt{1-\frac{a^2}{M^2}}\rightarrow 0
\end{equation}
combined with the near-horizon limit $\rhat\to\rhat_+=M(1+\lambda)$. Suitable coordinates in the near-horizon region are defined as (for a review, see Ref. \cite{Compere:2012jk})
\begin{align}
T&=\frac{\that}{2M}\lambda^{2/3},\qquad R =\frac{\rhat-\rhat_+}{M \lambda^{2/3}},\label{eq:cvn}\qquad
\Phi=\phihat-\frac{\that}{2M} .
\end{align}
Plugging \eqref{eq:cvn} into the Kerr metric \eqref{eq:kerr_bl} and expanding the result in powers of $\lambda$ gives the NHEK spacetime in Poincar\'e coordinates:
\begin{equation}
\dd s^2=2M^2\Gamma(\theta)\qty(-R^2\dd T^2+\frac{\dd R^2}{R^2}+\dd\theta^2+\Lambda^2(\theta)\qty(\dd\Phi+R\,\dd T)^2)+\mathcal{O}(\lambda^{2/3})\label{eq:NHEK_metric}
\end{equation}
where
\begin{align}
\Gamma(\theta)&\triangleq\frac{1+\cos^2\theta}{2},\qquad \Lambda(\theta) \triangleq\frac{2\sin\theta}{1+\cos^2\theta}.
\end{align}
The NHEK geometry admits a $SL(2,\mathbb R) \times U(1)$ symmetry generated by $\p_\Phi$ and 
\bea
H_0 = T \p_T - R \p_R,\qquad H_+ = \p_T,\qquad H_- = (T^2 + \frac{1}{R^2})\p_T -2 T R \p_R - \frac{2}{R}\p_\Phi. 
\eea
The Killing tensor $K_{\mu\nu}$ \eqref{Kmunu} becomes reducible \cite{AlZahrani:2010qb,Galajinsky:2010zy} and can be expressed as
\bea
K^{\mu\nu} = M^2 g^{\mu\nu} + \mathcal C^{\mu\nu} + (\p_\Phi)^\mu  (\p_\Phi)^\nu
\eea
where the $SL(2,\mathbb R)$ Casimir is given by
\bea
\mathcal C^{\mu\nu} \p_\mu \p_\nu = -H_0 H_0  + \frac{1}{2} (H_+  H_-  + H_- H_+).\label{Cas}
\eea

We are interested in the Kerr geodesics that exist in the near-extremal limit within the NHEK geometry at leading order in $\lambda$. The NHEK angular momentum $\ell$ and Carter constant $Q$ are identical to their values defined in Boyer-Lindquist coordinates. The NHEK energy is related to the Boyer-Lindquist energy $\hat E$ as
\begin{align}
\ehat&=\frac{\ell}{2M}+\frac{\lambda^{2/3}}{2M}E. \label{rel1}
\end{align}
From now on, we will consider the leading high-spin limit; i.e. we will neglect all $\mathcal O(\lambda^{2/3})$ corrections in \eqref{eq:NHEK_metric}.

\subsubsection{Geodesics}

In the NHEK geometry, Mino time is defined as $\lambda\triangleq\int^\tau\frac{\dd\tau'}{2M^2\Gamma(\theta(\tau'))}$, and the geodesic equations of motion simplify to 
\begin{align}
\dv{T}{\lambda}&=\frac{E}{R^2}+\frac{\ell}{R},\label{eq:nhekT}\\
\dv{R}{\lambda}&=\pm_R\sqrt{v_R(R)},\label{eq:nhekR}\\
\dv{\cos \theta}{\lambda}&=\pm_\theta \sqrt{v_\theta(\cos^2\theta)},\label{eq:nhektheta}\\
\dv{\Phi}{\lambda}&=\frac{\ell}{\Lambda^2}-\frac{E}{R}- \ell,\label{eq:nhekphi}
\end{align}
with
\begin{align}
v_R(R)&\triangleq E^2+2E \ell R+\frac{R^2}{4}\qty(3\ell^2-4(Q+\mu^2 M^2)),\label{eq:vR}\\
v_\theta(\cos^2\theta)&\triangleq Q \sin^2\theta +\cos^2\theta  \sin^2\theta \qty(\frac{\ell^2}{4}-M^2\mu^2 )-\ell^2\cos^2\theta\label{eq:vtheta}.
\end{align}
The limitation that $E$ remains real and finite implies from  \eqref{rel1} that we are only considering orbits with energy close to the extremal value $\ell/(2M)$. The ISSO angular momentum at extremality is denoted as 
\begin{empheq}[box=\ovalbox]{align}
\ell_* = \frac{\hat E_{\text{ISSO}}}{\Omega_{ext}} = \frac{2}{\sqrt{3}} \sqrt{M^2 \mu^2 + Q}\label{ls}
\end{empheq}
where the energy of the ISSO will be defined in \eqref{ISSOext}. It will play a key role in the following. On the equatorial plane ($Q=0$), the definition reduces to $\ell_*$ used in Refs. \cite{Compere:2017hsi,Chen:2019hac}. We can also write down more simply 
\bea
v_R(R)= E^2+2E \ell R-\mathcal C R^2
\eea
where $\mathcal C$ is the conserved quantity obtained from the $SL(2,\mathbb R)$ Casimir,
\bea
\mathcal C \triangleq \mathcal C^{\mu\nu} P_\mu P_\nu = Q - \frac{3}{4}\ell^2 + \mu^2 M^2 = \frac{3}{4} (\ell_*^2 -\ell^2).\label{eqn:Carter}
\eea
We also have 
\begin{equation}
v_R(R)=\left\lbrace\begin{array}{ll}
    -\mathcal C (R-R_+)(R-R_-), & \mathcal{C}\neq 0 ;\\
    2E\ell (R-R_0), & \mathcal{C}=0 ,
\end{array}\right.
\end{equation}
with
\bea
R_\pm \triangleq \frac{E}{\mathcal C} \ell \pm \frac{\vert E \vert }{\vert \mathcal C\vert } \sqrt{\ell^2+\mathcal C},\qquad R_0\triangleq -\frac{E}{2\ell}  . 
\eea
The non-negative Carter constant $k$ is $k = Q+\frac{\ell^2}{4} > 0$, which implies that $\mathcal C > -\ell^2$ and that $R_- < R_+$ with $R_\pm$ both real. These equations all agree with Ref. \cite{Kapec:2019hro}. 

Similarly, defining $z\triangleq\cos^2\theta$, one can rewrite the polar potential as
\begin{equation}
    v_\theta(z)= -\ell^2 z+\qty(Q+\mathcal C_\circ z)(1-z)=\left\lbrace
    \begin{array}{ll}
        (Q+\ell^2)(z_0-z) & \text{ for } \mathcal{C}_\circ=0\\
        \mathcal{C}_\circ(z_+-z)(z-z_-) & \text{ for } \mathcal{C}_\circ\neq 0
    \end{array}\right.
\end{equation}
where $\mathcal{C}_\circ$ is defined through the critical value of the angular momentum $\ell_\circ$:
\begin{equation}
    \mathcal{C}_\circ\triangleq\frac{\ell^2-\ell^2_\circ}{4}, \qquad \ell_\circ\triangleq 2M\mu.
\end{equation}
The roots of the polar potential are given by
\begin{equation}
    z_0\triangleq\frac{Q}{Q+\ell^2}\,\qquad z_\pm\triangleq\Delta_\theta\pm \text{sign}(\mathcal C_\circ) \sqrt{\Delta_\theta^2+\frac{Q}{\mathcal{C}_\circ}},\qquad\Delta_\theta\triangleq\frac{1}{2}\qty(1-\frac{Q+\ell^2}{\mathcal{C}_\circ}).
\end{equation}

Future orientation of the geodesic is equivalent to $dT/d\lambda > 0$ or 
\bea
E+\ell R > 0. 
\eea
Future-oriented geodesics with $\ell = 0$ have $E >0$. For $\ell \neq 0$, we define the critical radius as in \cite{Kapec:2019hro}:
\bea
R_c = -\frac{E}{\ell}. \label{defRc}
\eea
Future-orientation of the orbit requires 
\bea\label{consRc}
R < R_c \quad \text{for} \quad \ell < 0, \quad \text{and} \quad R > R_c \quad \text{for} \quad \ell > 0. 
\eea

\subsubsection{Solving the equations of motion}

Using the same reasoning as for the Kerr geometry, the formal solutions to the geodesic equations are given by 
\begin{align}
\lambda_f  -\lambda_i&=  T^{(0)}_R(R_f)-T^{(0)}_R(R_i)  = \lambda_\theta(\theta_f)-\lambda_\theta(\theta_i)  , \label{eq:12}\\
T(\lambda_f) -T(\lambda_i)&= E \qty( T^{(2)}_R(R(\lambda_f))-T^{(2)}_R(R(\lambda_i))) \nonumber\\&~+ \ell \qty(  T^{(1)}_R(R(\lambda_f))-T^{(1)}_R(R(\lambda_i))), \\
\Phi(\lambda_f)-\Phi(\lambda_i) &=-\frac{3}{4} \ell (\lambda_f-\lambda_i) - E \qty( T^{(1)}_R(R(\lambda_f))-T^{(1)}_R(R(\lambda_i))) \nonumber\\
&~+ \ell \big( \Phi_\theta(\theta(\lambda_f))-\Phi_\theta(\theta(\lambda_i))\big)\label{eq:NHEKphi}
\end{align}
where the three radial integrals are 
\bea
T^{(i)}_R(\lambda) & \triangleq &  \sint \frac{ \dd R}{\pm_RR^i\sqrt{E^2+2E \ell R-\mathcal C R^2}} ,\qquad i = 0,1,2\label{eqn:radialNHEK}
\eea
and the two polar integrals are 
\bea
\lambda_\theta(\lambda) & \triangleq & \sint \frac{ \dd\cos \theta}{\pm_\theta\sqrt{v_\theta(\theta)}}, \\
\Phi_\theta(\lambda) & \triangleq &  \sint \frac{ \dd\cos \theta}{\pm_\theta\sqrt{v_\theta(\theta)}}\left( \frac{1}{\Lambda^2(\theta)} - \frac{1}{4}\right). 
\eea
The notation was explained previously. We defined $\Phi_\theta$ such that it is zero for equatorial orbits (since $\Lambda(\pi/2)=2$). After integration, the equation \eqref{eq:12} can be inverted to give $R(\lambda)$ and $\theta(\lambda)$. We need to solve these five integrals as a function of the geodesic parameters $\mu,E,\ell,Q,s_\theta^i,s_R^i,T_i,R_i,\theta_i,\Phi_i$.

\paragraph{Polar behavior.}

The results derived in Sec. \ref{sec:polarKerr} in the context of general Kerr still hold in the near-horizon high-spin limit which is obtained by the scaling limit $\lambda \rightarrow 0$ taken in the near-horizon coordinates \eqref{eq:cvn}. We anticipate that the results also hold in the distinct near-NHEK limit $\lambda \rightarrow 0$ taken in the near-horizon coordinates \eqref{eq:cvnn}. Due to the high-spin limit, the following substitution can be made:
\begin{align}
&a \mapsto M, \qquad \hat E \mapsto \frac{\ell}{2M},\qquad  \eps_0 \mapsto \mathcal{C}_\circ\triangleq\frac{\ell^2-\ell_\circ^2}{4}, \\
&\Theta(z) \mapsto v_\theta(z),\qquad  \hat\Phi_\theta-\frac{1}{4}\hat{T}_\theta \mapsto \Phi_\theta.
\end{align}
Notice that the dependence on $\hat E$ of $\eps_0$ has been changed into a dependence in $\ell$, the Kerr energy being the same at zeroth order on $\lambda$ for all trajectories. Therefore, the quadratic term of the polar potential vanishes at the critical value $\ell_\circ$ of the angular momentum $\ell$.

\begin{figure}[!hbt]
    \centering \vspace{-0.2cm}
    \includegraphics[width=11.8cm]{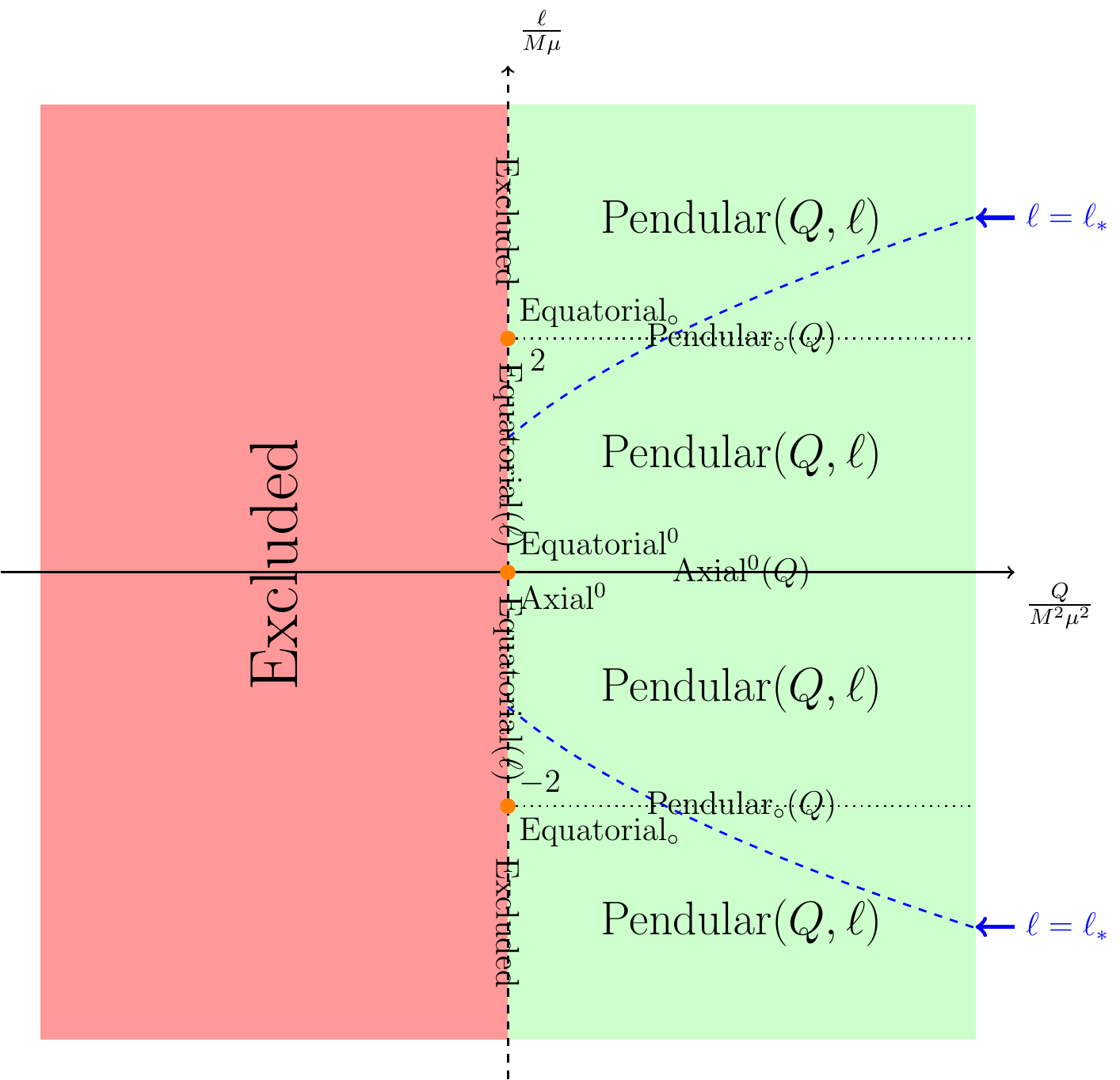}
    \vspace{-0.1cm}\caption{Polar taxonomy of near-horizon geodesics. For clarity, the scale is not respected on the horizontal axis. The dashed blue curves $\ell^2=\ell^2_*$ represent the position of the spherical orbits in parameter space. This figure contrasts with Figure \ref{fig:angularKerrTaxonomy}.}\vspace{-1cm}
    \label{fig:angularTaxonomy}
\end{figure}

One of the most striking features of the near-horizon polar motion is that $Q$ is non-negative as a consequence of the reality of polar motion, as noticed in Ref. \cite{Kapec:2019hro}: 
\begin{prop}\label{propQ}
\begin{equation}
    \forall z\in[0,1] :v_\theta(z)\geq0\Rightarrow Q\geq0.
\end{equation}
\end{prop}
\noindent \emph{Proof.} This property is a consequence of the dependence on $Q$ of $\mathcal{C}$ defined in \eqref{eqn:Carter}. Indeed,  using the fact that $z=\cos^2\theta\in[0,1]$ one can write
\bea 
 Q = \mathcal C + \frac{3}{4}\ell^2 -M^2 \mu^2 \geq \mathcal C + (1-\Lambda^{-2})\ell^2 -M^2 \mu^2 \geq v_\theta (z) \geq 0. 
\eea 
A direct consequence is that the near-horizon polar motion cannot be vortical and is consequently either equatorial, pendular, polar or axial. 
We note that the condition $\eps_0\geq\ell^2$ is never obeyed in the near-horizon case after using the definition \eqref{defeps0}, $a=M$ and $\hat E=\frac{\ell}{2M}$. The equator-attractive class is therefore discarded. The resulting polar classes are listed in Table \ref{table:taxPolar} and the phase space is represented in Fig.\ref{fig:angularTaxonomy}.

\paragraph{Radial behavior.} 
The radial behavior for generic inclined orbits can be solved using the equatorial results \cite{Compere:2017hsi} thanks to the following observation:
\begin{prop}\label{thm:equivNHEK}
The radial integrals  $T_R^{(i)}(R)$ $(i=0,1,2)$ only depend upon the NHEK energy $E$ and angular momentum $\ell$ while all the dependence upon the mass $\mu$ and Carter constant $Q$ is through $\ell_* = \frac{2}{\sqrt{3}}\sqrt{M^2 \mu^2 +Q}$. 
\end{prop}
This simple observation has far-reaching consequences. For any timelike geodesic with $Q \neq 0$, one could directly reuse the classification established in Ref. \cite{Compere:2017hsi}, modulo the substitution $\frac{2}{\sqrt{3}}M \mu \to \ell_*$ in every expression encountered. Moreover, null geodesics with $\mu = 0$ have $Q \geq 0$ from Proposition \ref{propQ}. We can therefore reuse the classification established in Ref. \cite{Compere:2017hsi} to classify null geodesics modulo the substitution $M \mu \to Q$ in every expression encountered. Overall, all radial integrals can be described in closed form for all cases by keeping the dependence upon $\ell_*$ or, equivalently, upon the Casimir invariant $\mathcal C$.

Since the equatorial taxonomy of Ref. \cite{Compere:2017hsi} did not consider bounded orbits and only considered $\ell > 0$, we will expand the taxonomy to the general case. The general classification can be achieved by studying the roots of $v_R$ and the range of $R$ where $v_R \geq 0$. We only consider orbits outside the horizon, $R>0$. There are three broad categories depending on the angular momentum: the supercritical case $\vert \ell  \vert > \ell_*$ or equivalently $\mathcal C < 0$, the critical case $\vert \ell \vert= \ell_*$ or equivalently $\mathcal C = 0$ and the subcritical case $0 \leq \vert \ell \vert  < \ell_*$ or $\mathcal C > 0$. 

The relative position of the critical radius \eqref{defRc} with respect to the roots of $v_R$ may restrict the allowed classes of future-oriented orbits. As a result of \eqref{consRc}, subcritical $\ell^2 <\ell^2_*$ orbits have either $R_+ < R_c$ for $\ell < 0$ or $R_c < 0$ for $\ell > 0$, and all orbits are future oriented. Critical orbits $\ell^2 =\ell^2_*$ have either $R_c < 0$ for $\ell = \ell_*$ or $R_c > R_0$ for $\ell = -\ell_*$. This restricts the classes of orbits. Supercritical orbits $\ell^2 > \ell^2_*$ with $E,\ell>0$ are future directed. Supercritical orbits with $E>0$, $\ell<0$ admit $R_- < R_c < R_+$, and only bounded orbits with $R \leq R_-$ are admissible. Finally, supercritical orbits with $E<0$ and $\ell > 0$ obey $R  \geq R_+ > R_c $ and are therefore deflecting. 

\begin{figure}[!hbt]
    \centering
    \includegraphics[width=0.8\textwidth]{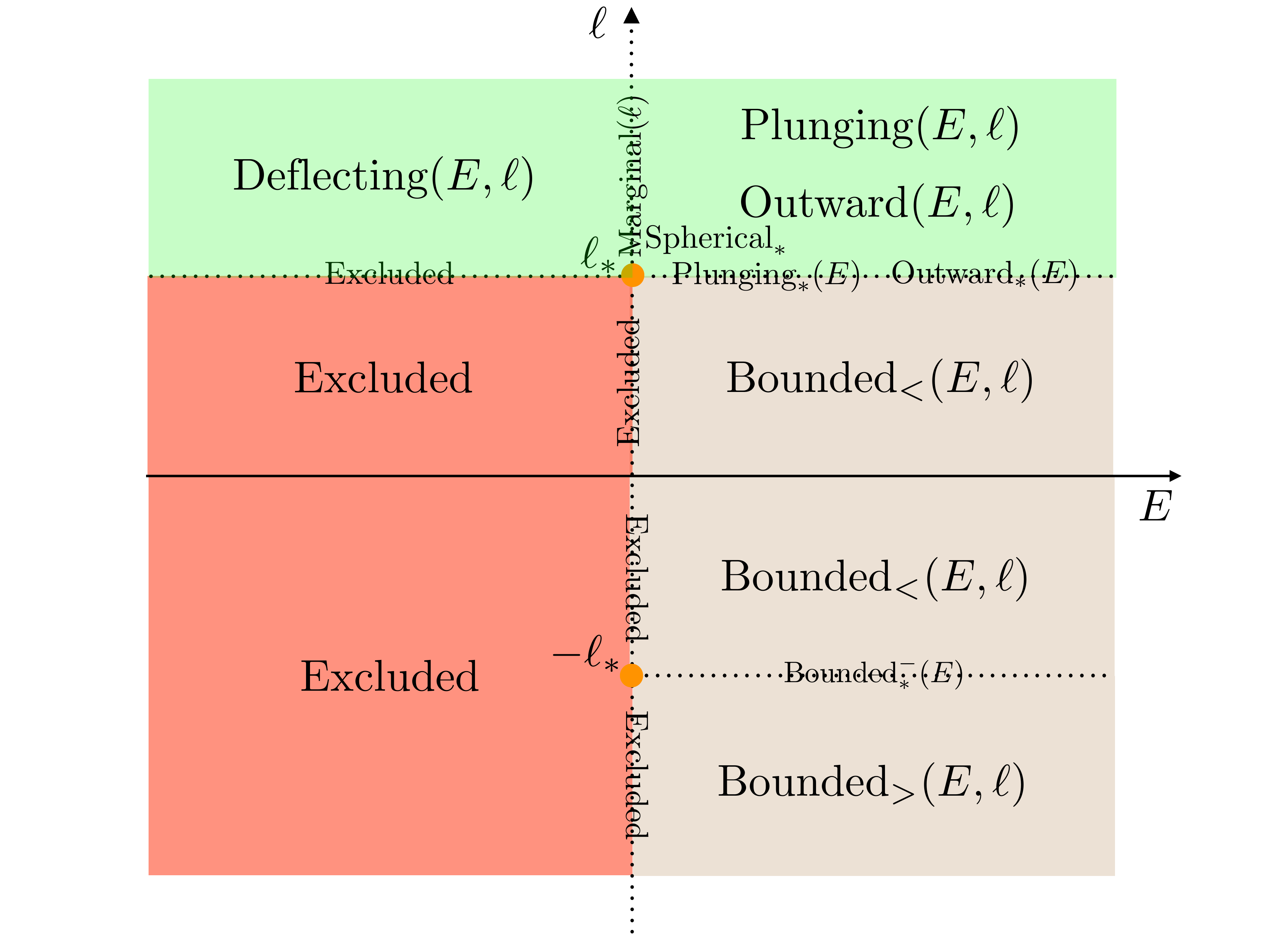} 
    \caption{Radial taxonomy of future oriented geodesics in NHEK. For equatorial geodesics, $\ell_*=\frac{2}{\sqrt{3}}M \mu$, while for orbits with inclination, $\ell_* = \frac{2}{\sqrt{3}} \sqrt{M^2\mu^2 + Q}$.}
    \label{fig:taxonomyEquator}
\end{figure}

After a simple analysis, we reach the following taxonomy, displayed in Table \ref{table:taxNHEK} and in Fig. \ref{fig:taxonomyEquator}. In comparison with Ref. \cite{Compere:2017hsi},  the classes Outward$(E,\ell)$, Outward$_*(E)$,  Bounded$_>(E,$ $\ell)$, Bounded$^-_*(E)$, and Bounded$_<(E,\ell)$ are new, while all other classes with $\ell > 0$ appeared in Ref. \cite{Compere:2017hsi}. The class $\text{Osculating}(E,\ell)$ is now better called $\text{Def}\text{lecting}(E,\ell)$. The classes with $\ell = \pm \ell_*$ will be denoted with a subscript $_*$. The Spherical$_*$ orbit with $\ell = \ell_*$ is also the prograde ISSO. For $\ell\geq 0$, the conformal diagrams corresponding to those orbits are depicted in Fig. \ref{fig:penrose_NHEK} and their explicit forms are given in Appendix \ref{app:equatorial}. Past-oriented geodesics (not depicted) are obtained from a central symmetry around the origin $E=\ell=0$ as a result of the $\uparrow\!\downarrow$-flip \eqref{PTflip}. 

\begin{table}[!htb]
    \centering
    \begin{tabular}{|c|c|c|c|}
    \hline
        \textbf{Angular momentum} & \textbf{Carter constant} & \textbf{Polar range} & \textbf{Denomination}\\\hline
        $\ell=0$ ($\mathcal{C}_\circ=-\ell_\circ^2/4$) &  $Q=0$ & $z=0,~1$ & \begin{tabular}{c} \rule{0pt}{13pt}Equatorial${^0}$\\\rule{0pt}{13pt}Axial${}^0$\end{tabular}\\\cline{2-4}
         & $Q> 0$ & $z=1$ & \rule{0pt}{13pt}Axial${}^0(Q)$\\\hline
         $0<\ell<\ell_\circ$ ($-\frac{\ell_\circ^2}{4}<\mathcal{C}_\circ<0$) & $Q=0$ & $z=0$ & \rule{0pt}{13pt}Equatorial$(\ell)$\\\cline{2-4}
          & $Q>0$ & $0\leq z \leq z_+$ & \rule{0pt}{13pt}Pendular$(Q,\ell)$\\\hline
        $\ell=\ell_\circ$ ($\mathcal{C}_\circ=0$) &  $Q=0$ & $z=0$ &\rule{0pt}{13pt} \rule{0pt}{13pt}Equatorial${}_\circ$\\\cline{2-4}
         & $Q>0$ & $0\leq z \leq z_0$ & \rule{0pt}{13pt}Pendular${}_\circ(Q)$\\\hline
         $\ell>\ell_\circ$ ($\mathcal{C}_\circ>0$) & $Q>0$ & $0\leq z \leq z_+$ & \rule{0pt}{13pt}Pendular$(Q,\ell)$\\\hline
         \end{tabular}
    \caption{Polar taxonomy of near-horizon  geodesics  with $\ell\geq 0$. The orbits with $\ell<0$ are obtained from $\ell>0$ by flipping the sign of $\ell$ with the rest unchanged.}
    \label{table:taxPolar}
\end{table}

\clearpage

\begin{table}[!tbh]    \centering
\begin{tabular}{|c|c|c|c|}\hline
\rule{0pt}{13pt}\textbf{Angular momentum (and Casimir)} & \textbf{Energy} & \textbf{Radial range} & \textbf{Denomination} \\ \hline
Supercritical: $\ell > \ell_*$ ($-\ell^2 < \mathcal C < 0$)& $E > 0$ & $0 \leq R \leq \infty$ & \begin{tabular}{c} \rule{0pt}{13pt}Plunging$(E,\ell)$\\\rule{0pt}{13pt}Outward$(E,\ell)$\end{tabular}\\\cline{2-4}
 & $E = 0$ & $0 < R \leq \infty$ & \rule{0pt}{13pt}Marginal$(\ell)$ \\ \cline{2-4}
& $E < 0$ & $R_+ \leq R \leq \infty$ & \rule{0pt}{13pt}Def\mbox{}lecting$(E,\ell)$ \\ \hline
Critical: $\ell = \ell_*$ ($\mathcal C = 0$) & $E > 0$ & $0 \leq R \leq \infty$ & \begin{tabular}{c} \rule{0pt}{13pt}Plunging$_*(E)$\\\rule{0pt}{13pt}Outward$_*(E)$\end{tabular} \\ \cline{2-4}
& $E = 0$ & $0 < R \leq \infty$ & \rule{0pt}{13pt}Spherical$_*$ (ISSO) \\  \hline
Subcritical: $0 \leq \ell^2 < \ell_*^2$ ($0 < \mathcal C \leq \frac{3 \ell_*^2}{4}$) & $E > 0$ & $0 \leq R \leq R_+$ & \rule{0pt}{13pt}Bounded$_<(E,\ell)$ \\ \hline
Critical: $\ell = -\ell_*$ ($\mathcal C = 0$) & $E > 0$ & $0 \leq R \leq R_0$ &\rule{0pt}{13pt} Bounded$^-_*(E)$ \\ \hline
Supercritical: $\ell < - \ell_*$ ($-\ell^2 < \mathcal C < 0$)& $E > 0$ & $0 \leq R \leq R_-$ &\rule{0pt}{13pt} Bounded$_>(E,\ell)$\\\hline 

\end{tabular}\caption{Radial taxonomy of future-oriented geodesics in NHEK. }\label{table:taxNHEK}
\end{table}

\begin{figure}[!htbp]
    \centering
   \begin{tabular}{cccc}
    \includegraphics[width=3.5cm]{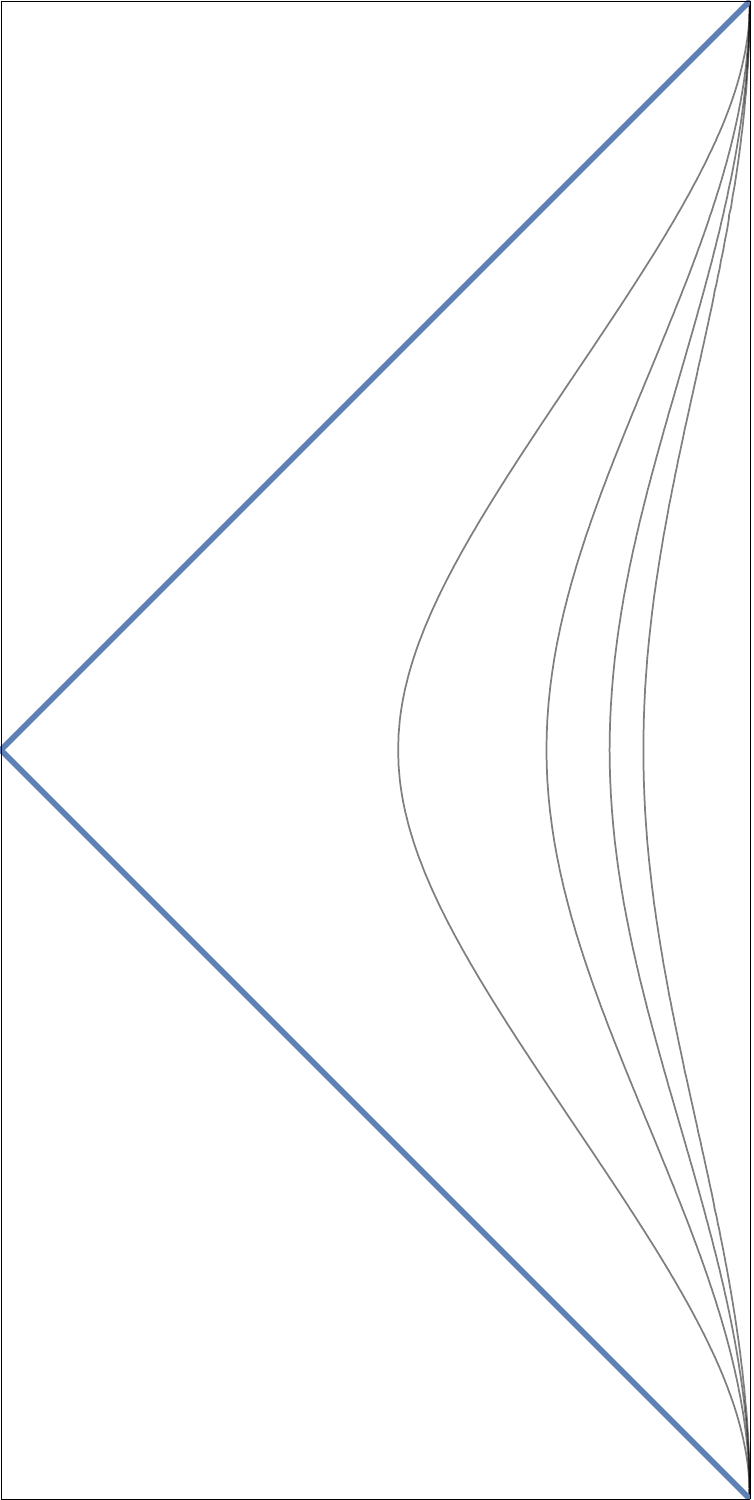} & \includegraphics[width=3.5cm]{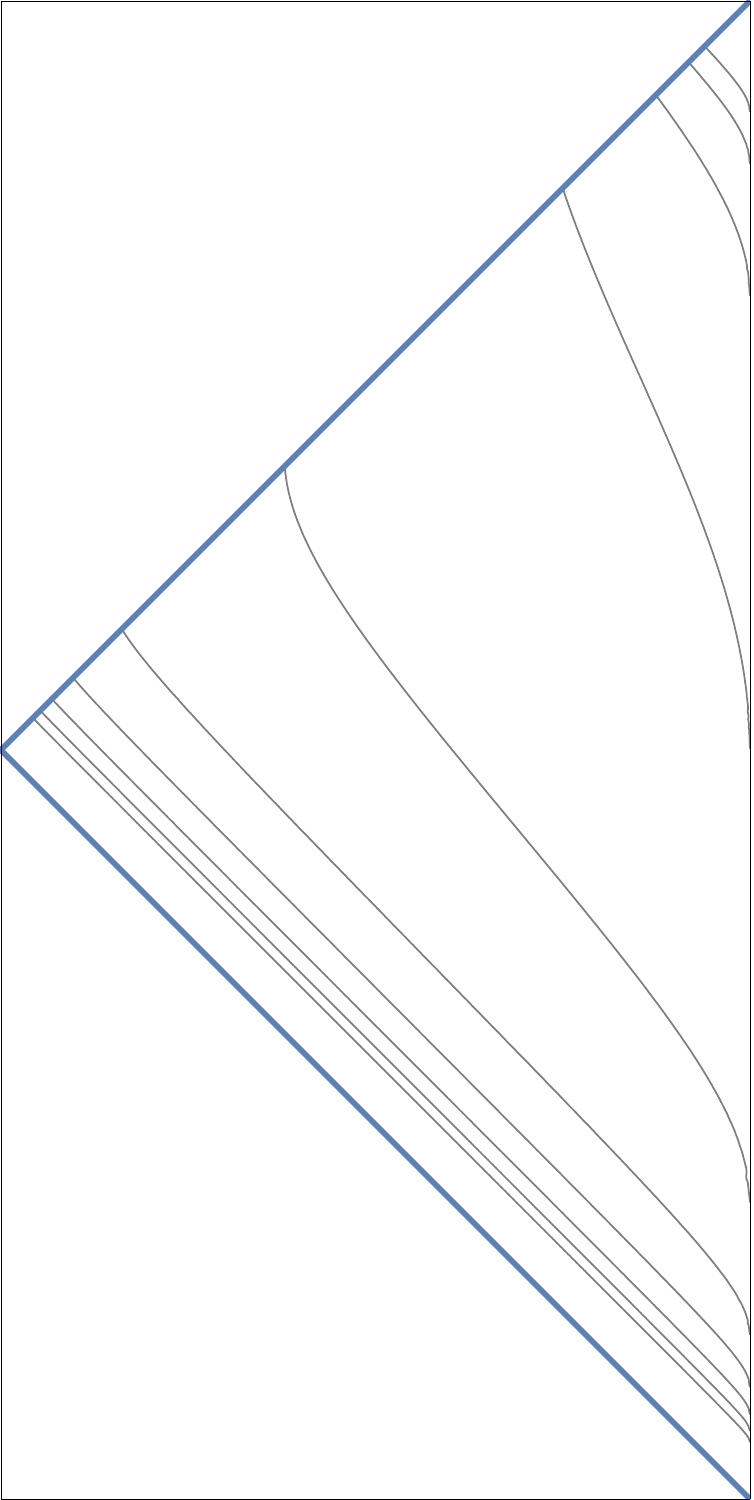} & \includegraphics[width=3.5cm]{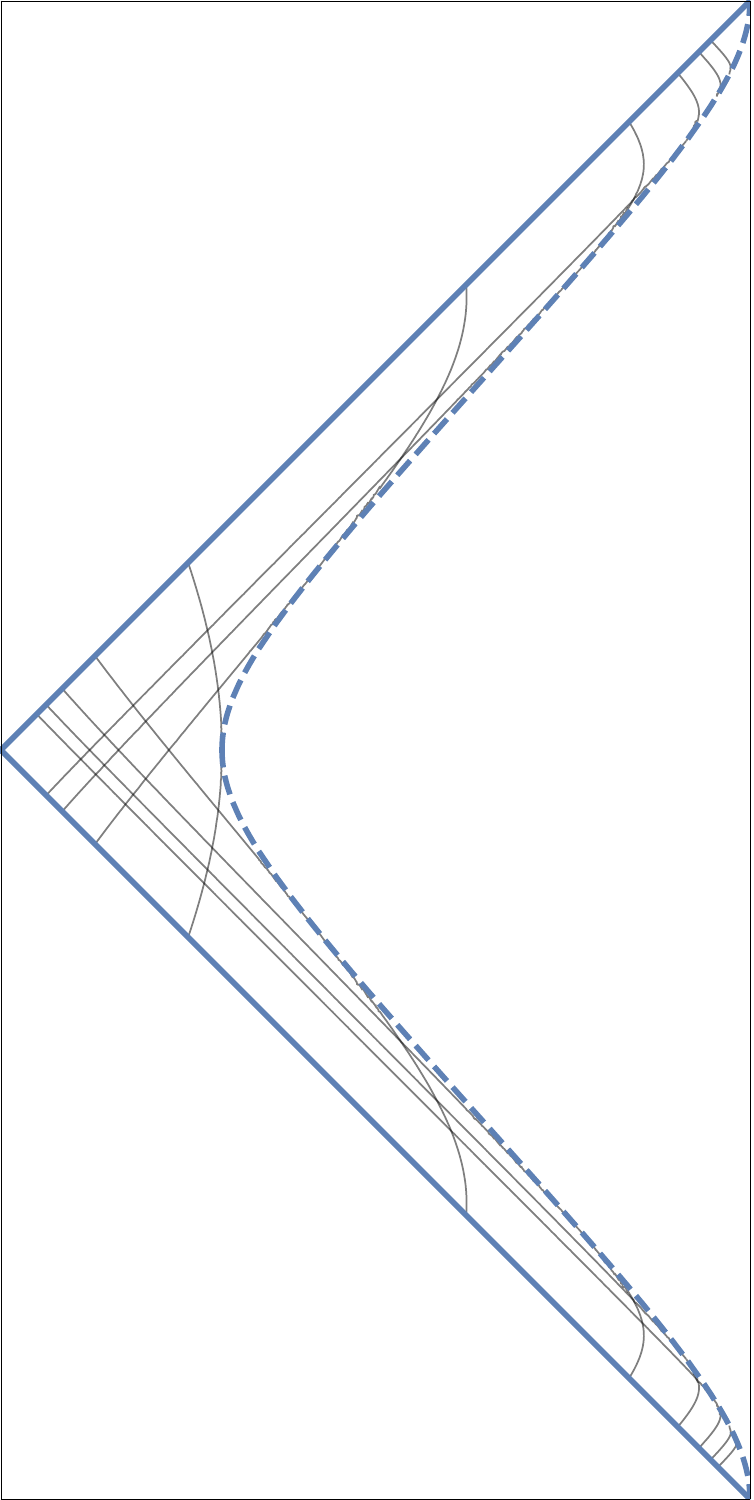} &
    \includegraphics[width=3.5cm]{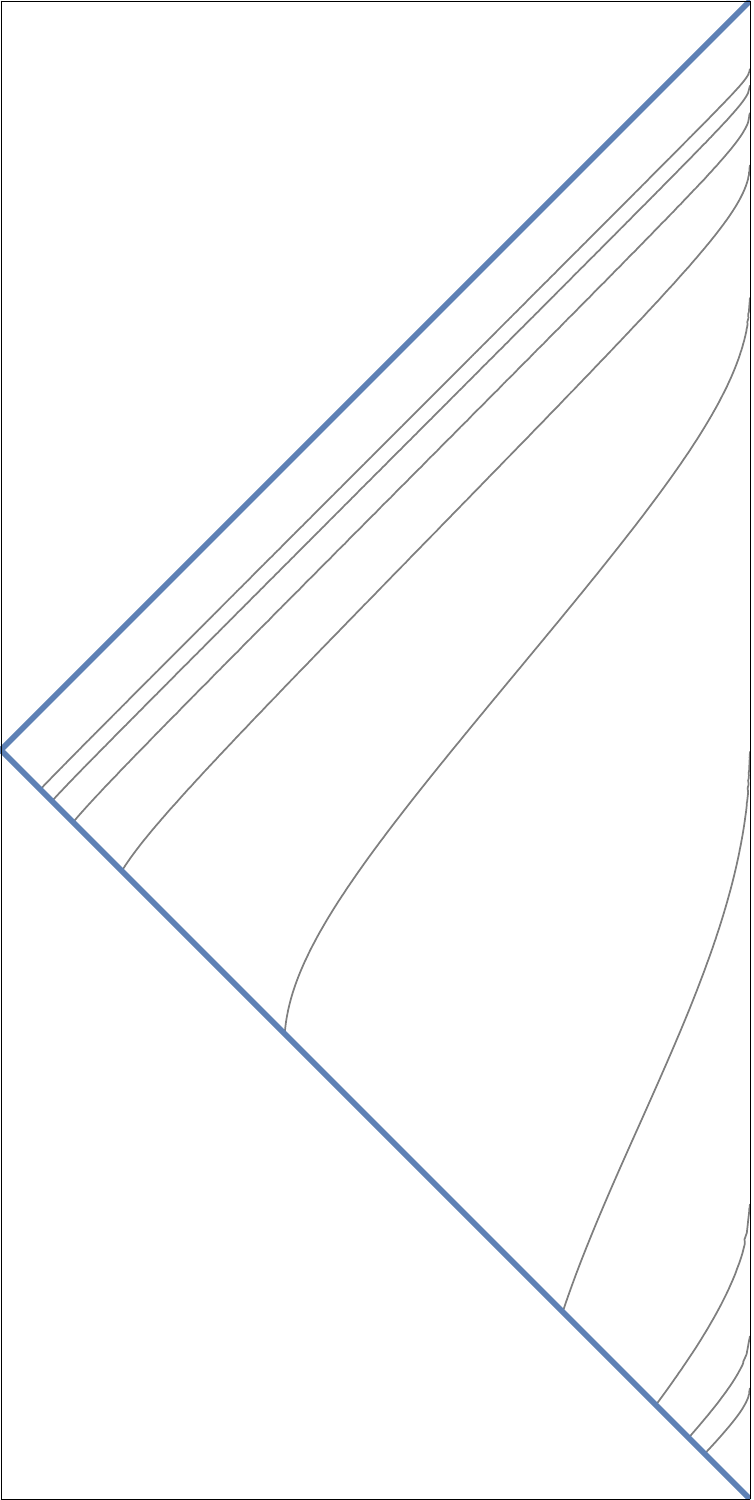}\\
    (a) Spherical${}_*(ISSO)$ & (b) Plunging${}_*(E)$ & (c) Bounded${}^-_*(E)$ & \begin{tabular}{c}
    (d) Outward$_*(E)$,\\
    Outward$(E,\ell)$
    \end{tabular} \\
    \rule{0pt}{30pt} & & & \\
    \includegraphics[width=3.5cm]{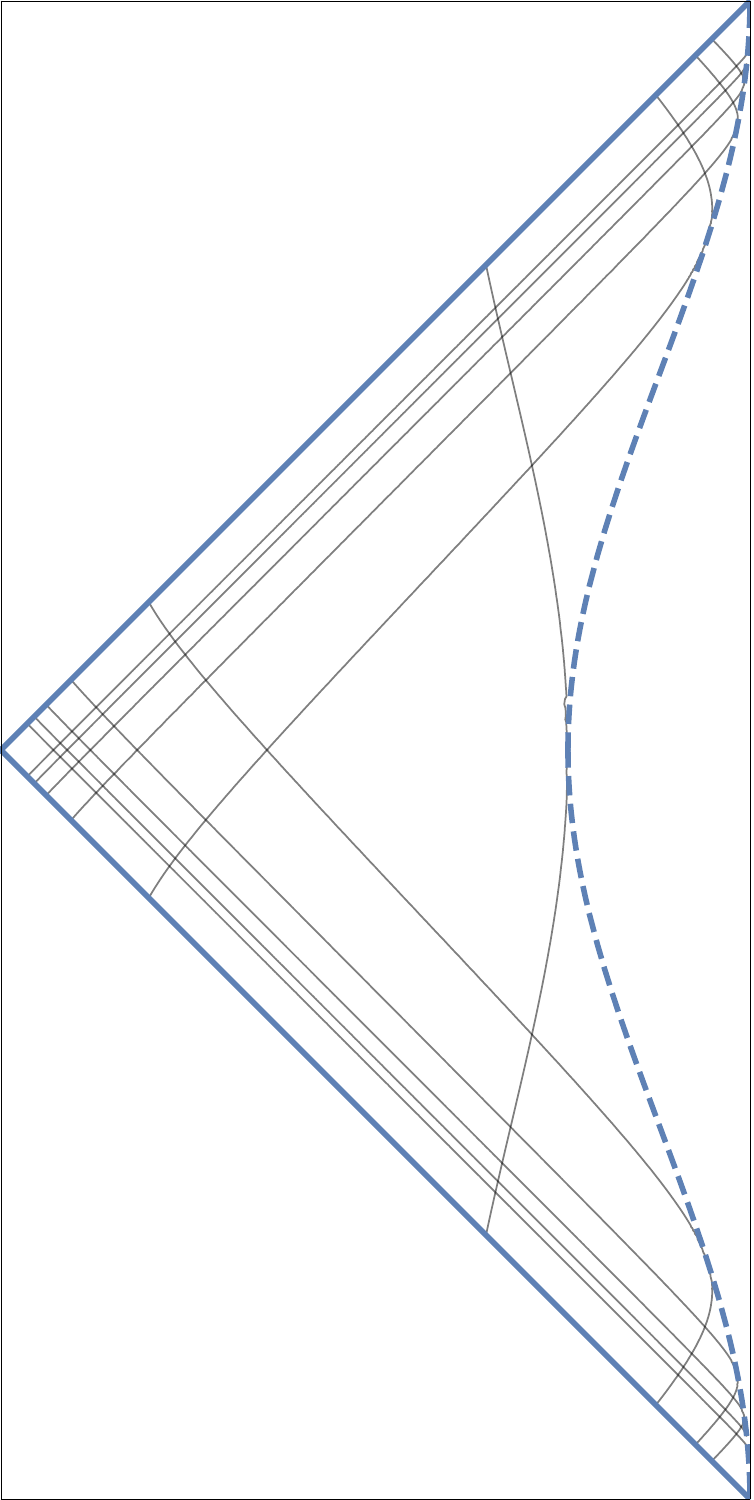} &
    \includegraphics[width=3.5cm]{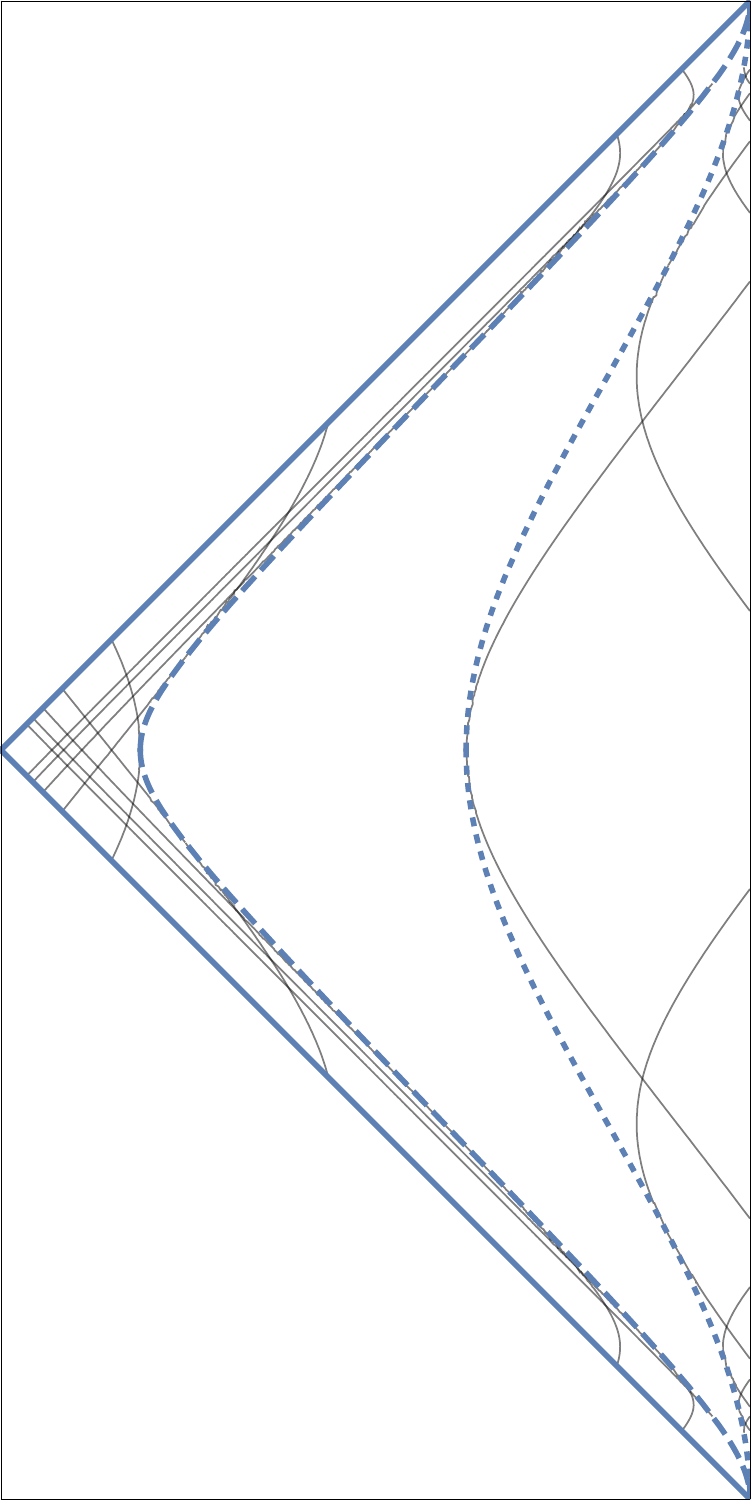} & \includegraphics[width=3.5cm]{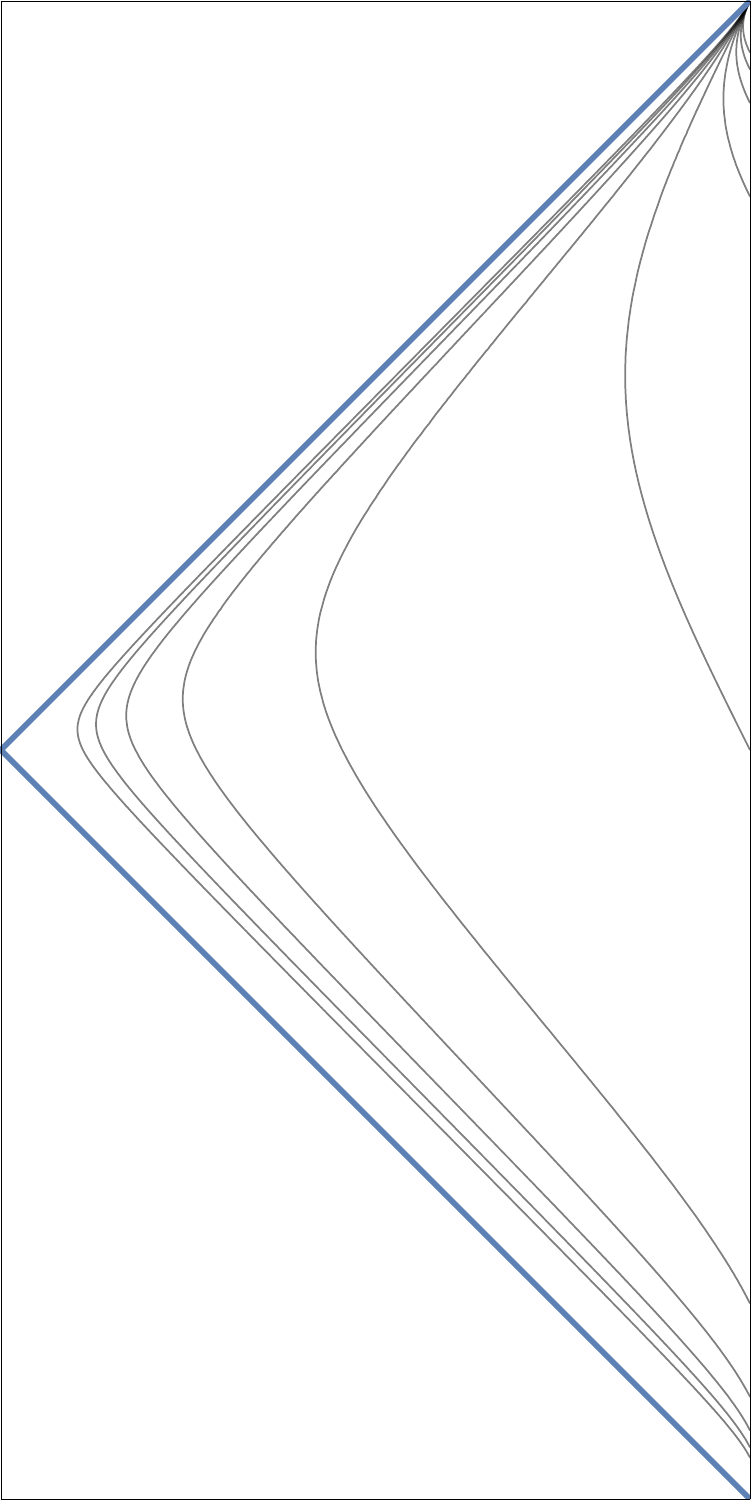} & \includegraphics[width=3.5cm]{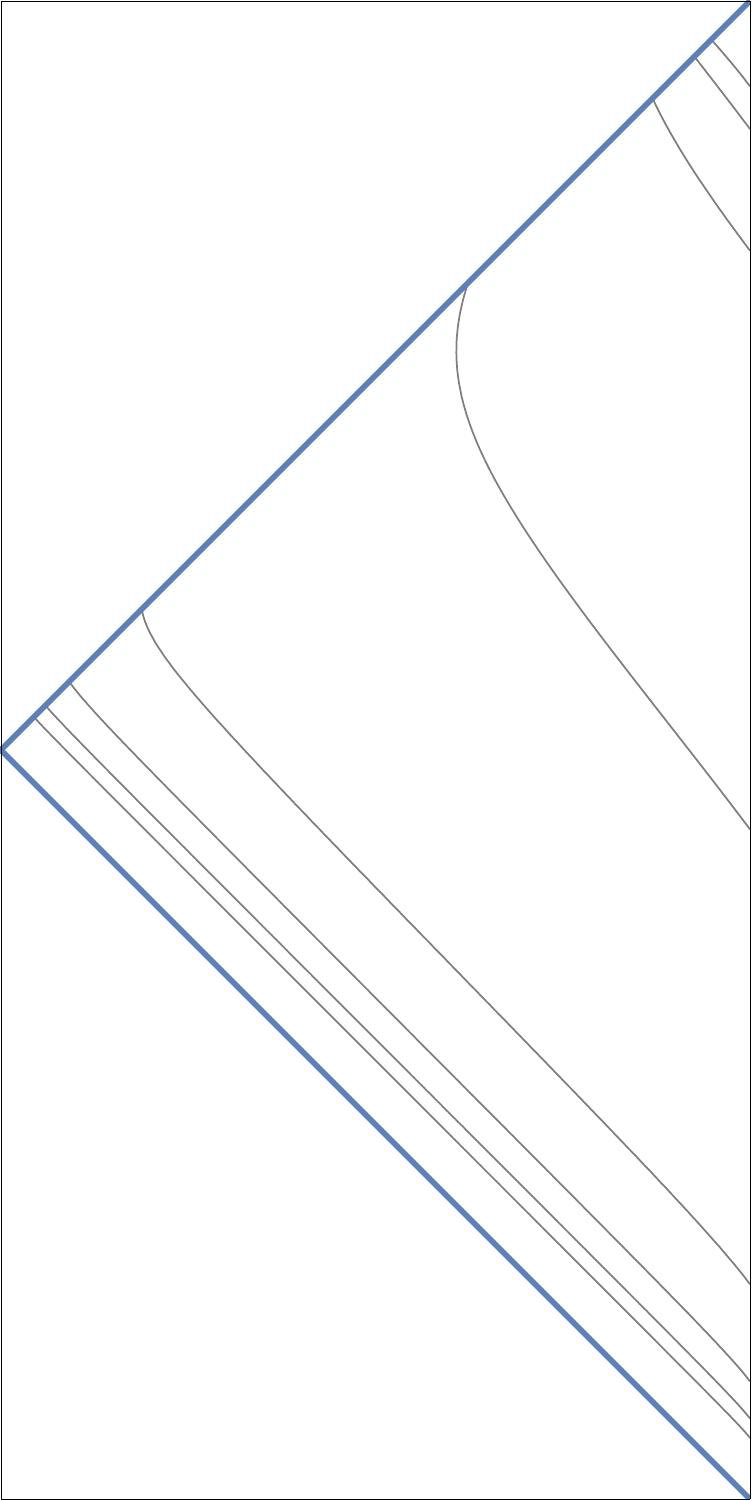} \\
     (e) Bounded${}_<(E,\ell)$ & \begin{tabular}{@{}c@{}}(f) Bounded${}_>(E,\ell)$, \\  Def\mbox{}lecting$(E,\ell)$\end{tabular} & (g) Marginal$(\ell)$ & (h) Plunging$(E,\ell)$\\
   \end{tabular}
    \caption{Taxonomy of NHEK geodesics depicted in the global NHEK conformal diagram. The upper (or respectively, lower) blue line represent the future (respectively, past) event horizon $R=0$ and the dashed/dotted lines are the roots of the radial potential. We used $M=1$, $E=\pm1$ and $\ell=\pm 2\ell_*$ (and $\ell=\pm \frac{1}{2}\ell_*$, respectively) for supercritical (and subcritical, respectively) trajectories.}
    \label{fig:penrose_NHEK}
\end{figure}

\clearpage
\subsection{Near-NHEK}\label{sec:nn}
\subsubsection{Metric and geodesic equations}
We completed the study of the inclined geodesic motion in NHEK spacetime, which describes in particular the motion of test particles near the ISSO and the separatrix at leading order in the high-spin limit, as we will further expand upon in Sec. \ref{sec:spherical_properties}. We now turn to the study of such a motion in a closer vicinity of the black hole event horizon. We introduce the so-called \textit{near-NHEK} coordinates ($t,r,\theta,\phi$), related to Boyer-Lindquist coordinates through the relations
\begin{align}
t&=\frac{\that}{2M\kappa}\lambda,\qquad 
r=\frac{\kappa}{M}\qty(\rhat-\rhat_+)\lambda^{-1},\label{eq:cvnn}\qquad
\phi=\phihat-\frac{\that}{2M}.
\end{align}
At leading order in $\lambda$, the metric becomes
\begin{equation}
\dd s^2=2M^2\Gamma(\theta)\qty(-r(r+2\kappa)\dd t^2+\frac{\dd r^2}{r(r+2\kappa)}+\dd\theta^2+\Lambda^2(\theta)\qty(\dd\phi+(r+\kappa)\dd t)^2)+\mathcal{O}(\lambda).
\end{equation}
Though the metric explicitly depends upon $\kappa$, no physical quantity depends upon $\kappa$, since it is introduced from a coordinate transformation. The corresponding geodesic equations of motion are now given by
\begin{align}
\dv{t}{\lambda}&=\frac{e+\ell (r+\kappa)}{r(r+2\kappa)},\label{eq:nn1}\\
\dv{\phi}{\lambda}&=-\frac{e(r+\kappa)+\ell \kappa^2}{r(r+2\kappa)}+\ell\qty(\frac{1}{\Lambda^2}-1),\label{eq:nn2}\\
\dv{\theta}{\lambda}&=\pm_\theta\sqrt{v_\theta(\cos^2 \theta)},\label{eq:nn3}\\
\dv{r}{\lambda}&=\pm_r\sqrt{v_{r;\kappa}(r)}\label{eq:nn4}
\end{align}
where the radial potential can be written as
\begin{align}
v_{r;\kappa}(r)&\triangleq (e+\ell \kappa)^2+2e \ell r+\frac{3}{4}\qty(\ell^2-\ell^2_*)r(r+2\kappa )
\end{align}
and the angular potential is still as given in \eqref{eq:vtheta}. Although $r$ is a meaningful radial coordinate (because of the horizon location at $r=0$), it is convenient to introduce the shifted radial variable $R\triangleq r+\kappa$ to get more elegant expressions. The coordinate $R$ is also used in NHEK, but the context allows us to distinguish them. The generators of $SL(2,\mathbb R)\times U(1)$ are $\partial_\phi$ and 
\begin{equation}
    H_0=\frac{1}{\kappa}\partial_t,\qquad H_\pm=\frac{\text{exp}({\mp\kappa t})}{\sqrt{R^2-\kappa^2}}\qty[\frac{R}{\kappa}\partial_t\pm(R^2-\kappa^2)\partial_R-\kappa\partial_\phi].\label{vecs}
\end{equation}
The $SL(2,\mathbb R)$ Casimir $\mathcal C^{\mu\nu}\p_\mu \p_\nu$ takes the form \eqref{Cas}, where the vectors are now given by \eqref{vecs}. 

The radial potential can be recast as 
\bea
v_{R;\kappa}(R)=\left\lbrace\begin{array}{ll}
    -\mathcal C(R-R_+)(R-R_-), & \mathcal C\neq 0 ;\\
    2e \ell_* (R-R_0), & \mathcal{C}=0
\end{array}\right.
\eea
where 
\bea
R_\pm \triangleq \frac{e \ell}{\mathcal C} \pm \frac{\sqrt{(\mathcal C+\ell^2)(e^2+\kappa^2\mathcal{C})}}{|\mathcal{C}|},\qquad R_0\triangleq-\frac{e^2+\kappa^2\ell_*^2}{2e\ell_*}.
\eea
The near-NHEK energy $e$ is related to Boyer-Lindquist energy by
\begin{equation}
\hat{E}=\frac{\ell}{2M}+\frac{\lambda}{2M\kappa}e.\label{eN}
\end{equation}
The (near-)NHEK and Boyer-Lindquist angular momenta and Carter constants $Q$ are equal. We define again the critical radius 
\bea
R_c = -\frac{e}{\ell}. 
\eea
and the future orientation of the orbit again requires \eqref{consRc}. 

\subsubsection{Solutions to the equations of motion}
The (near-)NHEK geodesic equations being very similar, this section will only briefly point out the similarities and the differences between the two cases. The formal solutions to the geodesic equations are
\begin{align}
    \lambda_f-\lambda_i&=t^{(0)}_{R;\kappa}(R_f)-t^{(0)}_{R;\kappa}(R_i)=\lambda_\theta(\theta_f)-\lambda_\theta(\theta_i),\\
    t(\lambda_f)-t(\lambda_i)&=e\qty(t^{(2)}_{R;\kappa}(R(\lambda_f))-t^{(2)}_{R;\kappa}(R(\lambda_i)))\nonumber\\
    &~+\ell\qty(t^{(1)}_{R;\kappa}(R(\lambda_f))-t^{(1)}_{R;\kappa}(R(\lambda_i))),\label{eqn:nnLambda}\\
    \phi(\lambda_f)-\phi(\lambda_i)&=-\frac{3}{4}\ell(\lambda_f-\lambda_i)-e \qty(t^{(1)}_{R;\kappa}(R(\lambda_f))-t^{(1)}_{R;\kappa}(R(\lambda_i)))\nonumber\\
    &~-\kappa^2\ell\qty( t^{(2)}_{R;\kappa}(R(\lambda_f))- t^{(2)}_{R;\kappa}(R(\lambda_i)))\nonumber\\
    &~+\ell\qty(\Phi_\theta(\theta(\lambda_f))-\Phi_\theta(\theta(\lambda_i))) \label{eqn:nnPhi} 
\end{align}
where the polar integrals are the same as in NHEK (see above) and the radial ones are defined by
\begin{align}
    t^{(0)}_{R;\kappa}(\lambda)&\triangleq\sint\frac{\dd R}{\pm_R\sqrt{v_{R;\kappa}(R)}},\\
    t^{(i)}_{R;\kappa}(\lambda)&\triangleq\sint\frac{\dd R}{\pm_R\sqrt{v_{R;\kappa}(R)}}\frac{R^{2-i}}{R^2-\kappa^2},\qquad i=1,2.
\end{align}
Notice that NHEK geodesics equations can be recovered from near-NHEK ones by taking the formal limit $\kappa\to 0$; the normalization of the radial integrals has been chosen to satisfy $\lim_{\kappa\to 0} t^{(i)}_{R;\kappa}(R)=T^{(i)}(R)$ ($i=0,1,2$) as defined in \eqref{eqn:radialNHEK}. Therefore, the formal solutions to NHEK geodesic equations can also be recovered by taking the limit $\kappa\to 0$ in \eqref{eqn:nnLambda} and \eqref{eqn:nnPhi}.

\paragraph{Radial behavior.} The only difference between NHEK and near-NHEK geodesic solutions lies in the terms involving the radial coordinate. The proposition stating the equivalence relation between the equatorial and inclined radial parts of the geodesic motion takes the same form as in NHEK:
\begin{prop}\label{thm:equivnearNHEK}
For a given normalization $\kappa$, the radial integrals  $t^{(i)}_{R;\kappa}(R)$ $(i=0,1,2)$ only depend upon the near-NHEK energy $e$ and angular momentum $\ell$ while all the dependence upon the mass $\mu$ and Carter constant $Q$ is through $\ell_* = \frac{2}{\sqrt{3}}\sqrt{M^2 \mu^2 +Q}$. 
\end{prop}

\begin{figure}[!bth]
    \centering
    \includegraphics[width=\textwidth]{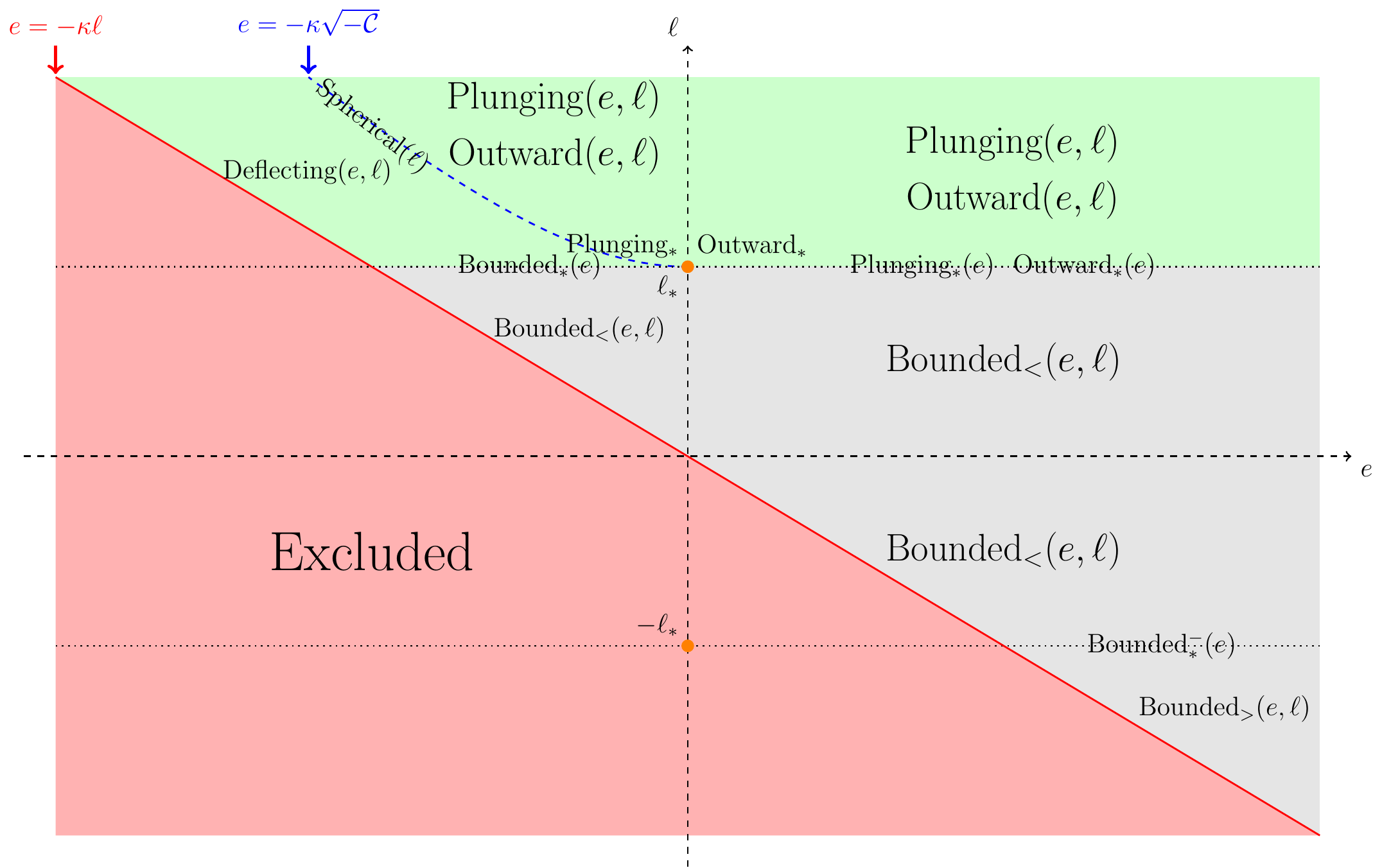} 
    \caption{Radial taxonomy of geodesics in near-NHEK. For equatorial geodesics, $\ell_*=\frac{2}{\sqrt{3}}M \mu$, while for orbits with inclination, $\ell_* = \frac{2}{\sqrt{3}} \sqrt{M^2\mu^2 + Q}$.}
    \label{fig:taxonomyNN}
\end{figure}

As in NHEK, the radial taxonomy of Ref. \cite{Compere:2017hsi} is easily extended to bounded, outward and/or retrograde orbits by studying the roots and the sign of $v_{R;\kappa}(R)$. This leads to the classification displayed in Table \ref{table:taxNN} and Figure \ref{fig:taxonomyNN}. 

\begin{table}[!bht]    \centering
\begin{tabular}{|c|c|c|c|}\hline
\begin{tabular}{c} \rule{0pt}{13pt} \textbf{Angular momentum}\\\textbf{(and Casimir)}\end{tabular} & \textbf{Energy} & \textbf{Radial range} & \textbf{Denomination} \\ \hline
Supercritical: $\ell > \ell_*$  & $e>-\kappa\sqrt{-\mathcal C}$ & $\kappa\leq R\leq\infty$ & \begin{tabular}{c}
\rule{0pt}{13pt}Plunging$(e,\ell)$ \\\rule{0pt}{13pt} Outward$(e,\ell)$ 
\end{tabular}\\ \cline{2-4}
($-\ell^2 < \mathcal C < 0$) & $e=-\kappa\sqrt{-\mathcal C}<0$ & $R=\frac{\kappa\ell}{\sqrt{-\mathcal C}}$ &\rule{0pt}{13pt} Spherical$(\ell)$ \\ \cline{2-4}
 & $-\kappa \ell < e<-\kappa\sqrt{-\mathcal C}<0$ & $R_+\leq R\leq\infty$ &\rule{0pt}{13pt} Def\mbox{}lecting$(e,\ell)$ \\ \hline
Critical: $\ell = \ell_*$ ($\mathcal C = 0$) & $e > 0$ & $\kappa \leq R \leq \infty$ & \begin{tabular}{c}
\rule{0pt}{13pt}Plunging$_*(e)$ \\
\rule{0pt}{13pt}Outward$_*(e)$
\end{tabular} \\ \cline{2-4}
 & $e = 0$ & $\kappa \leq R \leq \infty$ & \begin{tabular}{c}
\rule{0pt}{13pt}Plunging$_*$ \\
\rule{0pt}{13pt}Outward$_*$
 \end{tabular} \\ \cline{2-4}
 & $-\kappa\ell < e < 0$ & $\kappa \leq R \leq R_0$ &\rule{0pt}{13pt} Bounded$_*(e)$ \\ \hline
\begin{tabular}{c}
Subcritical: $0 \leq \ell^2 < \ell_*^2$\\($0 < \mathcal C \leq \frac{3 \ell_*^2}{4}$)\end{tabular} & $e>-\kappa\ell$ & $\kappa \leq R \leq R_+$ &\rule{0pt}{13pt} Bounded$_<(e,\ell)$ \\ \hline
Critical: $\ell = - \ell_*$ ($\mathcal C = 0$) & $e > -\kappa\ell>0$ & $\kappa \leq R \leq R_0$ &\rule{0pt}{13pt} Bounded$_*^-(e)$\\\hline
 \begin{tabular}{c}Supercritical: $\ell <- \ell_*$ \\ ($-\ell^2<\mathcal{C}<0$)\end{tabular}  & $e>-\kappa\ell>0$ & $\kappa\leq R\leq R_-$ & \rule{0pt}{13pt} Bounded$_>(e,\ell)$ \\\hline
\end{tabular}\caption{Taxonomy of future-directed geodesics in near-NHEK.}\label{table:taxNN}
\end{table}

The future-orientation condition \eqref{consRc} implies $e> -\kappa \ell$ for each orbit that reaches the horizon at $R=\kappa$. In the case $\ell > \ell_*$ and $e < 0$, the condition $e \leq -\kappa \sqrt{-\mathcal C}$ implies $e+\kappa \ell \geq 0$ and therefore the parabola does not intersect the line. Past-oriented geodesics (not depicted here) are obtained from a central symmetry around the origin $e=\ell=0$ as a result of the $\uparrow\!\downarrow$-flip \eqref{PTflip}.  The explicit expressions of all near-NHEK geodesics are listed in Appendix \ref{app:explicitNN}.

\subsection{High-spin features of geodesic motion}
\label{sec:univ}
Let us now discuss a few generic and universal features of near-horizon geodesic motion holding in the high-spin case.

\paragraph{Radial motion.}
A first straightforward conclusion one can derive from the analysis of the near-horizon radial geodesic motion is that
\begin{prop}\label{prop4}
All radially unbounded NHEK or near-NHEK geodesics are prograde and either critical or supercritical; \emph{i.e.}, they satisfy $\ell\geq\ell_*$.
\end{prop}
This feature of the near-horizon radial motion is directly visible in Figs. \ref{fig:taxonomyEquator} and \ref{fig:taxonomyNN} and leads to remarkable consequences concerning the polar behavior of such trajectories that we will derive in the following section.

The separatrix between bound and unbound motion is clearly visible in Figs. \ref{fig:taxonomyEquator} and \ref{fig:taxonomyNN}. It consists of the geodesic classes Plunging$_*(E)$ and Outward$_*(E)$ for NHEK and the geodesic classes Plunging$_*(e)$, Outward$_*(e)$, and Bounded$_*(e)$ for near-NHEK that each lie at the critical angular momentum line $\ell = \ell_*$.

\paragraph{Polar motion.} The polar motion of both NHEK and near-NHEK trajectories is bounded in an interval around the equator, $\theta_{\text{\text{min}}}\leq \theta \leq \pi - \theta_{\text{\text{min}}}$, where $\cos\theta_{\text{\text{min}}}=\sqrt{z_+}$ or  $\cos\theta_{\text{\text{min}}}=\sqrt{z_0}$. The maximal polar angle is determined for $\ell^2 \neq \ell^2_\circ = 4 M^2 \mu^2$ as
\begin{equation}
z_+(\ell ,Q ) =\frac{3\ell^2+4(Q+M^2\mu^2)-\sqrt{9\ell^4+16(M^2\mu^2-Q)^2+8\ell^2(3M^2\mu^2+5Q)}}{2(4M^2\mu^2-\ell^2)}
\end{equation}
and for $\ell = \pm \ell_\circ$ as
\bea
z_0(Q) = \lim_{\ell \rightarrow \pm 2 M \mu} z_+ = \frac{Q}{Q+4 M^2 \mu^2}. 
\eea
Remember that $Q \geq 0$ by consistency of polar motion. The asymptotic values are
\begin{align}
\lim_{\scriptsize{\begin{array}{l} Q\to 0 \\ \ell \text{ fixed} \end{array}}}z_+(\ell,Q)&=0, \qquad\qquad\qquad\qquad\qquad\;\;\;\;\;\;\;\; \lim_{\scriptsize{\begin{array}{l} Q\to \infty \\ \ell \text{ fixed} \end{array}}}z_+(\ell,Q)=1,\label{eq:asymptotics}\\
\lim_{\scriptsize{\begin{array}{l} \ell \to 0 \\ Q \text{ fixed} \end{array}}}z_+(\ell,Q)&=\left\lbrace
\begin{array}{cl}
\frac{Q}{M^2\mu^2} &\text{ if }Q<M^2\mu^2\\
1 & \text{ if }Q\geq M^2\mu^2
\end{array}\right. ,\qquad \lim_{\scriptsize{\begin{array}{l} \ell \to \infty \\ Q \text{ fixed} \end{array}}}z_+(\ell,Q) = 0.
\end{align}
 For fixed $\ell$, $z_+$ is a monotonic function of $Q$, and reciprocally $z_+$ is monotonic in $\ell$ at fixed $Q$. The pendular oscillation around the equatorial plane will explore a larger range of $\theta$ when $\theta_\text{\text{min}}$ is smallest or $z_+$ closer to 1, which occurs either for small $\ell$ and $Q \geq M^2 \mu^2$ or large $Q$. 
 
Now, one can check that for critical or supercritical angular momentum $\ell^2 \geq \ell^2_*(Q)$, one has $z_+ < 2 \sqrt{3}-3$ for $\ell \neq \ell_\circ(Q)$ and $z_0 < 2 \sqrt{3}-3$ for $\ell^2 = \ell^2_\circ$. The special angle 
\bea
\theta_{\text{VLS}} \triangleq \arccos{\sqrt{2\sqrt{3}-3}} \approx 47^\circ
\eea
is in fact the velocity-of-light surface in the NHEK geometry \eqref{eq:NHEK_metric} (or near-NHEK geometry) defined as the polar angle such that $\p_T$ is null. It obeys $\Lambda(\theta_{\text{VLS}}) = 1$. The polar region closer to either the north or south poles admits a timelike Killing vector, namely $\p_T$. On the contrary, the polar region around the equator $\theta\in ]\theta_{\text{VLS}},\pi-\theta_{\text{VLS}}[$ does not admit a timelike Killing vector. The velocity-of-light surface separates these two polar regions. We have therefore proven the following property:
\begin{prop}\label{prop5}
All critical or supercritical orbits $\ell^2 \geq \ell^2_*$ in (near-)NHEK geometry lie in the polar region $\theta\in ]\theta_{\text{VLS}},\pi-\theta_{\text{VLS}}[$ where there is no timelike Killing vector. This applies in particular to all spherical orbits.
\end{prop}

The subcritical orbits $\ell^2 < \ell^2_*$ can explore all polar regions of the (near)-NHEK geometry. As a consequence of Propositions \ref{prop4} and \ref{prop5}, we have 
\begin{prop}\label{prop6}
All radially unbounded geodesics in (near-)NHEK geometry lie in the polar region $\theta\in ]\theta_{\text{VLS}},\pi-\theta_{\text{VLS}}[$ bounded by the velocity-of-light surface. 
\end{prop}
In particular, for null geodesics, this feature provides the ``NHEKline'' in the imaging of light sources around a nearly extreme Kerr black hole \cite{Bardeen:1973aa,Gralla:2017ufe}. In \cite{AlZahrani:2010qb,Porfyriadis:2016gwb}, Proposition \ref{prop6} was proven for null geodesics. Here, we show that it is a generic property of all timelike geodesics as well.

\section{Spherical geodesics}
\label{sec:spherical_properties}

The spherical (near-)NHEK geodesics take a distinguished role among all geodesics. First, a subclass of spherical geodesics in NHEK and near-NHEK constitute the innermost stable spherical orbits (ISSOs) and the innermost spherical bound orbits (ISBOs) in the high-spin limit, respectively.  Our first motivation is to fully characterize the ISSO, in order to generalize the analysis of the inspiral/merger transition performed around the equatorial plane in the high-spin limit  \cite{Compere:2019cqe,Burke:2019yek} to inclined orbits.

Second, as noticed in Ref. \cite{Compere:2017hsi}, the equatorial NHEK (resp. near-NHEK) orbits are the simplest representatives for each equivalence class of prograde incoming critical (respectively, supercritical) equatorial orbits under $SL(2,\mathbb R) \times U(1) \times \mathbb Z_2$ symmetry. We will show in Sec. \ref{sec:classes} that the spherical (near-)NHEK orbits are the simplest representatives for each equivalence class of arbitrary timelike (near-)NHEK geodesics under $SL(2,\mathbb R) \times U(1) \times (\mathbb Z_2)^3$ symmetry without any restriction. These two reasons justify the comprehensive study of the spherical geodesics. 

\subsection{Innermost stable spherical orbits} 

The ISSOs are defined as the last stable spherical orbits of Kerr. They are defined from the solutions to 
\bea\label{defISSO}
\hat R(\hat r) =\p_{\hat r} \hat R(\hat r)=\p_{\hat r}\p_{\hat r}\hat R(\hat r) = 0
\eea
where $\hat R$ is defined in \eqref{eq:kerr_vr}. They admit a constant radius $\rhat$ and a fixed $\hat E$ and $\hat\ell$, which can be obtained as solutions of polynomial equations which we will not give explicitly. There are two branches at positive $\hat E$ corresponding to prograde ($\ell \geq 0$) and retrograde orbits  ($\ell < 0$). For the Schwarzschild black hole, the parameters on the two branches of the ISSO are
\bea
\hat r_{\text{ISSO}}=6M,\qquad \frac{\hat E_{\text{ISSO}}}{\mu}= \frac{2\sqrt{2}}{3},\qquad \frac{\ell_{\text{ISSO}}}{\mu M} = \pm \sqrt{12 - \frac{Q}{M^2\mu^2}},
\eea
which implies the bound $Q \leq 12 M^2 \mu^2$.

For arbitary spin, the \textit{innermost stable circular orbit} (ISCO) is defined as the prograde ISSO equatorial orbit, i.e. restricted to $Q=0$ ($\theta = \frac{\pi}{2}$). The parameters are \cite{Bardeen:1972fi}
\bea
\frac{\hat E_{\text{ISCO}}}{M \mu} = \frac{1-2 /\tilde r_{\text{ISCO}}-\tilde a/\tilde r_{\text{ISCO}}^{3/2}}{\sqrt{1-3/\tilde r_{\text{ISCO}}-2\tilde a / \tilde r_{\text{ISCO}}^{3/2}}},\qquad \frac{\ell_{\text{ISCO}}}{\mu M} = \frac{2}{\sqrt{3 \tilde r_{\text{ISCO}}}} (3 \sqrt{\tilde r_{\text{ISCO}}} +2\tilde a),
\eea 
where $\tilde a = a/M$ and 
\bea\label{ISCO}
\tilde r_{\text{ISCO}} &\triangleq & \frac{\hat r_{\text{ISCO}}}{M} =3+Z_2-\sqrt{(3-Z_1)(3+Z_1+2Z_2)}, \\
Z_1 & \triangleq & 1+(1-\tilde a^2)^{1/3}[(1+\tilde a)^{1/3}+(1-\tilde a)^{1/3}],\qquad Z_2 \triangleq \sqrt{3\tilde a^2+(Z_1)^2}.
\eea

\paragraph{Minimal polar angle.} 

\begin{figure}[!hbt]
    \centering
    \includegraphics[width=13cm]{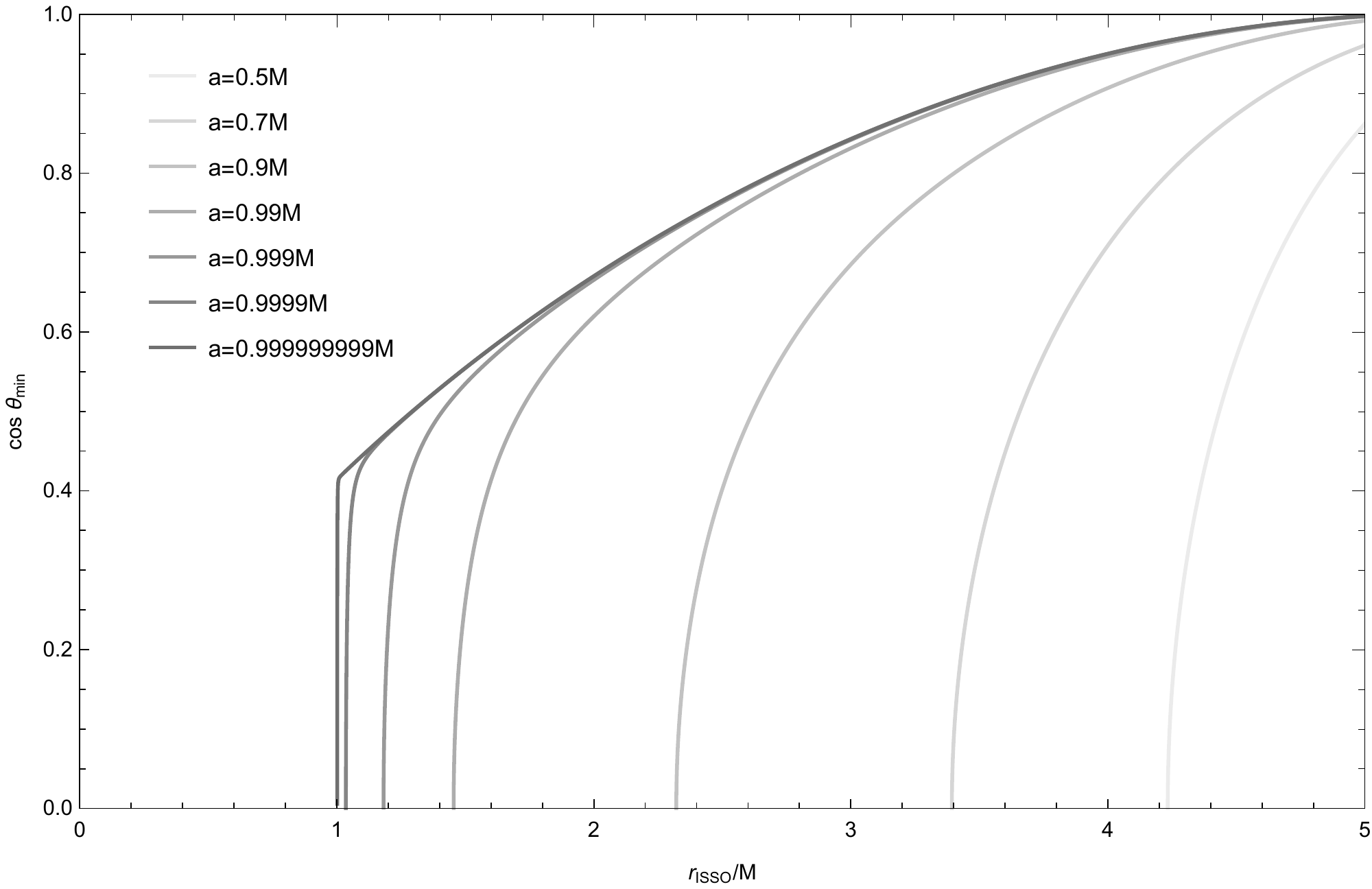}
    \caption{$\cos\theta_\text{\text{min}}$ as a function of ISSO radius for several black hole spins $a$.}
    \label{fig:hugues}
\end{figure}

In the generic case $\ell \neq 0$, the polar motion is pendular -- i.e., oscillating around the equator in the interval $[\theta_\text{\text{min}},\pi-\theta_\text{\text{min}}]$. The minimal angle as a function of the spin $a$ and ISCO radius $\hat r_{\text{ISSO}}$ can simply be found by solving numerically the three equations \eqref{defISSO} that define the ISSO together with the condition that there is a polar turning point, $\Theta(\cos \theta_{\text{\text{min}}}) = 0$ where $\Theta(\cos^2\theta)$ is defined in \eqref{eq:kerr_vtheta}. The resulting minimal angle is displayed in Fig. \ref{fig:hugues} for a large range of spins including nearly extremal. This completes a similar plot drawn in Ref. \cite{Apte:2019txp} for spins far from extremality.

\begin{figure}[!hbt]
    \centering
    \includegraphics[scale=0.5]{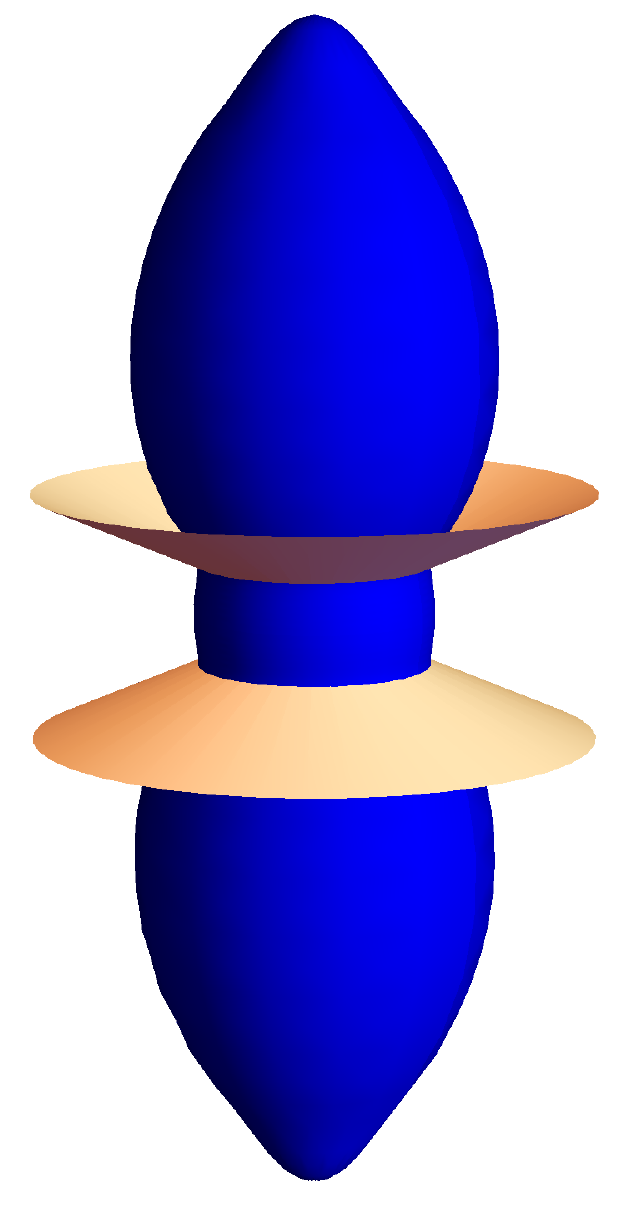}
    \caption{Euclidean embedding of the ISSO using the Boyer-Lindquist radius $\hat r$, azimuthal angle $\phi$ and polar angle $\theta$ for $a=0.9999$. A cone is drawn at the critical polar angle beyond which the ISSO lies in the NHEK region. In the extremal limit, this polar angle is $\theta \approx 65^\circ$.}
    \label{fig:cone}
\end{figure}

We note that for high-spins, the radius asymptotes to $\hat r = M$ and the minimal angle reaches a critical value around $0.42$ radians or $65^\circ$. When the motion reaches regions sufficiently far from the equatorial plane, the ISSO radius increases steeply and leaves the near-horizon region $\rhat\simeq M$. Another graphical representation of this behavior is shown in Fig. \ref{fig:cone}. We will explain these features in the next section. 

\subsection{The NHEK spherical orbit and the high-spin ISSOs}

In the high-spin limit $a \rightarrow M$, the prograde ISSOs are characterized by the following Boyer-Lindquist energies and angular momentum:
\bea\label{ISSOext}
\hat E_{\text{ISSO}}= \frac{1}{\sqrt{3}M}\sqrt{M^2 \mu^2 +Q},\qquad \ell_{\text{ISSO}} = + 2M \hat E_{\text{ISSO}} 
\eea
and the following Boyer-Lindquist radius:
\bea 
\hat r_{\text{ISSO}} = M + M\left( \frac{Q+M^2 \mu^2}{-Q+\frac{M^2 \mu^2}{2}} \right)^{1/3} \lambda^{2/3}+\mathcal O(\lambda^{4/3}). \label{hatrISSO}
\eea
Given the scaling in $\lambda$, for the range 
\bea
0 \leq Q \leq \frac{M^2 \mu^2}{2},\label{rangeQ}
\eea
the ISSOs belong to the NHEK geometry and admit the NHEK radius 
\bea
R = R_{\text{ISSO}} \triangleq  \left( \frac{Q+M^2 \mu^2}{-Q+\frac{M^2 \mu^2}{2}} \right)^{1/3}. 
\eea
In particular, the ISCO has the minimal radius $R_{\text{ISCO}}=2^{1/3}$. In terms of NHEK quantities, the orbits admit a critical angular momentum and a vanishing NHEK energy, 
\bea
\ell = \ell_* \triangleq \frac{2}{\sqrt{3}}\sqrt{Q+M^2 \mu^2},\qquad E = 0. 
\eea
In the high-spin limit, the prograde ISSOs in the range \eqref{rangeQ} are therefore exactly the $\text{Spherical}_*(Q)$ orbits in the classification of Sec. \ref{sec:NHEK}.  The prograde ISSOs outside the range \eqref{rangeQ} and the retrograde ISSOs do not belong to the near-horizon geometry and will not be described here. 

In terms of polar behavior, $\text{Spherical}_*(Q)$ orbits are instances of $\text{Pendular}(Q,\ell_*)$ motion (except for $Q=0$, where they are just equatorial orbits). In the range \eqref{rangeQ}, they admit an $\eps_0$ as defined in \eqref{defeps0} given by $\eps_0 = \frac{Q-2 M^2 \mu^2}{3} < 0$, and the angular momentum lies below the value $\ell_\circ$:
\bea
\ell_* \leq \sqrt{2} M \mu < \ell_\circ. 
\eea

The main property of $\text{Pendular}(Q,\ell_*)$ motion is that the polar angle $\theta$ is bounded in an interval around the equator (see \eqref{eq:costh} and \eqref{defzpm}) :
\begin{equation}
\theta\in[\theta_{\text{min}},\pi-\theta_{\text{min}}]
\end{equation}
where
\begin{equation}
\cos\theta_{\text{min}}=\sqrt{z_+}=\sqrt{\frac{Q}{\frac{3}{4}\ell^2_*+\sqrt{\frac{9}{16}\ell_*^4-\frac{\ell_*^2 Q}{2}+Q^2}}}.
\end{equation}

\begin{figure}[!hbtp]\center
\begin{tabular}{ccccc}\vspace{-20pt}
\includegraphics[width=0.2\textwidth]{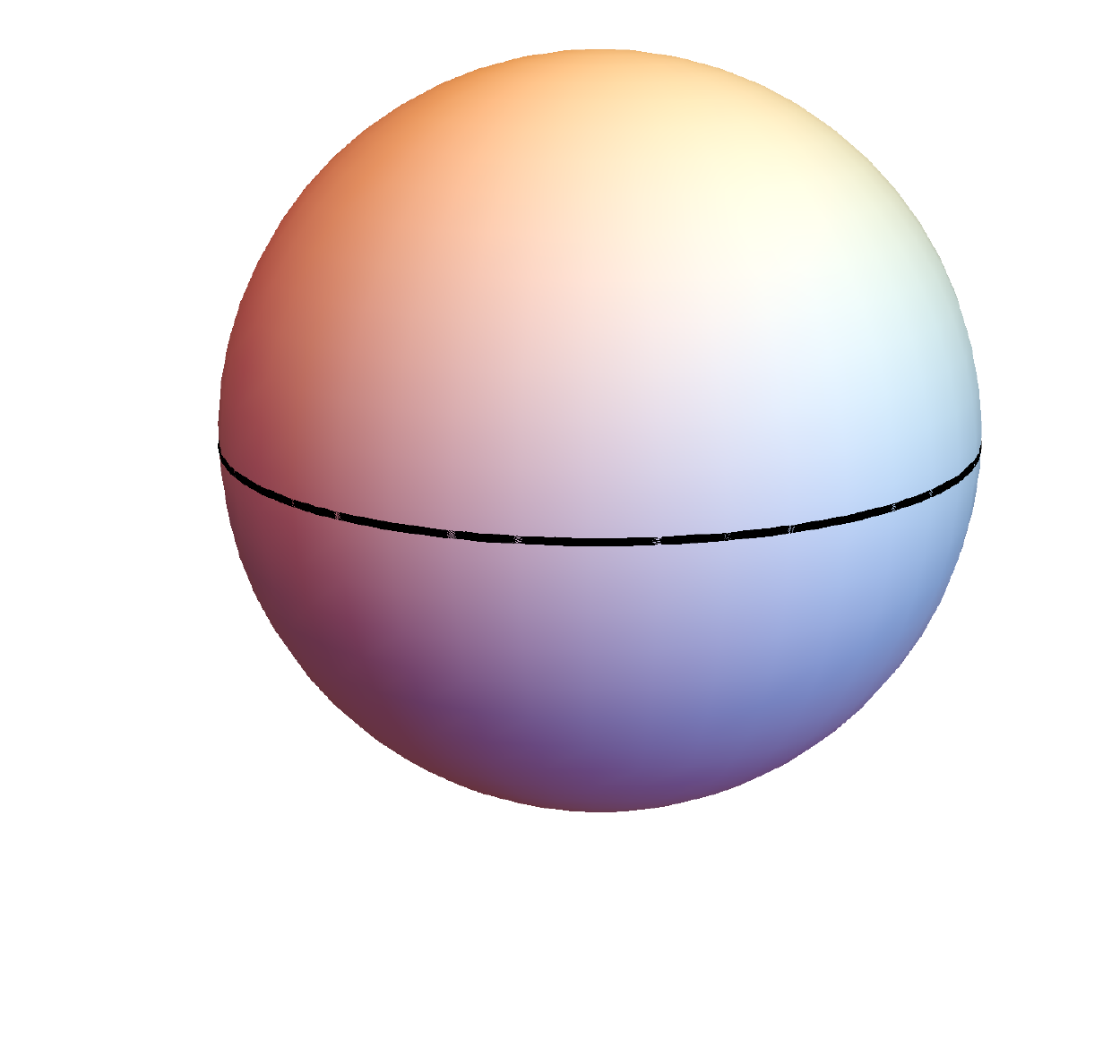} &  \includegraphics[width=0.2\textwidth]{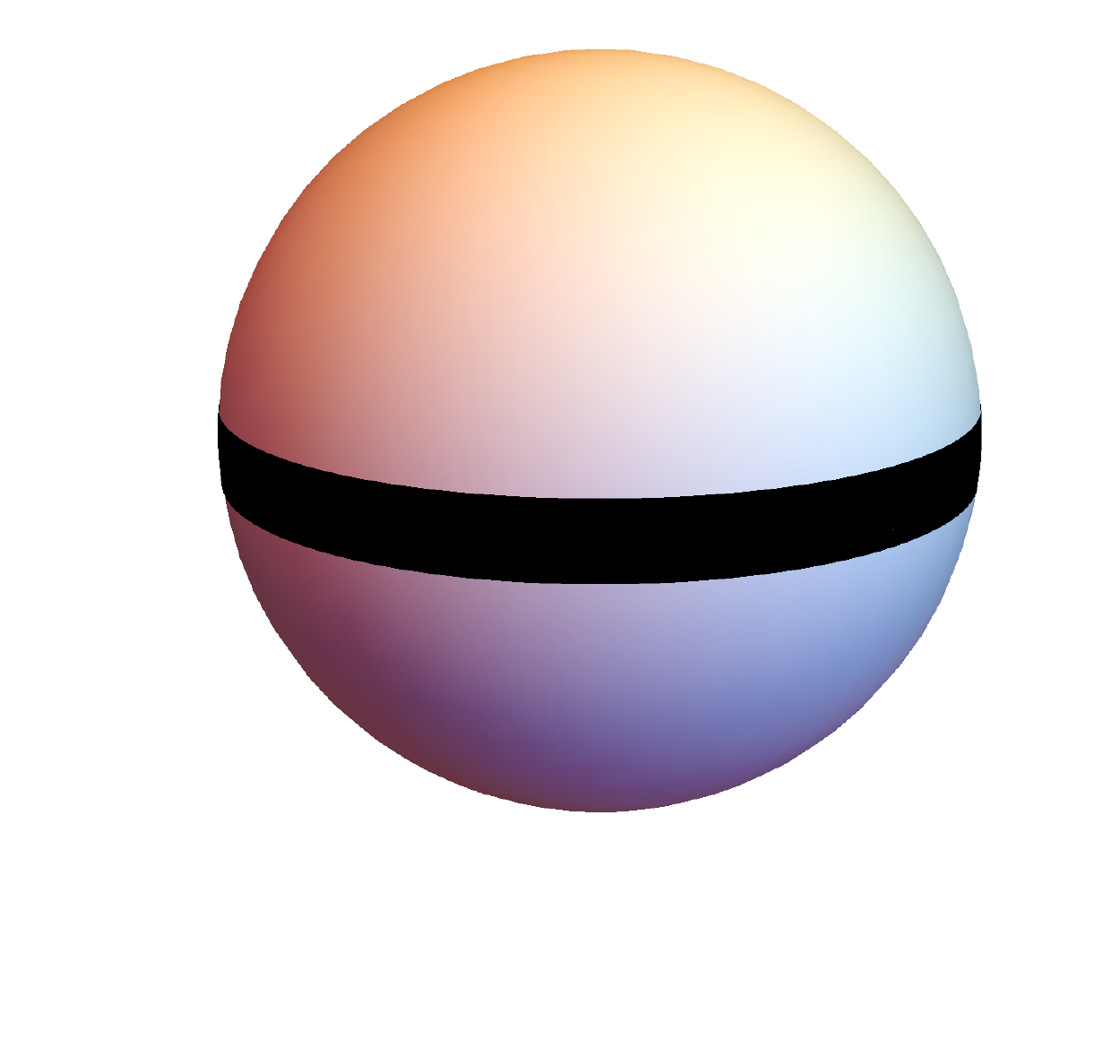} &  \includegraphics[width=0.2\textwidth]{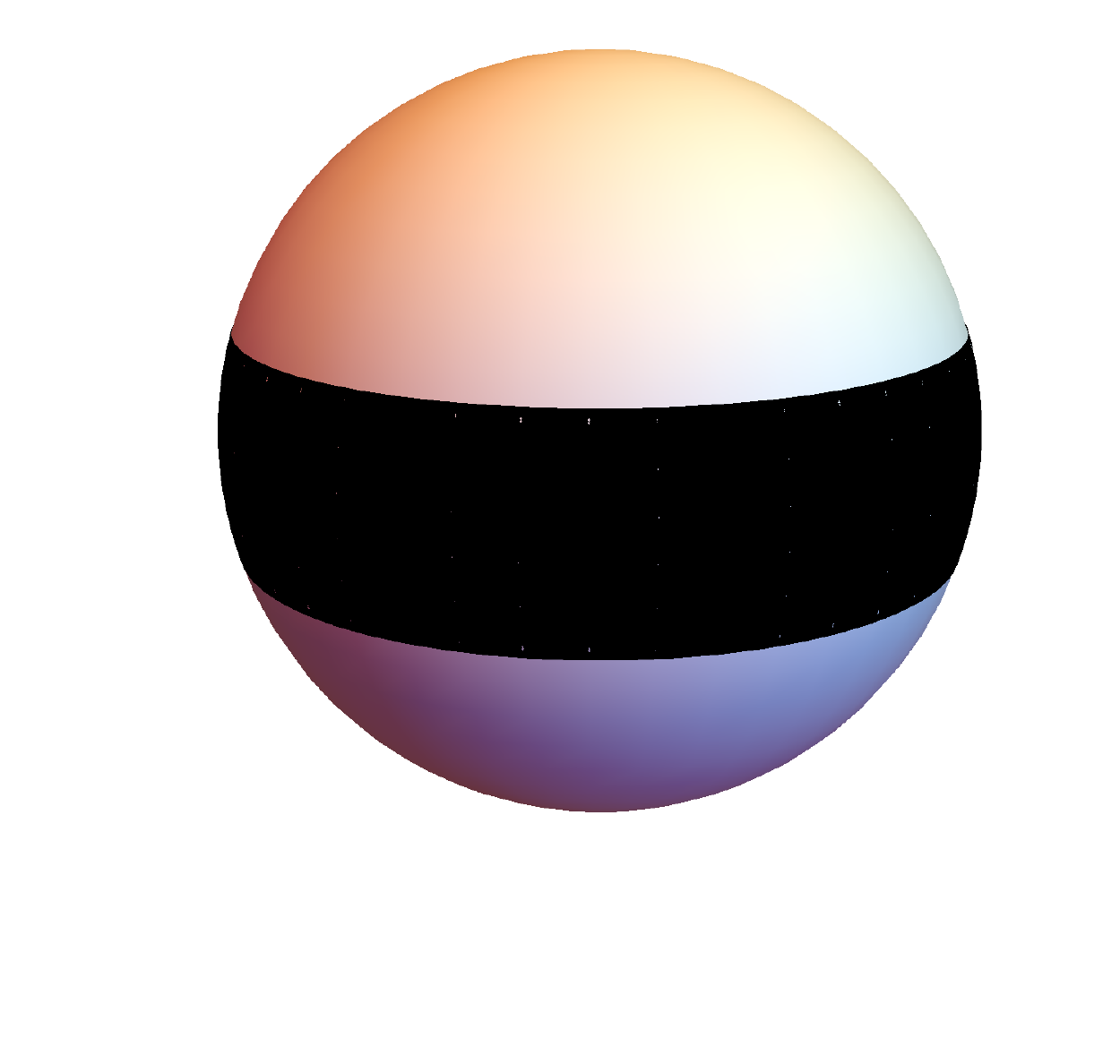} & \shortstack{\ldots\\\rule{0pt}{35pt}} &
\includegraphics[width=0.2\textwidth]{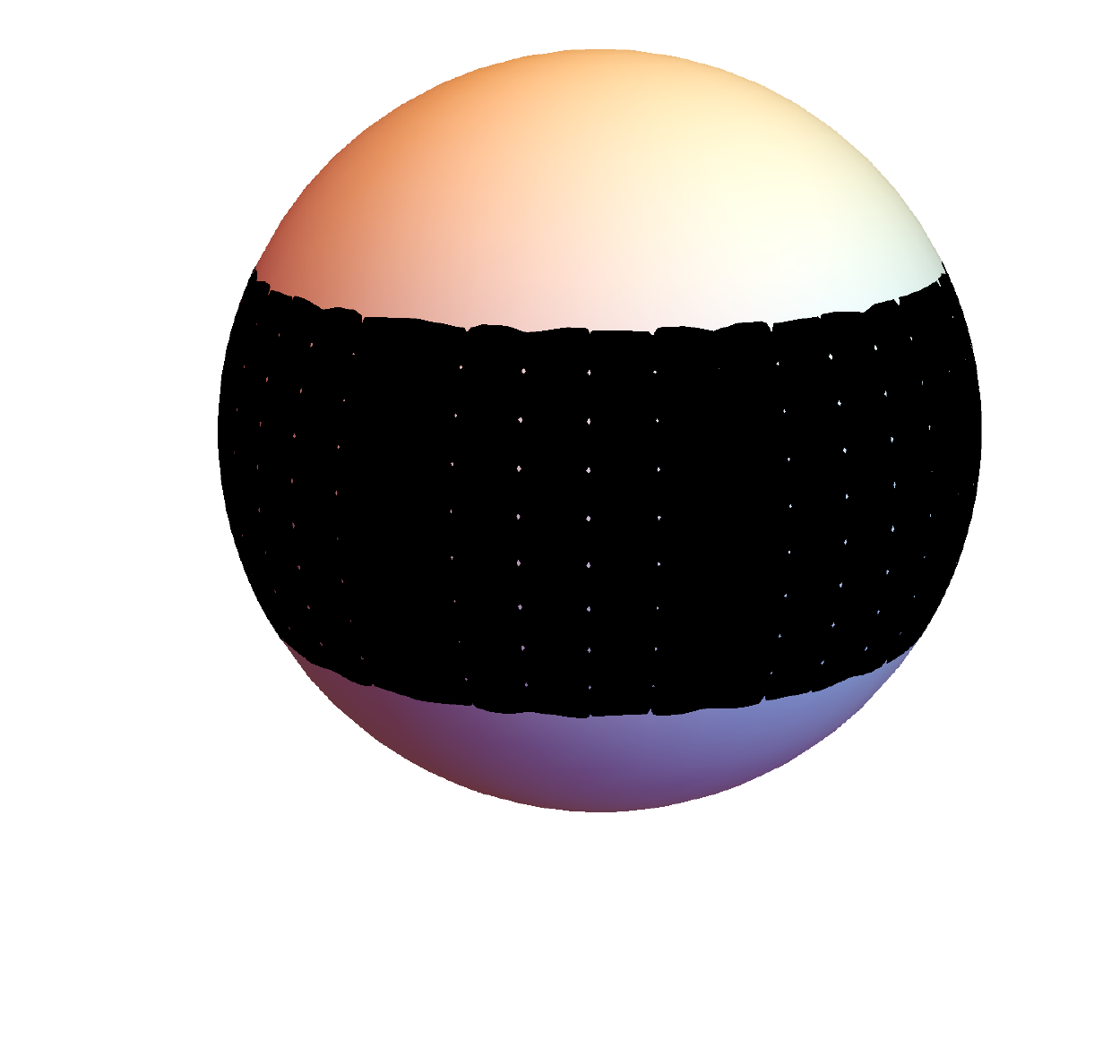}\\
\end{tabular}
\begin{tikzpicture}[scale=1.2]
\draw[thick] (0,0)--(7.5,0);
\draw[dashed,thick] (7.5,0)--(8.5,0);
\draw[->,thick] (8.5,0)--(10,0);
\node[right]() at (10,0) {$Q$};
\end{tikzpicture}
\caption{For increasing $Q \geq 0$, $\text{Spherical}_*(Q)$ orbits can explore a equator-centered band whose width becomes larger, finally reaching for $Q \rightarrow \infty$ the angular range $\theta\in[\theta_{\text{VLS}},\pi-\theta_{\text{VLS}}]$ bounded by the velocity-of-light surface. The prograde IBSOs lie in the near-NHEK region for $Q \leq 2 M^2\mu^2$, which further bounds the angular range.}
\label{fig:opening}
\end{figure}
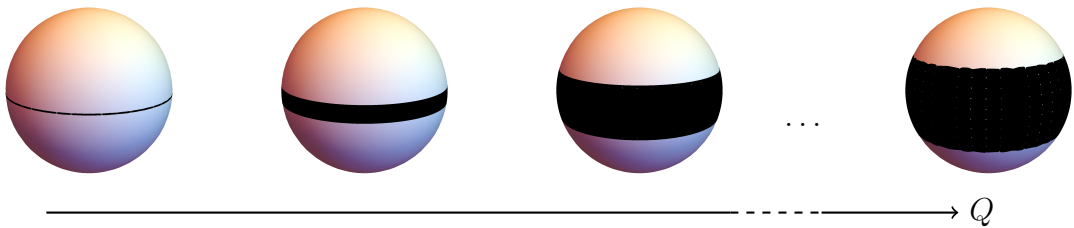

At fixed $M\mu$, $\theta_{\text{min}}(Q)$ is a monotonic function interpolating between the equator $\theta=90^\circ$ at $Q=0$ and $\theta_{\text{VLS}} \triangleq \arccos{\sqrt{2\sqrt{3}-3}} \approx 47^\circ$ for $Q \rightarrow \infty$. The special angle $\theta_{\text{VLS}}$ is the velocity-of-light surface in the NHEK geometry \eqref{eq:NHEK_metric} as described in Sec. \ref{sec:univ}. The ISSO therefore always lies in the region of NHEK spacetime around the equator, where there is no timelike Killing vector. This is depicted in Fig. \ref{fig:opening}.

However, since the ISSO admits the range \eqref{rangeQ} due to its relationship to the asymptotically flat Boyer-Lindquist radius \eqref{hatrISSO}, the limiting angle is reached first for $Q=\frac{M^2 \mu^2}{2}$ at $\arccos{\sqrt{3-2\sqrt{2}}}\approx 65^\circ$. This explains the behavior depicted in Fig. \ref{fig:hugues}. At the time of finalizing this draft, Ref. \cite{Stein:2019buj} found an identical result (see their Eq.(46)), the limiting angle of the ISSO is given by $\arcsin{\sqrt{2(\sqrt{2}-1)}} = \arccos{\sqrt{3-2\sqrt{2}}} \approx 65^\circ$. 

\subsection{The near-NHEK spherical orbits and the high-spin IBSOs}

The innermost bound spherical orbits (IBSOs) are determined by the equations
\bea
\hat R(\hat r) = \p_{\hat r} \hat R(\hat r) = 0,\qquad \hat E = \mu.
\eea
In the high-spin limit $\lambda \rightarrow 0$, the angular momentum and Boyer-Lindquist radius of the prograde IBSOs are given by 
\bea
\ell &=& \ell_\circ \left(1+\frac{\lambda}{\sqrt{2}}\sqrt{1-\frac{Q}{2M^2 \mu^2}}+\mathcal O(\lambda^2)\right), \\
\hat r &=& M \left(1+ \frac{\sqrt{2}\lambda}{\sqrt{1-\frac{Q}{2M^2 \mu^2}}}+\mathcal O(\lambda^2)\right) \eea
where $\ell_\circ \equiv 2M \mu$. In particular, for $Q=0$ we recover the scaling of the innermost bound circular orbit (IBCO) \cite{Bardeen:1972fi}. Given the scaling $\sim \lambda$, the prograde IBCOs therefore lie in the near-NHEK region for all $Q < 2M^2 \mu^2$. Using \eqref{eq:cvnn}--\eqref{eN}, the angular momentum, near-NHEK energy and near-NHEK radius are given in the high-spin limit by 
\bea
\ell &=&  \ell_\circ,\\
\frac{e}{\kappa} &=& -\sqrt{2M^2 \mu^2 -Q}, \\
\frac{r}{\kappa} &=& \frac{\sqrt{2}\lambda}{\sqrt{1-\frac{Q}{2M^2 \mu^2}}}. 
\eea
The prograde IBCOs in the range $0 \leq Q < 2M^2 \mu^2$ are described by instances of Spherical$(\ell)$ orbits. In terms of polar motion, $Q=0$ are equatorial and $Q>0$ are pendular of class Pendular$_\circ(Q)$; see Table \ref{table:taxPolar}. The polar range is determined as $\theta_{\text{\text{min}}} \leq \theta \leq \pi - \theta_{\text{\text{min}}}$ where
\bea
\theta_{\text{\text{min}}} = \arccos\sqrt{\frac{Q}{Q+\ell_\circ^2}}. 
\eea
The maximal polar angle reachable within the near-NHEK region by IBSOs is obtained for the limiting value $Q = 2M^2 \mu^2$ at
\bea
\theta_{\text{\text{min}}} = \arccos\sqrt{1/3} = \arcsin{\sqrt{2/3}} \approx 55^\circ.
\eea 
This critical angle was also previously obtained in Refs. \cite{Hod:2017uof,Stein:2019buj}. Finally, note that spherical photon orbits in the high-spin limit were also discussed in Refs. \cite{Yang:2012he,Hod:2012ax}.

\section{Conformal mappings between radial classes}\label{sec:classes}

The near-horizon region of near-extremal Kerr black holes admits four Killing vectors forming the group $SL(2,\mathbb R) \times U(1)$, hereafter denoted as the conformal group $G$. The geodesic equations are invariant under $G$ and the geodesics therefore transform under the action of $G$. Moreover, a group generated by four $\mathbb Z_2$ symmetries exists that preserve the geodesic equations. The subgroup preserving the domain $R > 0$ for NHEK (or $r > 0$ for near-NHEK) is generated by the $\uparrow\!\downarrow$-flip \eqref{PTflip}, which flips the geodesic orientation, and two additional $\mathbb Z_2$ transformations that preserve the geodesic orientation: namely, the parity flip
\bea
\theta \rightarrow \pi - \theta,\qquad  \Phi \rightarrow \Phi + \pi,\qquad s_\theta^i\rightarrow -s_\theta^i, \label{parity}
\eea
and the $\rightleftarrows$-flip
  \bea
    T \rightarrow -T,\qquad \Phi \rightarrow -\Phi,\qquad \lambda \rightarrow -\lambda,\qquad s_R^i \rightarrow -s_R^i,\qquad s_\theta^i \rightarrow -s_\theta^i .\label{Tphiflip}
\eea
The last discrete transformation that we use as a basis is the \rotatebox[origin=c]{-45}{$\rightleftarrows$}-flip
\bea
R \rightarrow -R, \qquad \Phi \rightarrow -\Phi, \qquad \ell \rightarrow -\ell, \qquad s^i_R \rightarrow -s^i_R. 
\eea

\begin{figure}[h!]
    \centering
    \begin{tabular}{c|c}
      \includegraphics[width=7cm]{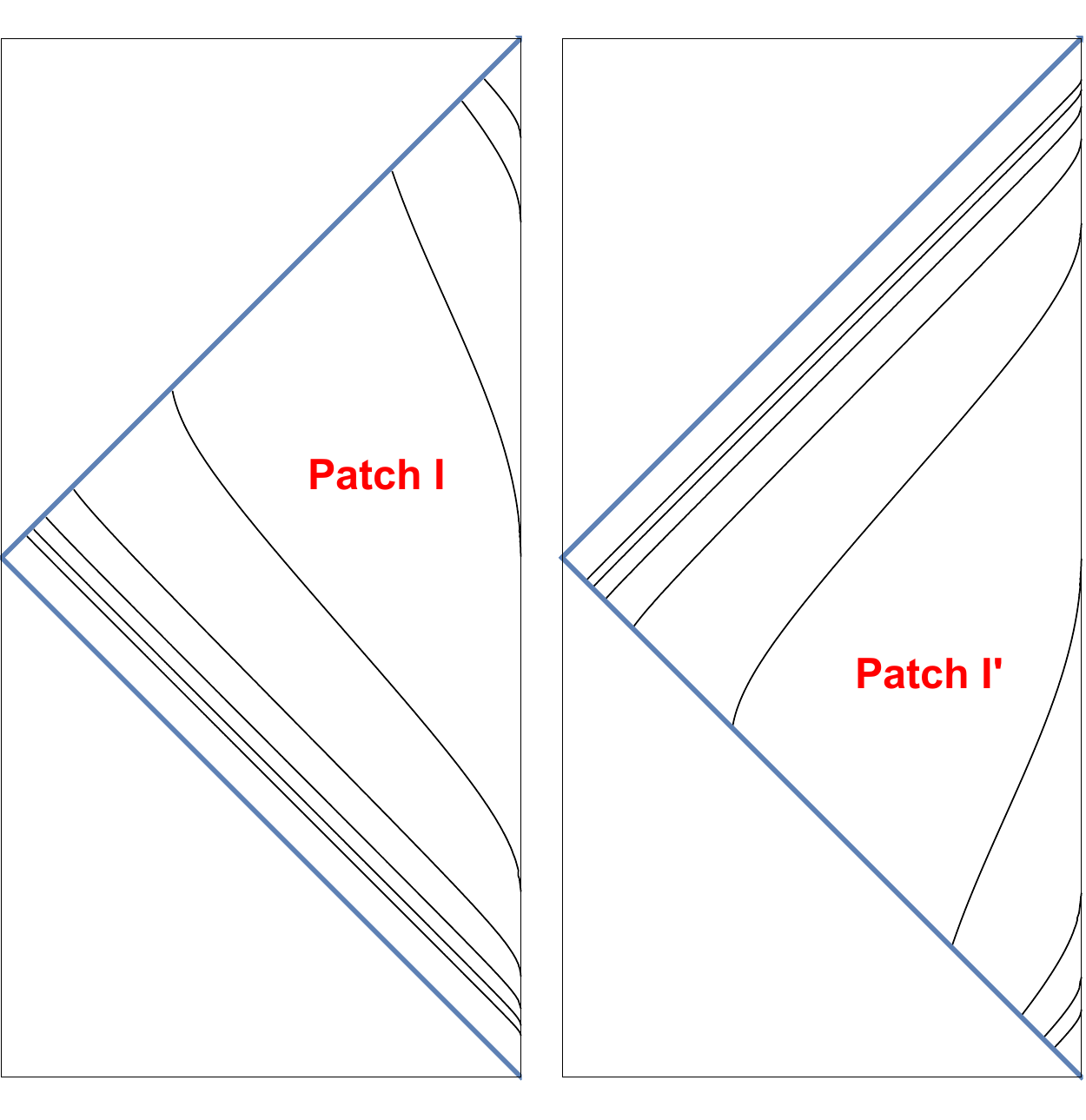}   & \includegraphics[width=7cm]{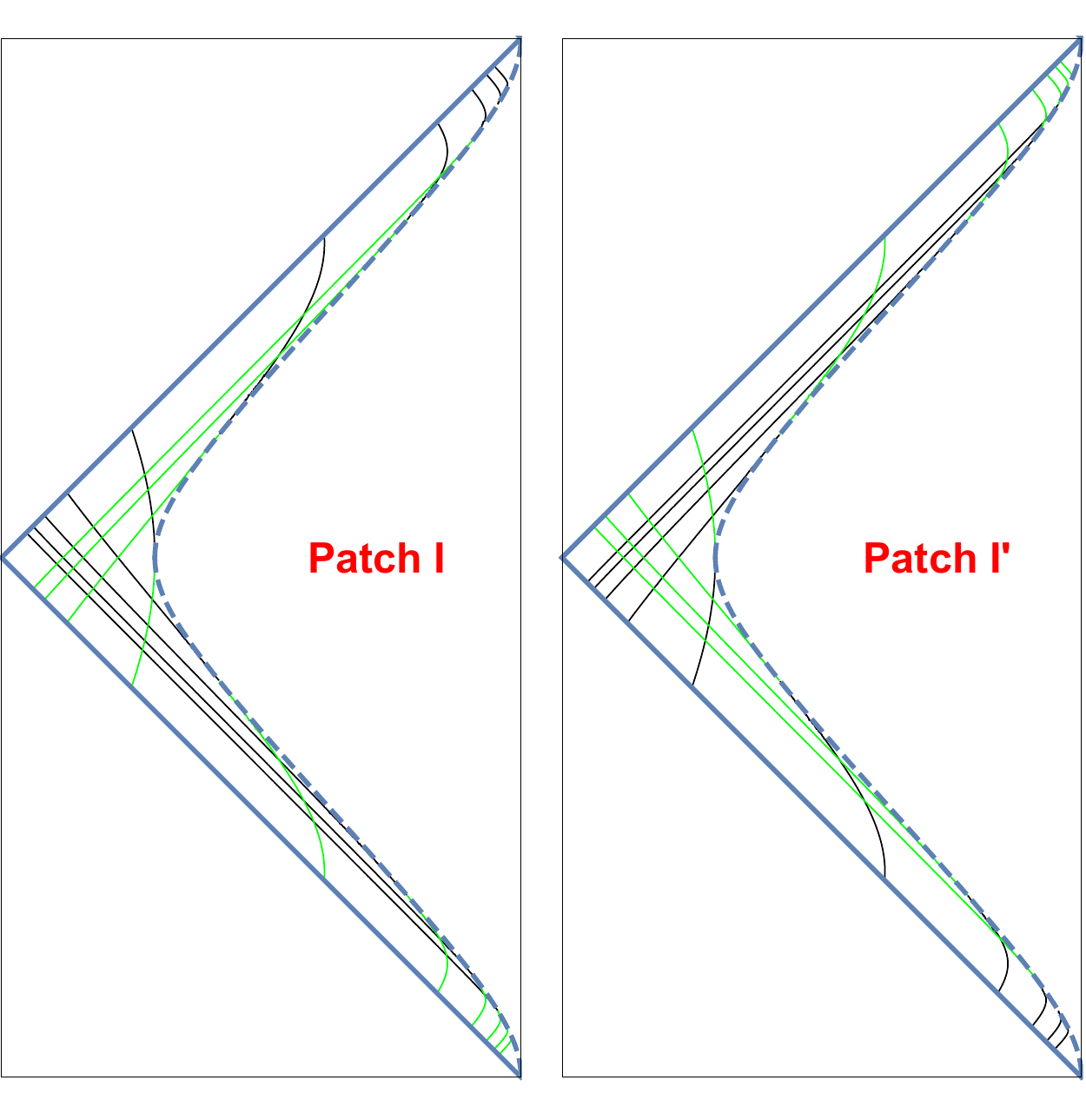}  \\
        (a) & (b)
    \end{tabular}
    \caption{Penrose diagram of NHEK spacetime depicting the action of the $\rightleftarrows$-flip on (a) plunging and (b) bounded geodesics. Under this transformation, a trajectory belonging to the patch I is mapped to an orbit of the patch I'. While plunging geodesics become outward ones, bounded motion remains bounded. The energy and angular momentum of the trajectory are unchanged.}
    \label{fig:flipTPhi}
\end{figure}

\begin{figure}
    \centering
    \begin{tabular}{ccc}
     \includegraphics[width=4cm]{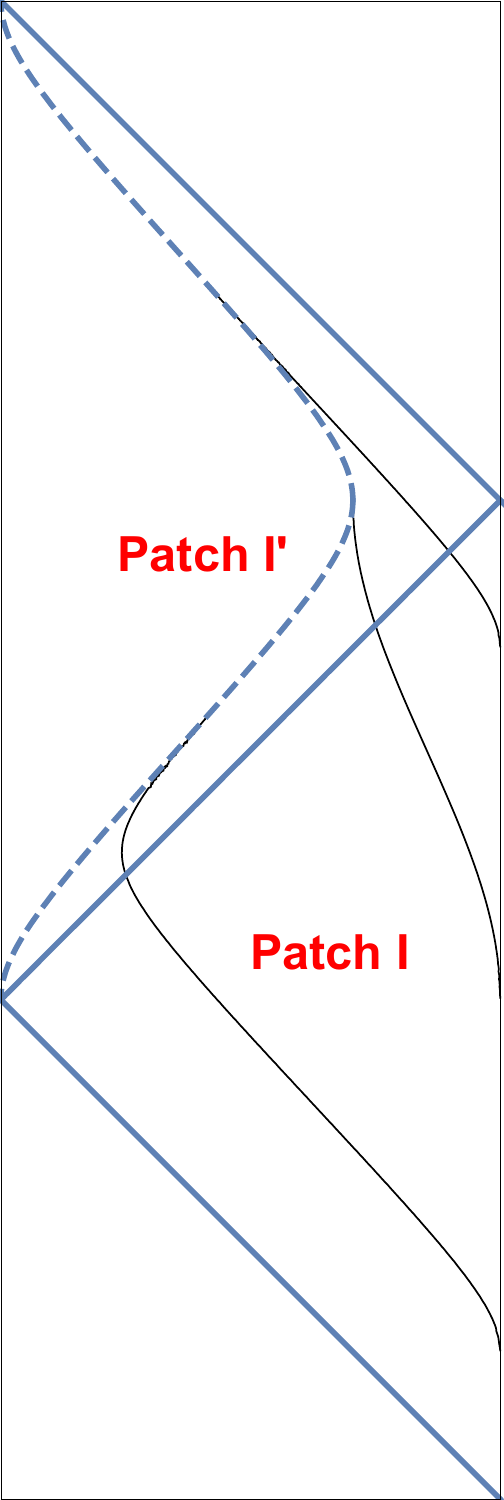} & \includegraphics[width=4cm]{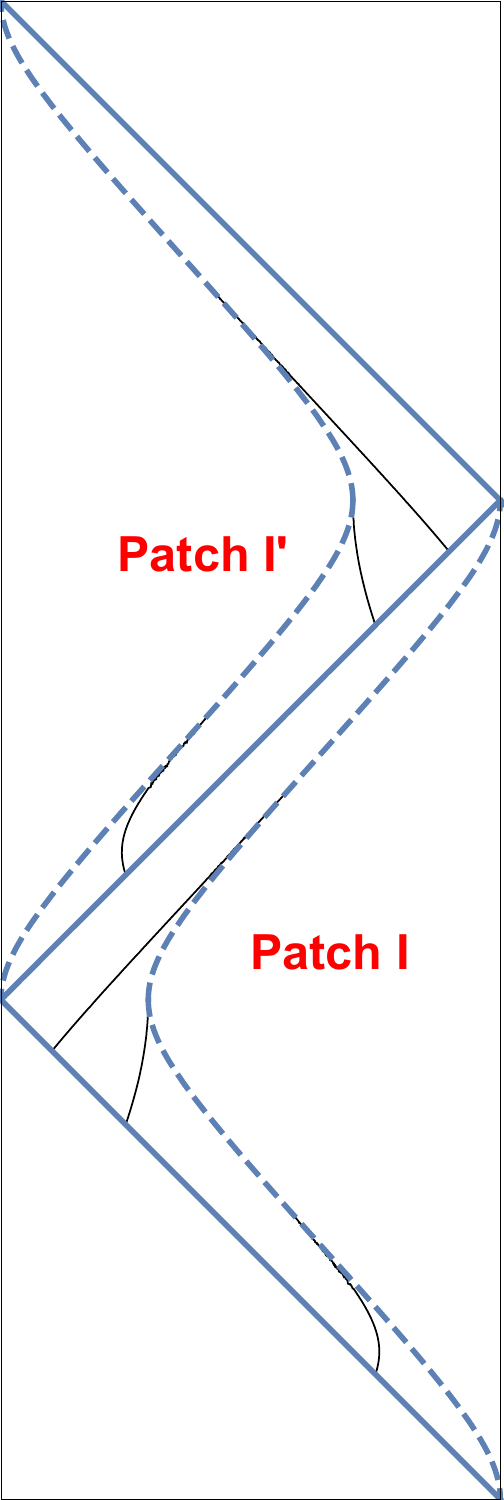} & \includegraphics[width=4cm]{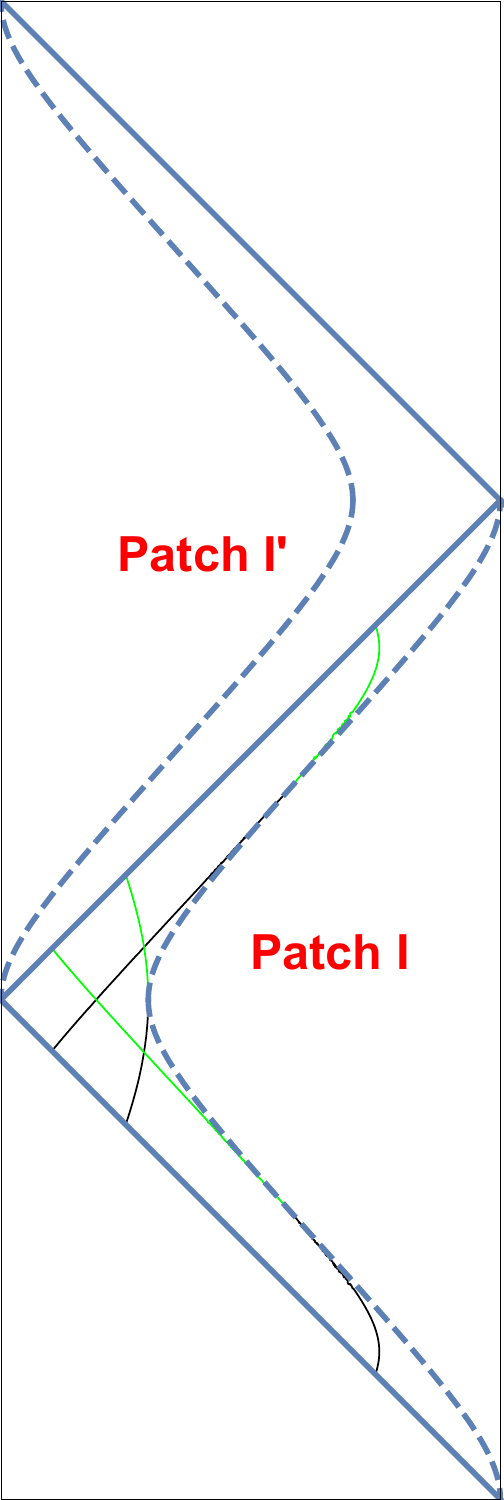}\\  (a) & (b) & (c) 
    \end{tabular}
    \caption{Penrose diagram representation of the construction of a critical NHEK bounded geodesic from a plunging one. (a) Continuation of the trajectory beyond the horizon in patch I' until the radial potential root (depicted with dashes); (b) \protect\rotatebox[origin=c]{-45}{$\rightleftarrows$}-flip which brings the part of the bounded geodesic before the turning point in the NHEK Poincar\'e patch I; (c) $\rightleftarrows$-flip, which maps the part of the bounded geodesic before the turning point to the part after the turning point.}
    \label{fig:bounded_from_plunging}
\end{figure}

The parity transformation defined in \eqref{parity} leaves each motion invariant and will not be considered further. The $\rightleftarrows$-flip changes the boundary conditions of the geodesics, which may affect their denomination. It maps bounded orbits to bounded orbits, and deflecting orbits to deflecting orbits, but plunging orbits to outward orbits, as illustrated in Fig. \ref{fig:flipTPhi}. For bounded orbits, the part before the turning point is mapped to the part after the turning point, and vice-versa. The \rotatebox[origin=c]{-45}{$\rightleftarrows$}-flip can be used as follows: one first continues a geodesic defined in $R > 0$ beyond the horizon $R = 0$ and the resulting geodesic with $R < 0$ is then mapped to a geodesic in the $R > 0$ region using the \rotatebox[origin=c]{-45}{$\rightleftarrows$}-flip. Together with the action of \eqref{Tphiflip}, it allows us to map plunging orbits with $\ell > 0$ to bounded orbits with $\ell < 0$. This process is illustrated in Fig. \ref{fig:bounded_from_plunging}.

The equivalence classes of equatorial critical and supercritical prograde timelike geodesics under the action of $SL(2,\mathbb R) \times U(1) \times \uparrow\!\downarrow$ symmetry were derived in  Ref. \cite{Compere:2017hsi} following earlier work \cite{Porfyriadis:2014fja,Hadar:2014dpa,Hadar:2015xpa,Hadar:2016vmk}. In this section, we will perform the decomposition of arbitrary geodesics into equivalence classes under the action of $SL(2,\mathbb R) \times U(1) \times \uparrow\!\downarrow \times \rightleftarrows \times$\rotatebox[origin=c]{-45}{$\rightleftarrows$}.

The Casimir $\mathcal{C}$ of $SL(2,\mathbb{R})$ cannot vary upon acting with $G \equiv SL(2,\mathbb{R}) \times U(1)$ transformations. Moreover, the action of the group $G$ acts trivially on the polar coordinate $\theta$. These two properties imply that both $Q$ and $\ell$ are invariant under the action of $G$. In particular, critical, supercritical or subcritical geodesics form distinct classes under $G$. On the contrary, the (near-)NHEK energy $E$ (or $e$) can vary under conformal transformations. Conformal transformations can map NHEK to near-NHEK orbits, and vice-versa. As a result of Propositions \ref{thm:equivNHEK} and \ref{thm:equivnearNHEK}, null geodesics can be treated on the same footing as timelike geodesics.

A conformal transformation belonging to $SL(2,\mathbb{R}) \times U(1)$ maps (near)-NHEK spacetime parametrized by $(T,R,\theta,\Phi)$ to (near-)NHEK spacetime parametrized by $(\bar T,\bar R,\theta,\bar \Phi)$\footnote{We denote here without distinction NHEK and near-NHEK coordinates with capital letters.} where 
\begin{align}
\overline{T}&=\overline{T}(T,R),\nonumber\\
\overline{R}&=\overline{R}(T,R),\label{eq:conformal_tfo}\\
\overline{\Phi}&=\Phi+\delta\bar{\Phi}(T,R)\nonumber.
\end{align}
The geodesic equations in (near)-NHEK imply $T=T(R)$. Therefore, the action of conformal symmetries reduces to an action on the radial motion, leaving the polar motion unchanged. More precisely, in the decomposition of $\Phi(\lambda)$ \eqref{eq:NHEKphi}--\eqref{eqn:nnPhi} in terms of a radial part and a polar part, the polar part will remain untouched by conformal transformations.

It was shown in Ref. \cite{Compere:2017hsi} that each equivalence class of equatorial prograde critical (respectively, supercritical) geodesics with incoming boundary conditions under $G \times \uparrow\!\downarrow$ admits a distinguished simple representative, namely the NHEK (respectively, near-NHEK) circular orbits. After analysis, we obtain that each geodesic equivalence class under $G \times \uparrow\!\downarrow \times \rightleftarrows \times$\rotatebox[origin=c]{-45}{$\rightleftarrows$} admits a spherical orbit as the simplest representative as illustrated in Fig. \ref{fig:conformal_mappings}. Past directed geodesics must be considered as intermediate steps in order to relate each future directed geodesic to spherical geodesics.  Supercritical orbits ($\ell^2 > \ell^2_*$) admit the near-NHEK Spherical$(\ell)$ orbit as a representative and critical orbits ($\ell = \pm\ell_*$) admit the NHEK Spherical$_*$ orbit as a representative. No subcritical spherical geodesic exists. However, we introduce an analytically continued complex subcritical geodesic by continuing the radius $R_0 \mapsto i R_0$ and show that it generates the subcritical class.

The explicit formulas for the three categories of equivalence classes of orbits under $G \times \uparrow\!\downarrow \times \rightleftarrows \times$\rotatebox[origin=c]{-45}{$\rightleftarrows$} are given in the following sections. We will denote the final coordinates and orbital parameters reached by the conformal mappings with bars.

\subsection{Critical $\mathcal{C}=0$}
\paragraph{Spherical$_*$ $\Leftrightarrow$ Plunging$_*(E)$ (NHEK/NHEK).} The conformal mapping is given by
\begin{align}
    \bar T&= - \frac{R^2T}{R^2T^2-1},\nonumber\\
    \bar R&=\frac{R^2T^2-1}{R},\label{ct1}\\
    \bar\Phi&=\Phi+\log\frac{RT+1}{RT-1}-i\pi.\nonumber
\end{align}
It maps a (future-directed) NHEK spherical trajectory of radius $R_0$ to a (future-directed)  critical plunge of energy $\bar E=\frac{2\ell_*}{R_0} > 0$.

\paragraph{Spherical$_*$ $\Leftrightarrow$ Plunging$_*$ (NHEK/near-NHEK).} One performs the NHEK/near-NHEK diffeomorphism $(T,R,\theta,\Phi)\to(\bar t,\bar R,\theta,\bar\phi)$, whose explicit form is
\begin{align}
T&=-\exp\qty(-\kappa \bar t)\frac{\bar R}{\sqrt{\bar R^2-\kappa^2}}\nonumber,\\
    R&=\frac{1}{\kappa}\exp\qty(\kappa \bar t)\sqrt{\bar R^2-\kappa^2},\label{ct2}\\
    \Phi&=\phi-\frac{1}{2}\log\frac{\bar R-\kappa}{\bar R+\kappa}.\nonumber
\end{align}
Its inverse is
\begin{align}
    \bar t &= \frac{1}{\kappa}\log\frac{R}{\sqrt{R^2T^2-1}},\nonumber\\
    \bar R &=-\kappa RT,\\
    \bar\phi &= \Phi+\frac{1}{2}\log\frac{RT+1}{RT-1}\nonumber
\end{align}
for $R>0$ and $RT<-1$. The orbital parameters are related as
\begin{align}
    R_0=\frac{1}{\kappa}\exp\qty(\kappa t_0),\qquad\Phi_0=\phi_0-\frac{3}{4}.
\end{align}

\paragraph{Plunging$_*$ $\Leftrightarrow$ Outward$_*$ (near-NHEK/near-NHEK).} The orbits are related by the $\rightleftarrows$-flip \eqref{Tphiflip}.

\paragraph{Plunging$_*$ $\Leftrightarrow$ Plunging$_*(e)$ (near-NHEK/near-NHEK).}
The two (future-directed) orbits are related via the diffeomorphism 
\begin{align}
\bar t &= \frac{1}{2\kappa}\log\frac{\sqrt{R^2-\kappa^2}\cosh{\kappa t}-R}{\sqrt{R^2-\kappa^2}\cosh{\kappa t}+R}-\frac{i\pi}{\kappa},\nonumber\\
    \bar R &= \sqrt{R^2-\kappa^2}\sinh{\kappa t},\label{ct3}\\
    \bar \phi &= \phi + \frac{1}{2}\log\frac{R\sinh\kappa t+\kappa\cosh\kappa t}{R\sinh\kappa t-\kappa\cosh\kappa t}\nonumber.
\end{align}
The energy of the new trajectory is a function of the initial time $t_0$ of the former one:
\begin{equation}
    \bar e=\kappa^2\ell_*\exp\qty(-\kappa t_0) > 0.
\end{equation}

\paragraph{Plunging$_*(e)$ $\Leftrightarrow$ Outward$_*(e)$ (near-NHEK/near-NHEK).} The orbits are related by the $\rightleftarrows$-flip.

\paragraph{Plunging$_*(E)$ $\Leftrightarrow$ Bounded$_*^-(E)$ (NHEK/NHEK).} The critical bounded orbit is obtained from the plunging orbit by a continuation of the trajectory beyond the horizon ($R<0$) combined with $\mathbb{Z}_2$ flips. One must proceed in three steps:
\begin{enumerate}
    \item Continue the plunge defined from the physical domain $0\leq R\leq \infty$ to its whole domain of definition $R_0\leq R\leq\infty$ (\textit{i.e.}, up to the root of the radial potential $R_0=-\frac{e}{2\ell_*}$) and consider now only the part of the trajectory located beyond the horizon $R_0 \leq R\leq 0$.
    \item Apply the \rotatebox[origin=c]{-45}{$\rightleftarrows$}-flip to the latter part of the solution. This transformation restores the positivity of the radial coordinate. It preserves the time orientation of the geodesic but flips the sign of its angular momentum $\ell_*\to-\ell_*$. The new domain of definition of the trajectory is consequently $0\leq R \leq \frac{E}{2\ell_*}$.
    \item The procedure outlined above only leads to the part of the geodesic with $R'(\lambda)>0$, which is located before the turning point. As outlined in Appendix \ref{app:equatorial}, the part of a bounded trajectory located after the turning point can be obtained from the one located before it by a $\rightleftarrows$-flip. 
\end{enumerate}
This whole procedure is represented in Fig. \ref{fig:bounded_from_plunging}.

\paragraph{Plunging$_*(e)$ $\Leftrightarrow$ Bounded$_*^-(e)$ (near-NHEK/near-NHEK).} The mapping is similar to the one outlined above using the \rotatebox[origin=c]{-45}{$\rightleftarrows$}-flip. One subtlety is that one should start with the Plunging$_*(e)$ orbit with $e > \kappa \ell_*$ in order to obtain the future-directed Bounded$_*^-(e)$ orbit.

\paragraph{Plunging$_*(e)$ $\Leftrightarrow$ Bounded$_*(-e)$ (near-NHEK/near-NHEK).} We apply the \rotatebox[origin=c]{-45}{$\rightleftarrows$}-flip as outlined in the previous paragraph, but now choosing $0< e < \kappa \ell_*$. This leads to a retrograde past-directed bounded orbit. The future-directed prograde geodesic is then reached using the $\uparrow\!\downarrow$-flip.

\subsection{Supercritical $\mathcal{C}<0$}
\paragraph{Spherical$(\ell)$ $\Leftrightarrow$ Marginal$(\ell)$ (near-NHEK/NHEK).}
One applies the NHEK/near-NHEK diffeomorphism
\begin{align}
    T&=-\exp\qty(-\kappa \bar t)\frac{\bar R}{\sqrt{\bar R^2-\kappa^2}}\nonumber,\\
    R&=\frac{1}{\kappa}\exp\qty(\kappa \bar t)\sqrt{\bar R^2-\kappa^2},\label{eqn:diffeo}\\
    \Phi&=\phi-\frac{1}{2}\log\frac{\bar R-\kappa}{\bar R+\kappa}\nonumber
\end{align}
which maps the orbit Spherical$(\ell)$ on the past-directed Marginal$(-\ell)$ orbit. The future-directed Marginal$(\ell)$ orbit is recovered by composing this transformation with a $\uparrow\!\downarrow$-flip.

\paragraph{Marginal$(\ell)$ $\Leftrightarrow$ Plunging$(E,\ell)$ or Def\mbox{}lecting$(E,\ell)$ (NHEK/NHEK).}
One performs the transformation ($\zeta\neq 0$)
\begin{align}
	\bar T &=\frac{1}{\bar R}\frac{2R^2T\cos\zeta-(1+R^2(1-T^2))\sin\zeta}{2 R},\nonumber\\
    \bar R &=\frac{R^2(1+T^2)-1+(1+R^2(1-T^2))\cos\zeta+2R^2T\sin\zeta}{2R},\label{eqn:NHEK_shift}\\
    \bar\Phi &= \Phi+\log\frac{\cos\frac{\zeta}{2}R+\sin\frac{\zeta}{2}(RT+1)}{\cos\frac{\zeta}{2}R+\sin\frac{\zeta}{2}(RT-1)}.\nonumber
\end{align}
As outlined in Ref. \cite{Compere:2017hsi}, this mapping can be viewed as the action on Poincaré NHEK coordinates of a shift of the global NHEK time $\tau\to\tau-\zeta$. The energy of the final orbit is
\begin{align}
    \bar E=\sqrt{-\mathcal C}\qty(\sin\zeta+T_0(\cos\zeta-1)).
\end{align}
We directly see that any energy $E\neq 0$ can be reached by conveniently choosing the values of $T_0$ and $\zeta$. 

\paragraph{Plunging$(E,\ell)$ $\Leftrightarrow$ Outward$(E,\ell)$ (NHEK/NHEK).} The orbits are related by the $\rightleftarrows$-flip.

\paragraph{Plunging$(E,\ell)$ $\Leftrightarrow$ Bounded$_>(E,-\ell)$ (NHEK/NHEK).} The mapping consists in extending the radial range of the plunging orbit beyond the horizon, $R<0$, then using the \rotatebox[origin=c]{-45}{$\rightleftarrows$}-flip, which leads to the Bounded$_>(E,-\ell)$ orbit.

\paragraph{Spherical$(\ell)$ $\Leftrightarrow$ Plunging$(e,\ell)$ or Def\mbox{}lecting$(e,\ell)$ (near-NHEK/ near-NHEK).}
One uses the diffeomorphism ($\chi\neq\pm 1$)
\begin{align}
	t &= \frac{1}{\kappa}\log\frac{\sqrt{\bar R^2-\kappa^2}\cosh \kappa\bar t-\bar R}{\sqrt{ R^2-\kappa^2}}\nonumber,\\
     R &=\sqrt{\bar R^2-\kappa^2}\qty(\sinh \kappa\bar t+\chi\cosh \kappa\bar t)-\chi\bar R,\label{eqn:nn_shift}\\
    \phi &= \bar \phi -\frac{1}{2}\log\qty[\frac{\sqrt{\bar R^2-\kappa^2}-\bar R\cosh \kappa\bar t+\kappa\sinh \kappa\bar t}{\sqrt{\bar R^2-\kappa^2}-\bar R\cosh \kappa\bar t-\kappa\sinh \kappa\bar t}\frac{R+\kappa}{R-\kappa}]\nonumber.
\end{align}
This mapping can be seen as a NHEK global time shift written in near-NHEK coordinates; see Refs. \cite{Compere:2017hsi,Hadar:2016vmk}. The explicit inversion formula can be found in Ref. \cite{Hadar:2016vmk}. The energy of the new trajectory reads as
\begin{align}
    \bar e=\kappa\sqrt{-\mathcal{C}}\,\chi.
\end{align}
For $-\frac{\ell}{\sqrt{-\mathcal{C}}}<\chi<-1$, the orbit reached is future-directed and deflecting. The trajectory becomes plunging for $\chi>-1$. Note that for $\abs{\chi}>1$, $\bar t_0=-\frac{1}{2\kappa}\log\frac{1+\chi}{1-\chi}$ is complex and one has to perform an additional shift on $\bar t$ to make it real.

\paragraph{Plunging$(e,\ell)$ $\Leftrightarrow$ Outward$(e,\ell)$ (near-NHEK/near-NHEK).} The orbits are related by the $\rightleftarrows$-flip.

\paragraph{Plunging$(e,\ell)$ $\Leftrightarrow$ Bounded$_>(e,-\ell)$ (near-NHEK/near-NHEK).} The mapping consists in extending the radial range of the plunging orbit with $e > \kappa \ell$ beyond the horizon, $r<0$, then using the \rotatebox[origin=c]{-45}{$\rightleftarrows$}-flip, which leads to the Bounded$_>(e,-\ell)$ orbit.

\subsection{Subcritical $\mathcal{C}>0$}

\paragraph{Complex class of spherical geodesics.} There is no near-NHEK spherical geodesic for $\mathcal{C}>0$. We can nevertheless introduce the formal class of \textit{complex} spherical trajectories
\begin{align}
    t(\lambda) &= -i\frac{\ell}{R_0}\lambda,\\
    R(\lambda) &= i R_0,\qquad R_0\triangleq\frac{\kappa\ell}{\sqrt{\mathcal{C}}},\\
    \phi(\lambda) &= \phi_0-\frac{3}{4}\ell\lambda+\ell\Phi_\theta(\lambda)
\end{align}
which is a formal (but nonphysical) solution of the near-NHEK geodesic equations, of complex near-NHEK ``energy'' $e=-i\kappa\sqrt{\mathcal{C}}$. We will denote this class of solutions as Spherical$_\mathbb{C}(\ell)$ and show that it can be used to generate all subcritical bounded trajectories by acting on it with properly chosen conformal transformations. The parametrized form of the orbit reads as
\begin{align}
    R &= i R_0,\\
    \phi(t) &= \phi_0-\frac{3}{4}iR_0t+\ell\Phi_\theta(\lambda(t)).
\end{align}

\paragraph{Spherical$_\mathbb{C}(\ell)$ $\Leftrightarrow$ Bounded$_<(E,\ell)$.}
 One has to proceed in two steps, mimicking the procedure used to obtain the NHEK Plunging$(E,\ell)$ class:
 \begin{itemize}[label=$\diamond$]
     \item  We apply the near-NHEK/NHEK diffeomorphism \eqref{eqn:diffeo} to a Spherical$_\mathbb{C}(\ell)$ orbit, leading to another complex NHEK geodesic of null energy parametrized by
\begin{align}
    T(R) &= -\frac{i\ell}{\sqrt{C}R},\\
    \Phi(R) &= \Phi_0-\frac{3i\ell}{8\sqrt{C}}\log\frac{\mathcal{C}R^2}{\mathcal{C}+\ell^2}
\end{align}
with the initial azimuthal angle $\Phi_0\triangleq\phi_0-\frac{3\pi\ell}{8\sqrt{C}}-\frac{1}{2}\log\qty(1-\frac{2\sqrt{C}}{\sqrt{C}+i\ell})$. We denote this class as Marginal$_\mathbb{C}(\ell)$.
    \item Second, we apply to the trajectory found above the global time shift \eqref{eqn:NHEK_shift}, but upgraded with an \textit{imaginary} parameter $\zeta\to i\zeta$. This leads to the Bounded$_<(E,\ell)$ class with orbital parameters
    \begin{align}
        \bar E &= \sqrt{\mathcal{C}}\sinh \zeta,\\
        \bar\Phi_0 &= \phi_0-\frac{3\pi\ell}{8 \sqrt{\mathcal C}}-\log\qty(\sqrt{\mathcal{C}}-i\ell)+\frac{3i\ell}{8\sqrt{\mathcal{C}}}\log\qty[\mathcal C(\mathcal{C}+\ell^2)\qty(1+\sqrt{\frac{\mathcal{C}+E^2}{\mathcal{C}}})^2]\nonumber\\
        &~-\frac{3\ell}{8\sqrt{C}}\log\qty[E^2(\mathcal{C}+\ell^2)]+\arctan\frac{\sqrt{\mathcal{C}}}{\ell}.
    \end{align}
    Note that choosing $\zeta>0$ is sufficient to reach the full range of energies allowed for such a geodesic ($E>0$). Any geodesic of orbital parameters ($T_0,\tilde\Phi_0$) can  finally be obtained by performing the transformation $T\to T+T_0$, $\Phi\to\Phi-\bar\Phi_0+\tilde{\Phi}_0$, which also removes the unphysical imaginary part of the azimuthal coordinate.
 \end{itemize}

\paragraph{Spherical$_\mathbb{C}(\ell)$ $\Leftrightarrow$ Bounded$_<(e,\ell)$.} We apply to the Spherical$_\mathbb{C}(\ell)$ class the near-NHEK global time shift \eqref{eqn:nn_shift} upgraded with an imaginary parameter $\chi\to i\chi$ ($\chi\neq\pm1$), leading to a Bounded$_<(e,\ell)$ orbit of parameters 
\begin{align}
\bar e &= \kappa\sqrt{\mathcal{C}}\,\chi,\\
\bar t_0 &= t_0+\frac{i}{\kappa}\arctan\frac{\kappa\sqrt{\mathcal{C}}}{e},\\
\bar\phi_0 &= \bar\phi_0(\phi_0,e,\ell,\mathcal{C},\kappa).
\end{align}
The explicit value of $\bar\phi_0$ is easily calculable, but too long to be reproduced here. To reach a manifestly real orbit of orbital parameters $(\tilde t_0,\tilde\phi_0)$, one has to perform the final shift
\begin{equation}
    t\to t-\bar{t}_0+\tilde t_0,\qquad\phi\to\phi-\bar\phi_0+\tilde\phi_0.
\end{equation}

\begin{landscape}
\begin{figure}
    \centering
    \begin{minipage}{0.95\textwidth}
    \includegraphics[width=\textwidth]{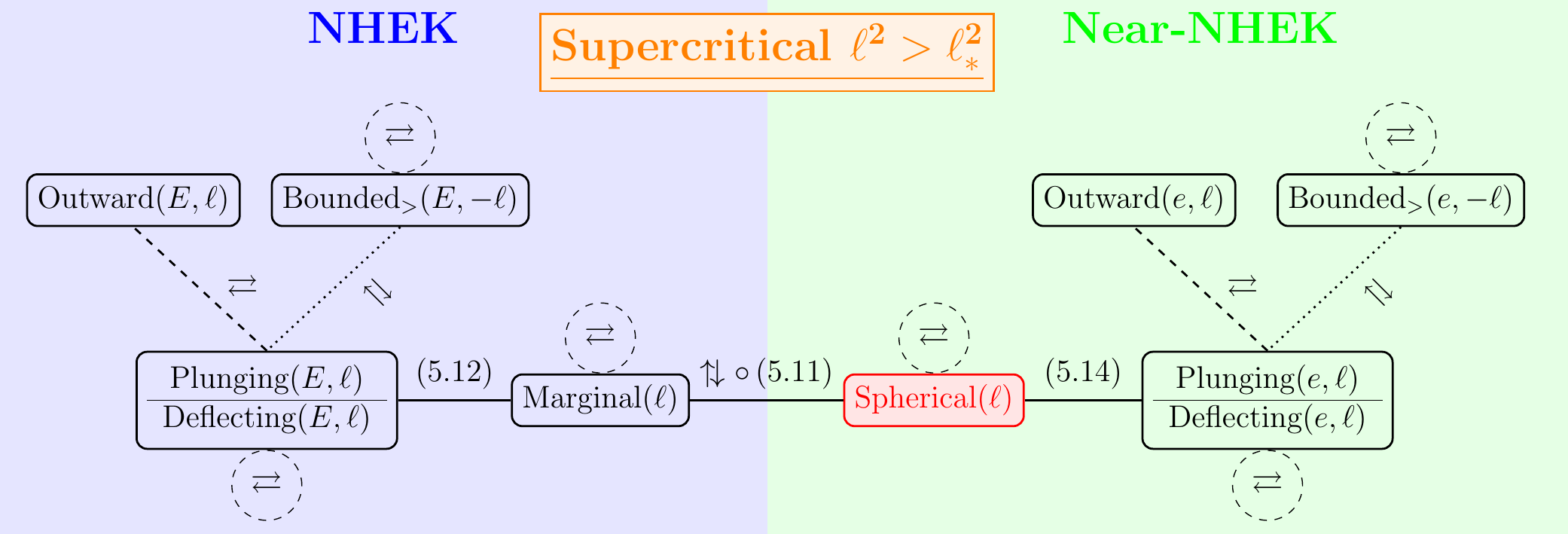}\\
    
    \includegraphics[width=\textwidth]{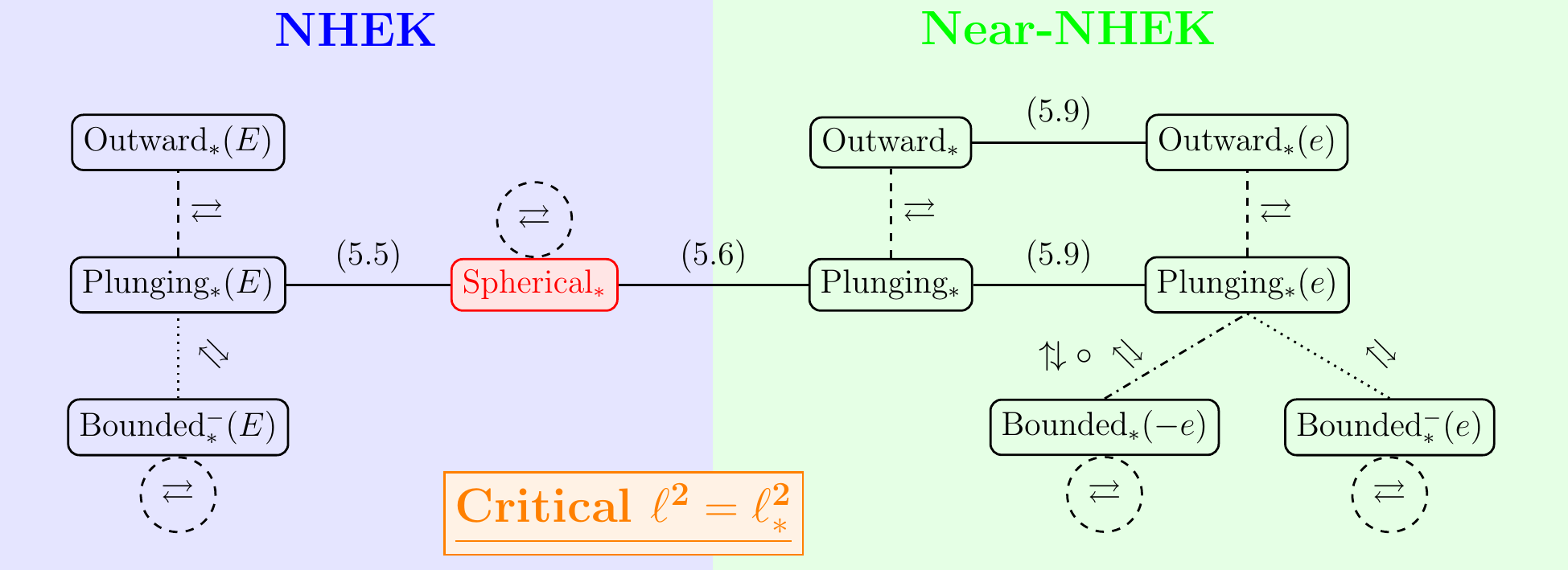}
    \end{minipage}
    \begin{minipage}{0.28\textwidth}
    \includegraphics[width=\textwidth]{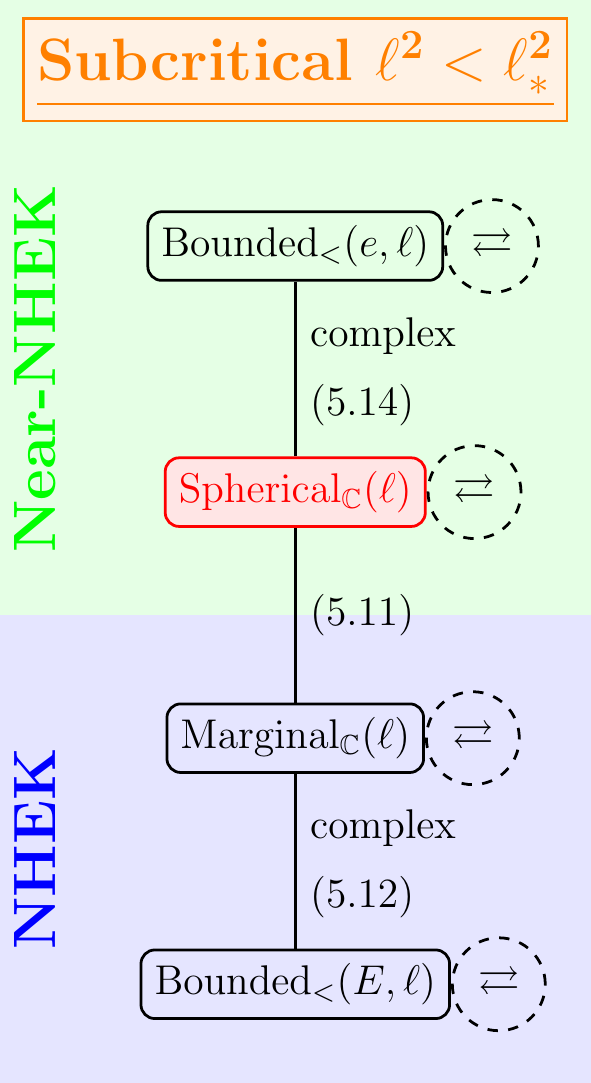}
    \end{minipage}
    
    \caption{Schematical overview of the action of the group  $SL(2,\mathbb{R})\times U(1) \times \uparrow\!\downarrow \times \rightleftarrows \times\!$\protect\rotatebox[origin=c]{-45}{$\rightleftarrows$} on near-horizon geodesics. The equation numbers refer to the conformal maps described in the text. 
    }
    \label{fig:conformal_mappings}
\end{figure}{}
\end{landscape}

\clearpage

\section*{Acknowledgements}
 We thank Alex Lupsasca and Edward Teo for pointing out mistakes in previous versions of this paper. A. D. acknowledges Kwinten Fransen for the help he provided computing complicated expressions using \textit{Mathematica}. A. D. is a Research Fellow and G. C. is a Research Associate of the F.R.S.-FNRS. G. C. acknowledges support from the FNRS research credit No. J003620F, the IISN convention No. 4.4503.15, and the COST Action GWverse No. CA16104.

\appendix
\addtocontents{toc}{\protect\setcounter{tocdepth}{1}}

\section{Elliptic integrals and Jacobi functions}\label{app:ellipticFunctions}

In this appendix, we set our conventions for the elliptic integrals and Jacobi functions used in the main text. 

The incomplete elliptic integrals of the first, second, and third kind are defined as
\bea
\!\!F(\varphi , m) &\!\!\!\triangleq \!\!\! & \int_0^\varphi \frac{\text{d} \theta}{\sqrt{1- m \sin^2 \theta}} = \int_0^{\sin \varphi} \frac{\text{d} t}{\sqrt{(1-t^2)(1-m t^2)}}, \\
\!\!E(\varphi , m) &\!\!\!\triangleq \!\!\!& \int_0^\varphi d\theta \sqrt{1- m \sin^2\theta}= \int_0^{\sin \varphi} \text{d} t \sqrt{\frac{1-m t^2}{1-t^2}}, \\
\!\!\!\!\!\!\!\!\!\!\Pi(n, \varphi ,m)& \!\!\!\triangleq\!\!\! & \int_0^\varphi \frac{1}{1- n \sin^2 \theta}\frac{\text{d} \theta}{\sqrt{1- m \sin^2 \theta}}= \int_0^{\sin\varphi} \frac{1}{1- n t^2} \frac{\text{d} t}{\sqrt{(1-m t^2)(1-t^2)}},
\eea
respectively. We also define $E'(\varphi, m ) = \p_m E(\varphi,m) = \frac{1}{2m} [E(\varphi,m) - F(\varphi , m)] $. 

The complete elliptic integrals of the first, second, and third kind are defined as
\bea
K(m) &\triangleq & F(\frac{\pi}{2},m), \\
E(m) &\triangleq & E(\frac{\pi}{2},m), \\
\Pi(n,m) &\triangleq & \Pi(n,\frac{\pi}{2},m),
\eea
respectively, and $E'(m)=\p_m E(m)$. 

Jacobi functions are defined as the inverse of the incomplete elliptic integrals of the first kind. More precisely, one can invert $u = F(\varphi,m)$ into $\varphi =\text{am}(u,m)$ in the interval $-K(m) \leq u \leq K(m)$. The elliptic sinus, elliptic cosinus, and delta amplitude are defined as 
\bea
\text{sn}(u,m) &\triangleq &\sin\text{am}(u,m), \qquad \text{cn}(u,m) \triangleq \cos\text{am}(u,m),\\
 \text{dn}(u,m) &\triangleq & \sqrt{1-m (\sin\text{am}(u,m))^2},
\eea
respectively. They have the periodicity ($k,l \in \mathbb Z$)
\begin{align}
\text{sn}(u + 2 k K(m)+2ilK(1-m) ,m)  &= (-1)^k \text{sn}(u,m) ,\label{per}\\
\text{dn}(u + 2 k K(m)+2ilK(1-m) ,m)  &= (-1)^l \text{sn}(u,m)\label{eqn:perDn}
\end{align}
and obey the properties
\bea
\text{cn}^2(u,m)+\text{sn}^2(u,m) &=&1 ,\qquad \text{dn}^2(u,m)+m\,\text{sn}^2(u,m) =1, \\
\frac{\p}{\p u}\text{am}(u,m) &=& \text{dn}(u,m), \\
\frac{\p}{\p u}\text{sn}(u,m) &=& \text{cn}(u,m)\text{dn}(u,m),\\
\frac{\p}{\p u}\text{cn}(u,m) &=& - \text{sn}(u,m)\text{dn}(u,m),\\
\frac{\p}{\p u}\text{dn}(u,m) &=& - m\, \text{sn}(u,m)\text{cn}(u,m).
\eea

\section{Elementary Polar Integrals}\label{app:basicIntegrals}

For the pendular and equator-attractive cases, one has to compute the following integrals:
\begin{align}
    &\hat I^{(0)}(x)\triangleq\int_{x_0}^x\frac{\dd t}{\sqrt{\Theta(t^2)}},\quad\hat I^{(1)}(x)\triangleq\int_{x_0}^x\frac{t^2\dd t}{\sqrt{\Theta(t^2)}}, \quad \hat I^{(2)}(x)\triangleq\int_{x_0}^x\frac{\dd t}{\sqrt{\Theta(t^2)}} \frac{1}{1-t^2} \label{eqn:polarInt}
\end{align}
where $-1 \leq x \leq 1$. In the main text, $x$ will be substituted by $\cos\theta$, where $\theta$ is a polar angle.

\subsection{$\eps_0=0$}
In this case, $\Theta(t^2)=\sqrt{\frac{z_0}{Q}}(z_0-t^2)$. Choosing $x_0=0$, one finds directly
\begin{align}
    \hat I^{(0)}(x)&=\sqrt{\frac{z_0}{Q}}\arcsin{\frac{x}{\sqrt{z_0}}},\label{eqn:I0epsEq0}\\
    \hat I^{(1)}(x)&=\frac{1}{2}\sqrt{\frac{z_0}{Q}}\qty[z_0\arcsin{\frac{x}{\sqrt{z_0}}}-x\sqrt{z_0-x^2}],\\
    \hat I^{(2)}(x)&=\sqrt{\frac{z_0}{Q(1-z_0)}}\arcsin\sqrt{\frac{x}{z_0}\frac{1-z_0}{1-x}}
\end{align}
and the particular values
\begin{align}
    \hat I^{(0)}\qty(\sqrt{z_0})&=\frac{\pi}{2}\sqrt{\frac{z_0}{Q}},\\
    \hat I^{(1)}\qty(\sqrt{z_0})&=\frac{\pi}{4}\sqrt{\frac{z_0}{Q}},\\
    \hat I^{(2)}\qty(\sqrt{z_0})&=\frac{\pi}{2}\sqrt{\frac{z_0}{Q(1-z_0)}}.
\end{align}
One inverts \eqref{eqn:I0epsEq0} as
\begin{equation}
    x=\sqrt{z_0}\sin\qty(\frac{Q}{z_0}\hat I^{(0)}).
\end{equation}

\subsection{$\eps_0\neq0$, $z_-\neq0$}
In this case, $\Theta(t^2)=\eps_0(t^2-z_-)(z_+-t^2)$, where $t=\cos\theta$. Instead of solving separately the pendular and vortical cases as in Ref. \cite{Kapec:2019hro}, we will introduce a formal notation enabling us to treat both cases simultaneously. Let us define
\begin{equation}
    q\triangleq\sign Q,\qquad z_{\pm1}\triangleq z_\pm,\qquad m\triangleq\left\lbrace\begin{array}{cc}
    \frac{z_+}{z_-},  & q=+1 \\
    1-\frac{z_-}{z_+},     & q=-1
    \end{array}\right.
\end{equation}
and
\begin{equation}
    y(t)\triangleq\left\lbrace\begin{array}{cc}
    \frac{t}{\sqrt{z_+}},  & q=+1  \\
   \sign t \sqrt{\frac{z_+-t^2}{z_+-z_-}},     & q=-1
    \end{array}\right.,\qquad\Psi^q(x)\triangleq\arcsin{y(x)}.
\end{equation}
Pendular motion corresponds to $q = 1$, which has $0 \leq t^2 \leq z_+ <1$ and $\eps_0 z_- <0$, while vortical motion corresponds to $q = -1$, which has $0< z_- \leq t^2 \leq z_+ <1$ and $\eps_0 >0$. One can then rewrite
\begin{equation}
    \frac{\dd t}{\sqrt{\eps_0(t^2-z_-)(z_+-t^2)}}=\frac{q}{\sqrt{-q\eps_0z_{-q}}}\frac{\dd y}{\sqrt{(1-y^2)(1-my^2)}}.
\end{equation}
Both factors of the right side of this equation are real, either for pendular or for vortical motions. The lower bound of the integral will be chosen as $x_0=0$ for $Q\geq0$ and $x_0=\sign{x}\sqrt{z_+}$ for $Q<0$. Then $y(x_0)=0$ for both values of $q$. This allows us to solve directly the integrals in terms of elliptic integrals (see Appendix \ref{app:ellipticFunctions} for definitions and conventions used):
\begin{align}
    \hat I^{(0)}(x)&=\frac{q}{\sqrt{-q\eps_0z_{-q}}}F\qty(\Psi^q(x),m)\label{eqn:I0epsNeq0},\\
    \hat I^{(1)}(x)&=\left\lbrace\begin{array}{ll}
    \frac{-2z_+}{\sqrt{-\eps_0 z_-}}E'\qty(\Psi^+(x),m), & q=+1\\
   - \sqrt{\frac{z_+}{\epsilon_0}}E\qty(\Psi^-(x),m),& q=-1
    \end{array}\right., \label{eqn:I1epsNeq0}\\
    \hat I^{(2)}(x)&=\left\lbrace\begin{array}{ll}
    \frac{1}{\sqrt{-\epsilon_0 z_-}}\Pi\qty(z_+,\Psi^+(x),m), & q=+1,\\
    \frac{-1}{(1-z_+)\sqrt{\epsilon_0 z_+}}\Pi\qty(\frac{z_--z_+}{1-z_+},\Psi^-(x),m), & q=-1
    \end{array}\right. .\label{eqn:I2epsNeq0}
\end{align}
For $x=\sqrt{z_q}$, $\Psi^q(\sqrt{z_q})=\frac{\pi}{2}$ for both $q= \pm 1$ and the incomplete elliptic integrals are replaced by complete ones:
\begin{align}
    \hat I^{(0)}(\sqrt{z_q})&=\frac{q}{\sqrt{-q\eps_0z_{-q}}}K\qty(m),\\
    \hat I^{(1)}(\sqrt{z_q})&=\left\lbrace\begin{array}{ll}
    \frac{-2z_+}{\sqrt{-\eps_0 z_-}}E'\qty(m), & q=+1\\
  -  \sqrt{\frac{z_+}{\epsilon_0}}E\qty(m),& q=-1
    \end{array}\right.,\\
    \hat I^{(2)}(\sqrt{z_q})&=\left\lbrace\begin{array}{ll}
    \frac{1}{\sqrt{-\epsilon_0 z_-}}\Pi\qty(z_+,m), & q=+1,\\
    \frac{-1}{(1-z_+)\sqrt{\epsilon_0 z_+}}\Pi\qty(\frac{z_--z_+}{1-z_+},m), & q=-1
    \end{array}\right. .
\end{align}
For $q=-1$, all integrals \eqref{eqn:I0epsNeq0}, \eqref{eqn:I1epsNeq0}, and \eqref{eqn:I2epsNeq0} vanish when evaluated at $x=\pm\sqrt{z_+}$ because $\Psi^-(\sqrt{z_+})=0$. Finally, \eqref{eqn:I0epsNeq0} can be inverted as
\begin{equation}
    x=y^{-1}\qty(\sn\qty(\sqrt{-q\eps_0z_{-q}}\hat I^{(0)},m))\label{eqn:inversionEpsNeq0}
\end{equation}
with $y^{-1}$ the inverse function of $y$:
\begin{equation}
    y^{-1}(t)=\left\lbrace\begin{array}{ll}
        \sqrt{z_+}t, & q=+1  \\
         \sign t\sqrt{z_+(1-mt^2)},& q=-1
    \end{array}\right.
\end{equation}
leading to the explicit formula
\begin{equation}
    x=\left\lbrace\begin{array}{ll}
        \sqrt{z_+}\sn\qty(\sqrt{-\eps_0z_{-}}\hat I^{(0)},m), & q=+1 \\
        \sign x \sqrt{z_+}\dn\qty(\sqrt{\eps_0z_{+}}\hat I^{(0)},m),& q=-1.
    \end{array}\right.\label{eqn:inversionFormula}
\end{equation}
In the context of this paper, we will always have $x=\cos\theta$. 

\subsection{$\eps_0>0$, $z_-=0$}
This case is relevant for equator-attractive orbits. In this case, the potential reduces to $\Theta(t^2)=\eps_0\,t^2(z_+-t^2)$. One needs the following integrals (see also Ref. \cite{Kapec:2019hro}):
\begin{align}
    \hat{\mathcal{I}}^{(0)}&\triangleq\int_x^{\sqrt{z_+}}\frac{\dd t}{\sqrt{\Theta(t^2)}}=\frac{1}{\sqrt{\eps_0z_+}}\arctanh\sqrt{1-\frac{x^2}{z_+}},\\
    \hat{\mathcal{I}}^{(1)}&\triangleq\int_x^{\sqrt{z_+}}\frac{t^2\dd t}{\sqrt{\Theta(t^2)}}=\sqrt{\frac{z_+-x^2}{\eps_0}},\\
    \hat{\mathcal{I}}^{(2)}&\triangleq\int_x^{\sqrt{z_+}}\frac{\dd t}{\sqrt{\Theta(t^2)}}\qty(\frac{1}{1-t^2}-1)=\frac{1}{\sqrt{\eps_0(1-z_+)}}\arctan\sqrt{\frac{z_+-x^2}{1-z_+}}
\end{align}
for any $x\leq \sqrt{z_+}$. All such integrals are obviously vanishing when evaluated at $x=\sqrt{z_+}$. Note that $ \hat{\mathcal{I}}^{(2)}$ contains a $-1$ integrand for simplicity of the final answer.

\section{Explicit form of prograde NHEK geodesics}
\label{app:equatorial}

We provide here the explicit motion in Mino time and the parametrized form of all classes of future-oriented geodesics in NHEK in the region outside the horizon $R > 0$. Without loss of generality, we can use the existence of the $\rightleftarrows$ and \rotatebox[origin=c]{-45}{$\rightleftarrows$}-flips to restrict ourselves to the subclasses of NHEK future-oriented geodesics, determined as
\begin{itemize}[label=$\diamond$]
    \item {\textit{Prograde ($\ell\geq 0$) geodesics;}} 
    \item {\textit{Partially ingoing geodesics,}} i.e., trajectories whose radial coordinate is decreasing on at least part of the total motion.
\end{itemize}

Because the near-horizon motion admits at most one turning point, we will only encounter spherical, plunging, marginal, bounded and deflecting motions.

\paragraph{General integrals.} One must provide explicit solutions to equations \eqref{eq:nhekT} to \eqref{eq:nhekphi}. The main point one has to deal with consists of solving the radial integrals \eqref{eqn:radialNHEK}. Notice that the primitives \begin{equation}
\mathbb{T}^{(i)}(R)=\int\frac{\dd R}{R^i\sqrt{E^2+2E\ell R-\mathcal{C}R^2}},\qquad i=0,1,2
\end{equation}
can be directly integrated as
\begin{align}
\mathbb{T}^{(0)}(R)&=\left\lbrace
\begin{array}{ll}
\frac{1}{\sqrt{-\mathcal{C}}}\log\qty[E\ell-\mathcal{C} R+\sqrt{-\mathcal{C}}\sqrt{v_R(R)}],  & \qquad\mathcal{C}\neq 0 \\
\frac{\sqrt{E^2+2E\ell_*R}}{E\ell_*},  & \qquad\mathcal{C}=0,~E\neq 0,
\end{array}
\right.\\
\mathbb{T}^{(1)}(R)&=\frac{\log R-\log\qty[E+\ell R+\sqrt{v_R(R)}]}{E}\qquad E\neq 0,\\
\mathbb{T}^{(2)}(R)&=-\frac{1}{E}\qty(\frac{\sqrt{v_R(R)}}{ER}+\ell\,\mathbb{T}^{(1)}(R)),\qquad E\neq 0.
\end{align}

We treat each type of motion cited above separately. We consider a geodesic path linking two events as described in the main text. We consider here a future-oriented path, $\Delta T>0$ and $\Delta\lambda>0$. Let us proceed systematically:
\begin{itemize}[label=$\diamond$]
    \item \textit{Spherical} motion has $R$ constant, and the radial integrals are ill defined and irrelevant. One can directly integrate the basic geodesic equations in this case.
    
    \item For \textit{plunging} motion, we will fix the final conditions at the final event $(T_f,R_f,\theta_f,\Phi_f)$ such that $R_f<0$ (We will choose $R_f$ to be a root of the radial potential in all cases where they are real. Otherwise, we will chose it for convenience.) We will drop the subscript $i$ of the initial event. We denote here $\Delta\lambda=\lambda_f-\lambda$, $\Delta T=T_f-T(\lambda)$,\ldots and
    \begin{equation}
        T^{(j)}_R(R)=\mathbb{T}^{(j)}(R)-\mathbb T^{(j)}(R_f),\qquad j=0,1,2.
    \end{equation}
    
    \item For \textit{bounded} motion, we identify the physically relevant root of the radial potential $R_{\text{turn}}>0$ at $T_{\text{turn}}$ (which is an integration constant) that represents the turning point of the motion. We consider the motion either before or after the turning point: 
    \begin{itemize}[label=$\rightarrow$]
        \item $\underline{T(\lambda)>T_{\text{turn}}}:$ We choose the initial event as the turning point and the final one as $(T_>(\lambda),R_>(\lambda),$ $\theta_>(\lambda),\Phi_>(\lambda)).$ This leads to
        \begin{equation}
         T^{(j)}_R(R_>)=\mathbb{T}^{(j)}(R_{\text{turn}})-\mathbb T^{(j)}(R_>),\qquad j=0,1,2.   
        \end{equation}
        \item $\underline{T(\lambda)<T_{\text{turn}}}:$ We choose the initial event as $(T_<(\lambda),R_<(\lambda),\theta_<(\lambda),\Phi_<(\lambda))$ and the final event as the turning point. This leads to
        \begin{equation}
         T^{(j)}_R(R_<)=\mathbb{T}^{(j)}(R_{\text{turn}})-\mathbb T^{(j)}(R_<),\qquad j=0,1,2.
        \end{equation}
    \end{itemize}
    The motion being symmetric in $R$ with respect to $R_{\text{turn}}$, one must only determine $T_>(R)$ and $\Phi_>(R)$, because
    \begin{equation}
        T_<(R)=-T_>(R),\qquad\Phi_<(R)=-\Phi_>(R).\label{eqn:symmetry}
    \end{equation}
    \item For \textit{deflecting} motion, there is also a turning point and \eqref{eqn:symmetry} remains true. This case is treated similarly to the bounded case but with a minus-sign change. We get here
    \begin{equation}
        T_{R}^{(j)}(R_>)=\mathbb T^{(j)}(R_>)-\mathbb T^{(j)}(R_{\text{turn}}), \qquad
         T^{(j)}_R(R_<)=\mathbb T^{(j)}(R_<)-\mathbb{T}^{(j)}(R_{\text{turn}}).
    \end{equation}
    
    \item Finally, for \textit{marginal} motion, there is no turning point and the integral is immediate.

\end{itemize}
In what follows, for the motions with one turning point, we will only make explicit the motion after the turning point, $T>T_{\text{turn}}$, and we will denote $T=T_>$ and $\Phi=\Phi_>$ in order to simplify the notations.

\vspace{10pt}
\paragraph{\bf Spherical$_*$ (ISSO).} The explicit form is
\bea
 T(\lambda) &=& T_0 + \frac{\ell_*}{R_0}(\lambda-\lambda_0)  , \\
 R(\lambda) &=& R_0 , \\
\Phi(\lambda) &=&\Phi_0 - \frac{3}{4}\ell_* (\lambda-\lambda_0) + \ell_* \Phi_\theta(\lambda-\lambda_0). 
\eea
The parametrized form is 
\begin{empheq}[box=\ovalbox]{align}
R & = R_0 , \\
\Phi & =  \Phi_0 -\frac{3}{4}R_0 (T-T_0) + \ell_*  \Phi_\theta\qty( \frac{R_0}{\ell_*} (T-T_0)). 
\end{empheq}

\paragraph{\bf Marginal$(\ell)$.} The Casimir obeys here $\mathcal{C}<0$; we denote $q\triangleq\sqrt{-\mathcal{C}}$ and consider the initial condition $R(\lambda_m)=R_m$. The explicit form of the solution is given (as a function of the Mino time) by
\bea
 T(\lambda) &=& T_0+\frac{\ell}{R_m q}\exp\qty(q(\lambda-\lambda_m)), \\
 R(\lambda) &=& R_m \exp\qty(-q(\lambda-\lambda_m)) , \\
\Phi(\lambda) &=& \Phi_0 -\frac{3}{4}\ell(\lambda-\lambda_m)+\ell\Phi_\theta(\lambda-\lambda_m). 
\eea
The parametrized form is
\begin{empheq}[box=\ovalbox]{align}
T(R) & =  T_0+\frac{\ell}{q\,R}, \\
\Phi(R) & =  \Phi_0+\frac{3}{4}\frac{\ell}{q}\log\frac{R}{R_m}+\ell\Phi_\theta\qty(\lambda\qty(\frac{R}{R_m})). 
\end{empheq}
Here, the constants $T_0$ and $\Phi_0$ remain arbitrary.

\paragraph{\bf Plunging$_*(E)$.} The energy satisfies $E>0$ and the initial condition is $R(\lambda_0)=R_0=-\frac{E}{2\ell_*}$, leading to
\bea
\lambda(R) -\lambda_0= - \frac{1}{\ell_*}  \sqrt{1+\frac{2 \ell_* R}{E}}. 
\eea
The explicit form is
\begin{align}
T(\lambda) &= \frac{2  \ell_*^2 (\lambda - \lambda_0)}{E(1-\ell_*^2 (\lambda-\lambda_0)^2)}, \\
 R(\lambda) &= \frac{E}{2\ell_*} \left( \ell_*^2 (\lambda-\lambda_0)^2 - 1 \right) , \\
\Phi(\lambda) &= -\frac{3}{4}\ell_* (\lambda-\lambda_0) +2 \, \arctanh (\ell_* (\lambda - \lambda_0)) \nonumber\\
&+ \ell_* \Phi_\theta(\lambda-\lambda_0) . 
\end{align}
The parametrized form is 
\begin{empheq}[box=\ovalbox]{align}
T(R) & = \frac{1}{R} \sqrt{1+\frac{2 \ell_* R}{E}}, \\
\Phi(R) & = \frac{3}{4} \sqrt{1+\frac{2 \ell_* R}{E}}-2 \, \arctanh\sqrt{1+\frac{2 \ell_* R}{E}}+\ell_* \Phi_\theta (\lambda(R)-\lambda_0). 
\end{empheq}

\paragraph{\bf Plunging$(E,\ell)$.} The parameters obey $\mathcal C < 0$ and $E>0$. We denote $q = \sqrt{-\mathcal C}$ and $\lambda_-$ the value of the Mino time such that $R(\lambda_-)= R_-$. The orbit is given by
\bea
\lambda(R) - \lambda_- = \frac{1}{q} \log\left(  \frac{E \sqrt{\mathcal C+\ell^2}}{E \ell - \mathcal C R +q \sqrt{v_R}} \right)=\frac{i}{q}\arccos\qty(\frac{E\ell-\mathcal{C}R}{E\sqrt{\mathcal{C}+\ell^2}}).
\eea
The last form is not explicitly real, but it allows us to find easily
\begin{align}
T(\lambda) &= \frac{q}{E} \frac{\abs{\sinh (q(\lambda-\lambda_-))}}{\frac{\ell}{\sqrt{\mathcal C + \ell^2}}-\cosh (q(\lambda-\lambda_-))}, \\
R(\lambda) &= \frac{E}{\mathcal C} \left( \ell - \sqrt{\mathcal C + \ell^2} \cosh (q (\lambda - \lambda_-))\right), \\
\Phi(\lambda) &= - \frac{3 \ell}{4}(\lambda-\lambda_-) +2\, \arctanh \left( \frac{\ell + \sqrt{\mathcal C + \ell^2}}{q} \tanh (\frac{q}{2}(\lambda-\lambda_-))\right)\nonumber\\
&+\ell\, \Phi_\theta(\lambda-\lambda_-) . 
\end{align}
The parametrized form can be simplified as
\begin{empheq}[box=\ovalbox]{align}
T(R) & = \frac{\sqrt{v_R}}{E R},\label{eqn:paramNHEKR} \\
\Phi(R) & = \Phi_0-\log\frac{E+\ell R+\sqrt{v_R(R)}}  {R}\nonumber\\
&~+\frac{3\ell}{4q}\log\qty(E\ell-\mathcal{C}R+q\sqrt{v_R(R)})+\ell\Phi_\theta(\lambda(R)-\lambda_-)  \label{eqn:paramNHEKphi}
\end{empheq}
where $\Phi_0\triangleq \log\frac{E+\ell R_-}{R_-}-\frac{3\ell}{4q}\log\qty(E\ell-\mathcal C R_-)$. The orbit start at $R=+\infty$ at $\lambda=-\infty$ and reaches the black hole horizon $R=0$ at $\lambda = \lambda_- - \lambda_H$, where $\lambda_H \equiv \frac{1}{q}\text{arccosh}(\frac{\ell}{\sqrt{\mathcal C + \ell^2}})$. It never reaches $R_- < 0$. 

\paragraph{\bf Bounded$_<(E,\ell)$.}
We have $\mathcal C > 0$ and $E>0$, leading to
\bea
\lambda(R) -\lambda_+ = \frac{1}{q}\text{arccos}\left( \frac{\mathcal C R - E \ell}{E \sqrt{\mathcal C+\ell^2}} \right).
\eea
We normalized Mino time such that $R(\lambda_+) = R_+$ and denoted $q \triangleq \sqrt{\mathcal C}$. The orbit starts at $\lambda = \lambda_+$ at the turning point $R_+$ and plunges inside the black hole $R=0$ at $\lambda = \lambda_++\frac{1}{q}\text{arccos}(\frac{-\ell}{\sqrt{\mathcal C+\ell^2}})$. We have
\begin{align}
T (\lambda) &=  \frac{q}{E} \frac{\sin (q(\lambda-\lambda_+))}{\frac{\ell}{\sqrt{\mathcal C + \ell^2}}+\cos (q(\lambda-\lambda_+))} , \\
R(\lambda) &= \frac{E}{\mathcal C} \left( \ell + \sqrt{\mathcal C + \ell^2} \cos (q (\lambda-\lambda_+) )\right) , \\
\Phi(\lambda)  &= - \frac{3 \ell}{4}(\lambda-\lambda_+) +2\,\arctanh \left( \frac{\ell - \sqrt{\mathcal C + \ell^2}}{q} \tan (\frac{q}{2}(\lambda-\lambda_+))\right)\nonumber\\
&+\ell \Phi_\theta(\lambda-\lambda_+). 
\end{align}
The parametrized form is given by \eqref{eqn:paramNHEKR} and \eqref{eqn:paramNHEKphi}, but with $q\to iq$. One can write a manifestly real form of the azimuthal coordinate by shifting the initial value $\Phi_0$, leading to
\begin{empheq}[box=\ovalbox]{align}
    \Phi(R)=\Phi_0'-\log\frac{E+\ell R+\sqrt{v_R(R)}}  {R}+\frac{3\ell}{4q}\arctan\frac{q\sqrt{v_R}}{E\ell-\mathcal C R}+\ell\Phi_\theta(\lambda(R)-\lambda_-)
\end{empheq}
with $\Phi'_0\triangleq\log\frac{E+\ell R_-}{R_-}.$

\paragraph{Def\mbox{}lecting$(E,\ell)$.} We have $\mathcal C < 0$ and $E<0$, leading to ($q\triangleq\sqrt{-\mathcal{C}}$)
\begin{equation}
    \lambda(R)-\lambda_+=-\frac{i}{q}\arccos{\frac{E\ell-\mathcal{C}R}{E\sqrt{\mathcal{C}+\ell^2}}}.\label{eq45}
\end{equation}
The initial condition $R(\lambda_+)=R_+$ corresponds to the minimal radius reached by the trajectory. The orbit starts and ends at $R=+\infty$ at Mino time $\lambda=\pm\infty$. We have
\begin{align}
    T(\lambda)&=\frac{q}{E}\frac{\sinh{q(\lambda-\lambda_+)}}{\frac{\ell}{\sqrt{\mathcal{C}+\ell^2}}-\cosh{q(\lambda-\lambda_+)}}\\
    R(\lambda)&=\frac{E}{\mathcal{C}}\qty(\ell-\sqrt{\mathcal{C}+\ell^2}\cosh{q(\lambda-\lambda_+)})\\
    \Phi(\lambda)&=-\frac{3\ell}{4}(\lambda-\lambda_+)+2\,\arctanh \left( \frac{\ell - \sqrt{\mathcal C + \ell^2}}{q} \tanh (\frac{q}{2}(\lambda-\lambda_+))\right)
    \nonumber\\
&+\ell \Phi_\theta(\lambda-\lambda_+). 
\end{align}
We finally have the parametrized form 
\begin{empheq}[box=\ovalbox]{align}
    T(R) & = - \frac{\sqrt{v_R(R)}}{E R},\\
    \Phi(R)& = \Phi_0+\log\frac{E+\ell R+\sqrt{v_R(R)}}  {R}\nonumber\\
&~-\frac{3\ell}{4q}\log\qty(E\ell-\mathcal{C}R+q\sqrt{v_R(R)})+\ell\Phi_\theta(\lambda(R)-\lambda_+)
\end{empheq}
with $\Phi_0\triangleq-\log\frac{E+\ell R_+}{R_+}+\frac{3\ell}{4q}\log\qty(E\ell-\mathcal{C}R_+)$.

\section{Explicit form of prograde near-NHEK geodesics}
\label{app:explicitNN}

We follow the same procedure as the one in Appendix \ref{app:equatorial}. As before, we only focus, without loss of generality, on future-oriented partially ingoing prograde orbits. For convenience, we also include one class of retrograde bounded geodesics. 

\paragraph{Spherical$(\ell)$.} The explicit form reads as
\begin{align}
    R&=R_0=\frac{\kappa\ell}{\sqrt{-\mathcal C}},\\
    t(\lambda)&=t_0+\frac{\ell}{R_0}(\lambda-\lambda_0),\\
    \phi(\lambda)&=\phi_0-\frac{3}{4}\ell(\lambda-\lambda_0)+\ell\Phi_\theta(\lambda-\lambda_0).
\end{align}
The parametrized form is
\begin{empheq}[box=\ovalbox]{align}
    R&=R_0=\frac{\kappa\ell}{\sqrt{-\mathcal C}},\\
    \phi(t)&=\phi_0-\frac{3}{4}R_0(t-t_0)+\ell\Phi_\theta(\lambda-\lambda_0).\label{eq46}
\end{empheq}
Note that $R_0\geq\qty(\frac{2}{\sqrt{3}}-1)\kappa$.

\paragraph{Plunging$_*$.} One has
\begin{equation}
    \lambda-\lambda_i=-\frac{R-R_i}{\kappa\ell_*}.
\end{equation}
The explicit form is
\begin{align}
    R(\lambda)&=R_i-\kappa\ell_*(\lambda-\lambda_i),\\
    t(\lambda)&=-\frac{1}{2\kappa}\log\qty[1+\kappa\ell_*(\lambda-\lambda_i)\frac{\kappa\ell_*(\lambda-\lambda_i)-2R_i}{R_i^2-\kappa^2}],\\
    \phi(\lambda)&=-\frac{3}{4}\ell_*(\lambda-\lambda_i)+\frac{1}{2}\log\frac{1-\frac{\kappa\ell_*(\lambda-\lambda_i)}{R_i-\kappa}}{1-\frac{\kappa\ell_*(\lambda-\lambda_i)}{R_i+\kappa}}+\ell_*\Phi_\theta(\lambda-\lambda_i).
\end{align}
The parametrized form is
\begin{empheq}[box=\ovalbox]{align}
    t(R)&=-\frac{1}{2\kappa}\log\frac{R^2-\kappa^2}{R_i^2-\kappa^2},\\
    \phi(R)&= \phi_i+\frac{3}{4\kappa}R+\frac{1}{2}\log\frac{R-\kappa}{R+\kappa}+\ell_*\Phi_\theta(\lambda(R)-\lambda_i)
\end{empheq}
with $\phi_i\triangleq -\frac{3R_i}{4\kappa}-\frac{1}{2}\log\frac{R_i-\kappa}{R_i+\kappa}$. The geodesic starts from $R=+\infty$ at $\lambda=-\infty$ and reaches the horizon at Mino time $\lambda_H=\lambda_i+\frac{1}{\ell_*}\qty(\frac{R_i}{\kappa}-1)$.

\paragraph{Bounded$_*(e)$ and Plunging$_*(e)$.} The orbital parameters satisfy $\mathcal C=0$ and $e<0$ (bounded) or $e>0$ (plunging). The potential is simply $v_{R;\kappa}(R)=e^2+\kappa^2\ell_*^2+2e\ell_* R$ and the initial conditions are imposed at $R(\lambda_0)=R_0$, where $R_0$ is the (unique) root of the radial potential. One directly obtains
\begin{equation}
    \lambda(R)-\lambda_0=-\frac{\sqrt{v_{R;\kappa(R)}}}{e\ell_*},
\end{equation}
leading to
\begin{align}
    R(\lambda)&=R_0+\frac{e\ell_*}{2}(\lambda-\lambda_0)^2,\\
    t(\lambda)&= -\frac{1}{\kappa}\,\arctanh\frac{2\kappa e \ell_*^2(\lambda-\lambda_0)}{\kappa^2\ell_*^2+e^2\qty[\ell^2_*(\lambda-\lambda_0)^2-1]},\\
    \phi(\lambda)&=-\frac{3}{4}\ell_*(\lambda-\lambda_0)+\sign e\, \arctanh\frac{2e^2\ell_*(\lambda-\lambda_0)}{-\kappa^2\ell_*^2+e^2\qty[\ell_*^2(\lambda-\lambda_0)^2+1]}\nonumber\\
    &~+\ell_*\Phi_\theta(\lambda-\lambda_0).   
\end{align}
The parametric form is
\begin{empheq}[box=\ovalbox]{align}
    t(R)&=\frac{1}{\kappa}\,\text{arccosh} \frac{\abs{R+\frac{\kappa^2\ell_*}{e}}}{\sqrt{R^2-\kappa^2}},\\
    \phi(R)&=-\frac{3}{4e}\sqrt{v_{R;\kappa}(R)}+\arctanh\frac{\sqrt{v_{R;\kappa}(R)}}{e+\ell_* R}+\ell_*\Phi_\theta(\lambda(R)-\lambda_0).
\end{empheq}
Notice that the requirements $v_{R;\kappa}\geq0$ and $R\geq\kappa$ are sufficient to guarantee the reality of the inverse hyperbolic functions involved. The trajectory reaches the horizon at Mino time $\lambda_H=\lambda_0-\sign e \sqrt{\frac{2(\kappa-R_0)}{e\ell_*}}$, which is smaller than $\lambda_0$ for plunging motion and greater for bounded motion, as expected.

\paragraph{(Retrograde) Bounded$_>(e,\ell)$.} The geodesic parameters satisfy $\mathcal{C}<0$, $\ell < 0$, and $e>0$. Therefore, $R_-$ is positive, and we choose the initial condition as $R(\lambda_-)=R_-$. Defining $q\triangleq\sqrt{-\mathcal{C}}$, the explicit form reads as
\begin{align}
    R(\lambda)&= \frac{1}{\mathcal{C}}\qty[e\ell-\sqrt{(\mathcal{C}+\ell^2)(e^2+\kappa^2\mathcal{C})}\cosh{q(\lambda-\lambda_-)}].
\end{align}
The parametrized form is
\begin{empheq}[box=\ovalbox]{align}
    t(R)&=\frac{1}{4\kappa}\log\frac{F_+(R)}{F_-(R)},\label{eqn:paramNNt}\\
    \phi(R)&= \frac{3\ell}{4q}\log\qty[e\ell-\mathcal{C}R+q\sqrt{v_{R;\kappa}(R)}]-\frac{1}{4}\log\frac{G_+(R)}{G_-(R)}+\ell\Phi_\theta(\lambda(R)-\lambda_-)\label{eqn:paramNNphi}
\end{empheq}
where we define
\begin{align}
    F_\pm(R)&\triangleq \qty[eR+\kappa\qty(\kappa\ell\pm\sqrt{v_{R;\kappa}(R)})]^2,\\
    G_\pm(R)&\triangleq\qty(e+\ell R\pm\sqrt{v_{R;\kappa}(R)})^2.
\end{align}
Note that using the identities
\begin{align}
    F_+(R)F_-(R)&=(e^2+\kappa^2\mathcal C)^2(R^2-\kappa^2)^2,\\
    G_+(R)G_-(R)&=(\mathcal C+\ell^2)^2(R^2-\kappa^2)^2,
\end{align}
one can rewrite \eqref{eqn:paramNNt} and \eqref{eqn:paramNNphi} as
\begin{align}
    t(R)&=-\frac{1}{2\kappa}\log\frac{F_+(R)}{(e^2+\kappa^2\mathcal{C})(R^2-\kappa^2)},\\
    \phi(R)&= \frac{3\ell}{4q}\log\qty[e\ell-\mathcal{C}R+q\sqrt{v_{R;\kappa}(R)}]-\frac{1}{2}\log\frac{G_+(R)}{(\mathcal{C}+\ell^2)(R^2-\kappa^2)}+\ell\Phi_\theta(\lambda(R)-\lambda_-).
\end{align}
The geodesic motion starts from the past horizon, reaches $R_-$ at Mino time $\lambda_-$ and crosses the future horizon at  $\lambda_H-\lambda_-=\frac{1}{q}\text{arccosh}\frac{e\ell-\kappa\mathcal{C}}{\sqrt{(\mathcal{C}+\ell^2)(e^2+\kappa^2\mathcal{C})}}$.

\paragraph{Bounded$_<(e,\ell)$.} The parameters obey $\mathcal{C}>0$ and $e\neq0$; therefore, $R_+$ is positive and the initial condition is chosen as $R(\lambda_+)=R_+$. One defines $q\triangleq\sqrt{\mathcal{C}}$ and gets the explicit form:
\begin{align}
    R(\lambda)&= \frac{1}{\mathcal{C}}\qty[e\ell+\sqrt{(\mathcal{C}+\ell^2)(e^2+\kappa^2\mathcal{C})}\cos\qty(q(\lambda-\lambda_+))].
\end{align}
The parametrized form is given by \eqref{eqn:paramNNt} and \eqref{eqn:paramNNphi}, with the replacement rule $q\to iq$. A manifestly real form of $\phi$ is
\begin{empheq}[box=\ovalbox]{align}
    \phi(R)&=\frac{3\ell}{4q}\arctan\frac{q\sqrt{v_{R;\kappa}}}{e\ell-\mathcal C R}-\frac{1}{2}\log\frac{G_+(R)}{(\mathcal{C}+\ell^2)(R^2-\kappa^2)}+\ell\Phi_\theta(\lambda(R)-\lambda_-).
\end{empheq}
The geodesic motion starts from the white hole past horizon, reaches $R_+$ at Mino time $\lambda_+$ and crosses the future horizon at  $\lambda_H-\lambda_+=\frac{1}{q}\,\text{arccos}\frac{\kappa\mathcal{C}-e\ell}{\sqrt{(\mathcal{C}+\ell^2)(e^2+\kappa^2\mathcal{C})}}$.

\paragraph{Plunging$(e,\ell)$.} The parameters satisfy $\mathcal C<0$ and $e>-\kappa q$, where $q\triangleq\sqrt{-\mathcal C}$. The roots of the radial potential are consequently either complex or negative. In the complex case ($e^2+\kappa^2\mathcal{C} < 0$), we define the real quantity 
\begin{equation}
    R_f \triangleq\frac{1}{\mathcal{C}}\qty[e\ell-\sqrt{-(\mathcal{C}+\ell^2)(e^2+\kappa^2\mathcal{C})}]
\end{equation}
and impose the final condition $R({\lambda}_f)={R}_f$, leading to
\begin{align}
    R(\lambda)&= \frac{1}{\mathcal{C}}\qty[e\ell-\sqrt{-( \mathcal{C}+\ell^2)(e^2+\kappa^2\mathcal{C})}\qty(\cosh{q(\lambda-{\lambda}_f)}-\sqrt{2}\sinh{q(\lambda-{\lambda}_f)})].
\end{align}
Note that $\lim_{x\to\pm\infty}\cosh x -\sqrt{2}\sinh x=\mp\infty$ leads to the expected behavior. For negative roots, $R=R_-$ can be used as the final condition. In both cases, the parametrized form is again as given in \eqref{eqn:paramNNt} and \eqref{eqn:paramNNphi}. The orbit starts from $R=+\infty$ at $\lambda=-\infty$ and reaches the horizon at Mino time $\lambda_H-{\lambda}_f=\frac{1}{q}\,\text{arcsinh}\, 1+\log\frac{e\ell-\kappa\mathcal{C}+q((e+\kappa\ell)}{\sqrt{-(\mathcal{C}+\ell^2)(e^2+\kappa^2\mathcal{C})}}$.

\paragraph{Def\mbox{}lecting$(e,\ell)$.} One has $\mathcal{C}<0$ and $e<0$. Choosing the initial condition as $R(\lambda_-)=R_-$ and defining again $q\triangleq\sqrt{-\mathcal{C}}$, we get
\begin{align}
    R(\lambda)&=\frac{1}{\mathcal{C}}\qty[e\ell-\sqrt{(\mathcal{C}+\ell^2)(e^2+\kappa^2\mathcal{C})}\cosh\qty(q(\lambda-\lambda_-))].
\end{align}
The parametrized form is
\begin{empheq}[box=\ovalbox]{align}
    t(R)&=-\frac{1}{4\kappa}\log\frac{F_+(R)}{F_-(R)},\\
    \phi(R)&=- \frac{3\ell}{4q}\log\qty[e\ell-\mathcal{C}R+q\sqrt{v_{R;\kappa}(R)}]+\frac{1}{4}\log\frac{G_+(R)}{G_-(R)}+\ell\Phi_\theta(\lambda(R)-\lambda_-).
\end{empheq}
The orbit starts from $R=+\infty$ at $\lambda=-\infty$, reaches its minimal radial value $R_-$ at Mino time $\lambda_-$, and goes back to the asymptotic region at $\lambda\to+\infty$.

\section{Comparison with Kapec-Lupsasca}
\label{app:comparison}

An analysis of the (near-)NHEK geodesic motion with $\ell \neq 0$ was also performed by D. Kapec and A. Lupsasca in Ref. \cite{Kapec:2019hro}. This appendix aims to compare our radial taxonomy to the classification proposed in their paper, where the following classes of trajectories are defined:
\begin{itemize}[label=$\diamond$]
\item \textit{Type I }trajectories, for which $\mathcal{C}>0$.
\item \textit{Type II}, for which $-\ell^2<\mathcal{C}<0$.
\item \textit{Type III}, for which $\mathcal{C}=0$.
\end{itemize}

\paragraph{(Poincar\'e) NHEK.}
Two subclasses of Type II are defined: Type IIA, with $E\ell>0$ and Type IIB, with $E\ell<0$. The comparison is displayed in Table \ref{tab:Kapec_NHEK}. In Ref. \cite{Kapec:2019hro}, the cases $E=0$ were not considered.

\begin{table}[htb!]
    \centering
\begin{center}
\begin{tabular}{|c|c|}
\hline
\rule{0pt}{13pt}\textbf{This work} & \textbf{Ref. \cite{Kapec:2019hro}} \\
\hline
\rule{0pt}{13pt} $\text{Bounded}_<(E,\ell)$ & Type I\\
\hline
\rule{0pt}{13pt}$\text{Marginal}(\ell)$  & Type II, $E=0$ (discarded)\\ 
\hline
\begin{tabular}{c}
\rule{0pt}{13pt} $\text{Plunging}(E,\ell)$\\\rule{0pt}{13pt}$\text{Outward}(E)$
\end{tabular} & Type IIA\\ 
\hline
\rule{0pt}{13pt}$\text{Def\mbox{}lecting}(E,\ell)$ & Type IIB, $E<0$\\ 
\hline
\rule{0pt}{13pt}$\text{Bounded}_>(E,\ell)$ & Type IIB, $E>0$\\ 
\hline
\rule{0pt}{13pt}$\text{Spherical}_*$(ISSO) &  Type III, $E=0$ (discarded)\\ 
\hline
\begin{tabular}{c}
\rule{0pt}{13pt}$\text{Plunging}_*(E)$\\\rule{0pt}{13pt}$\text{Outward}_*(E)$
\end{tabular} & Type III, $E>0$, $\ell>0$\\ 
\hline
\rule{0pt}{13pt}$\text{Bounded}^-_*(E)$ & Type III, $E>0$, $\ell<0$\\ 
\hline
\end{tabular}
\end{center}
\caption{Comparison between our classification and Ref. \cite{Kapec:2019hro} for NHEK geodesics.}
    \label{tab:Kapec_NHEK}
\end{table}

\paragraph{Near-NHEK.}
For Type II, in addition to Types IIA and IIB, Ref. \cite{Kapec:2019hro} defines two additional subcases:
\begin{itemize}[label=$\diamond$]
\item Case 1: $-\qty(\frac{e}{\kappa})^2<\mathcal{C}$.
\item Case 2: $-\qty(\frac{e}{\kappa})^2>\mathcal{C}$.
\end{itemize}
This is relevant for distinguishing geodesic classes with $e<0$, $\ell > \ell_*$. The comparison is provided in Table \ref{tab:Kapec_NN}. The zero-measure cases $e=0$ and $e=-\kappa \sqrt{-\mathcal C}$ were not considered in Ref. \cite{Kapec:2019hro}. 

\begin{table}[htb!]
    \centering
\begin{center}
\begin{tabular}{|c|c|}
\hline
\rule{0pt}{13pt}\textbf{This work} & \textbf{Ref. \cite{Kapec:2019hro}} \\
\hline
\rule{0pt}{13pt} Bounded$_<(e,\ell)$ & Type I\\\hline
\begin{tabular}{c}
\rule{0pt}{13pt} Plunging$(e,\ell)$\\\rule{0pt}{13pt} Outward$(e,\ell)$
\end{tabular} & Type IIA, $-(e/\kappa)^2>\mathcal C$  \\\hline
\rule{0pt}{13pt} Spherical$(\ell)$ & Type IIA, $-(e/\kappa)^2=\mathcal{C}$ (discarded) \\\hline
\rule{0pt}{13pt} Def\mbox{}lecting$(e,\ell)$ & Type IIB, $-\kappa\ell<e<-\kappa\sqrt{-\mathcal{C}}$, $\ell > \ell_*$ \\\hline
\rule{0pt}{13pt} Bounded$_>(e,\ell)$ & Type IIB, $e>\kappa\ell_*$, $\ell < -\ell_*$  \\\hline
\rule{0pt}{13pt} Bounded$_*(e)$ & Type IIIB, $e<0$, $\ell = \ell_*$\\\hline
\rule{0pt}{13pt} Bounded$_*^-(e)$ & Type IIIB, $e>0$, $\ell=-\ell_*$\\\hline
\begin{tabular}{c}
\rule{0pt}{13pt} Plunging$_*$\\\rule{0pt}{13pt} Outward$_*$
\end{tabular} &  Type III, $e=0$ (discarded)\\\hline
\begin{tabular}{c}
\rule{0pt}{13pt} Plunging$_*(e)$\\\rule{0pt}{13pt} Outward$_*(e)$
\end{tabular} & Type IIIA\\\hline
\end{tabular}
\end{center}
    \caption{Comparison between our classification and \cite{Kapec:2019hro} for near-NHEK geodesics.}
    \label{tab:Kapec_NN}
\end{table}

\newpage

\providecommand{\href}[2]{#2}\begingroup\raggedright\endgroup


\begin{thebibliography}{10}

\bibitem{PhysRev.174.1559}
B.~Carter, ``Global structure of the {K}err family of gravitational fields,''
  {\em Phys. Rev.} {\bf 174} (Oct, 1968) 1559--1571.

\bibitem{Wilkins:1972rs}
D.~C. Wilkins, ``{Bound Geodesics in the Kerr Metric},'' {\em Phys. Rev.} {\bf
  D5} (1972)
814--822.

\bibitem{Bardeen:1973aa}
J.~M. Bardeen, ``{Timelike and null geodesics in the Kerr metric},'' {\em Black
  holes (Les Astres Occlus), C. Dewitt and B. S. Dewitt, eds. Gordon and Breach
  Science Publishers} (1973) 215--239.

\bibitem{Chandrasekhar:1983aa}
S.~Chandrasekhar, {\em {The mathematical theory of black holes}}.
\newblock The International Series of Monographs on Physics, Oxford: Clarendon
  Press, 1983.

\bibitem{Rauch:1994aa}
K.~P. Rauch and R.~D. Blandford, ``{Optical Caustics in a Kerr Spacetime and
  the Origin of Rapid X-Ray Variability in Active Galactic Nuclei},'' {\em
  Astrophys. J.} {\bf 421} (1994) 46.

\bibitem{Neill:1995aa}
B.~O'Neill, {\em {The geometry of Kerr black holes}}.
\newblock A. K. Peters, Ltd, 1995.

\bibitem{Schmidt:2002qk}
W.~Schmidt, ``{Celestial mechanics in Kerr space-time},'' {\em Class. Quant.
  Grav.} {\bf 19} (2002) 2743,
\href{http://www.arXiv.org/abs/gr-qc/0202090}{{\tt gr-qc/0202090}}.

\bibitem{Mino:2003yg}
Y.~Mino, ``{Perturbative approach to an orbital evolution around a supermassive
  black hole},'' {\em Phys. Rev.} {\bf D67} (2003) 084027,
\href{http://www.arXiv.org/abs/gr-qc/0302075}{{\tt gr-qc/0302075}}.

\bibitem{Vazquez:2003zm}
S.~E. Vazquez and E.~P. Esteban, ``{Strong field gravitational lensing by a
  Kerr black hole},'' {\em Nuovo Cim.} {\bf B119} (2004) 489--519,
\href{http://www.arXiv.org/abs/gr-qc/0308023}{{\tt gr-qc/0308023}}.

\bibitem{Kraniotis:2005zm}
G.~V. Kraniotis, ``{Frame-dragging and bending of light in Kerr and Kerr-(anti)
  de Sitter spacetimes},'' {\em Class. Quant. Grav.} {\bf 22} (2005)
  4391--4424,
\href{http://www.arXiv.org/abs/gr-qc/0507056}{{\tt gr-qc/0507056}}.

\bibitem{Dexter:2009fg}
J.~Dexter and E.~Agol, ``{A Fast New Public Code for Computing Photon Orbits in
  a Kerr Spacetime},'' {\em Astrophys. J.} {\bf 696} (2009) 1616--1629,
\href{http://www.arXiv.org/abs/0903.0620}{{\tt 0903.0620}}.

\bibitem{Fujita:2009bp}
R.~Fujita and W.~Hikida, ``{Analytical solutions of bound timelike geodesic
  orbits in Kerr spacetime},'' {\em Class. Quant. Grav.} {\bf 26} (2009)
  135002,
\href{http://www.arXiv.org/abs/0906.1420}{{\tt 0906.1420}}.

\bibitem{Kraniotis:2010gx}
G.~V. Kraniotis, ``{Precise analytic treatment of Kerr and Kerr-(anti) de
  Sitter black holes as gravitational lenses},'' {\em Class. Quant. Grav.} {\bf
  28} (2011) 085021,
\href{http://www.arXiv.org/abs/1009.5189}{{\tt 1009.5189}}.

\bibitem{Hackmann:2015ewa}
E.~Hackmann and C.~Lämmerzahl, ``{Analytical solution methods for geodesic
  motion},'' {\em AIP Conf. Proc.} {\bf 1577} (2015), no.~1, 78--88,
\href{http://www.arXiv.org/abs/1506.00807}{{\tt 1506.00807}}.

\bibitem{Hackmann:2015vla}
C.~Lämmerzahl and E.~Hackmann, ``{Analytical Solutions for Geodesic Equation
  in Black Hole Spacetimes},'' {\em Springer Proc. Phys.} {\bf 170} (2016)
  43--51,
\href{http://www.arXiv.org/abs/1506.01572}{{\tt 1506.01572}}.

\bibitem{Porfyriadis:2016gwb}
A.~P. Porfyriadis, Y.~Shi, and A.~Strominger, ``{Photon Emission Near Extreme
  Kerr Black Holes},'' {\em Phys. Rev.} {\bf D95} (2017), no.~6, 064009,
\href{http://www.arXiv.org/abs/1607.06028}{{\tt 1607.06028}}.

\bibitem{Compere:2017hsi}
G.~Comp{\`e}re, K.~Fransen, T.~Hertog, and J.~Long, ``{Gravitational waves from
  plunges into Gargantua},'' {\em Class. Quant. Grav.} {\bf 35} (2018), no.~10,
  104002,
\href{http://www.arXiv.org/abs/1712.07130}{{\tt 1712.07130}}.

\bibitem{Kapec:2019hro}
D.~Kapec and A.~Lupsasca, ``{Particle motion near high-spin black holes},''
  {\em Class. Quant. Grav.} {\bf 37} (2020), no.~1, 015006,
\href{http://www.arXiv.org/abs/1905.11406}{{\tt 1905.11406}}.

\bibitem{Gralla:2019ceu}
S.~E. Gralla and A.~Lupsasca, ``{The Null Geodesics of the Kerr Exterior},''
\href{http://www.arXiv.org/abs/1910.12881}{{\tt 1910.12881}}.

\bibitem{Rana:2019bsn}
P.~Rana and A.~Mangalam, ``{Astrophysically relevant bound trajectories around
  a Kerr black hole},'' {\em Class. Quant. Grav.} {\bf 36} (2019) 045009,
\href{http://www.arXiv.org/abs/1901.02730}{{\tt 1901.02730}}.

\bibitem{Stein:2019buj}
L.~C. Stein and N.~Warburton, ``{The location of the last stable orbit in Kerr
  spacetime},''
\href{http://www.arXiv.org/abs/1912.07609}{{\tt 1912.07609}}.

\bibitem{Luminet:1979nyg}
J.~P. Luminet, ``{Image of a spherical black hole with thin accretion disk},''
  {\em Astron. Astrophys.} {\bf 75} (1979)
228--235.

\bibitem{Falcke:1999pj}
H.~Falcke, F.~Melia, and E.~Agol, ``{Viewing the shadow of the black hole at
  the galactic center},'' {\em Astrophys. J.} {\bf 528} (2000) L13,
\href{http://www.arXiv.org/abs/astro-ph/9912263}{{\tt astro-ph/9912263}}.

\bibitem{James:2015yla}
O.~James, E.~von Tunzelmann, P.~Franklin, and K.~S. Thorne, ``{Gravitational
  Lensing by Spinning Black Holes in Astrophysics, and in the Movie
  Interstellar},'' {\em Class. Quant. Grav.} {\bf 32} (2015), no.~6, 065001,
\href{http://www.arXiv.org/abs/1502.03808}{{\tt 1502.03808}}.

\bibitem{Luminet:2019hfx}
J.-P. Luminet, ``{An Illustrated History of Black Hole Imaging : Personal
  Recollections (1972-2002)},''
\href{http://www.arXiv.org/abs/1902.11196}{{\tt 1902.11196}}.

\bibitem{Gralla:2019xty}
S.~E. Gralla, D.~E. Holz, and R.~M. Wald, ``{Black Hole Shadows, Photon Rings,
  and Lensing Rings},'' {\em Phys. Rev.} {\bf D100} (2019), no.~2, 024018,
\href{http://www.arXiv.org/abs/1906.00873}{{\tt 1906.00873}}.

\bibitem{Gralla:2019drh}
S.~E. Gralla and A.~Lupsasca, ``{Lensing by Kerr Black Holes},''
\href{http://www.arXiv.org/abs/1910.12873}{{\tt 1910.12873}}.

\bibitem{Akiyama:2019cqa}
{\bf Event Horizon Telescope} Collaboration, K.~Akiyama {\em et al.}, ``{First
  M87 Event Horizon Telescope Results. I. The Shadow of the Supermassive Black
  Hole},'' {\em Astrophys. J.} {\bf 875} (2019), no.~1, L1,
\href{http://www.arXiv.org/abs/1906.11238}{{\tt 1906.11238}}.

\bibitem{1973ApJ...185..635T}
S.~A. {Teukolsky}, ``{Perturbations of a Rotating Black Hole. I. Fundamental
  Equations for Gravitational, Electromagnetic, and Neutrino-Field
  Perturbations},'' {\em Astrophys. J.} {\bf 185} (Oct, 1973) 635--648.

\bibitem{Sasaki:1981sx}
M.~Sasaki and T.~Nakamura, ``{Gravitational Radiation From a Kerr Black Hole.
  1. Formulation and a Method for Numerical Analysis},'' {\em Prog. Theor.
  Phys.} {\bf 67} (1982)
1788.

\bibitem{Ryan:1995zm}
F.~D. Ryan, ``{Effect of gravitational radiation reaction on circular orbits
  around a spinning black hole},'' {\em Phys. Rev.} {\bf D52} (1995)
  R3159--R3162,
\href{http://www.arXiv.org/abs/gr-qc/9506023}{{\tt gr-qc/9506023}}.

\bibitem{Finn:2000sy}
L.~S. Finn and K.~S. Thorne, ``{Gravitational waves from a compact star in a
  circular, inspiral orbit, in the equatorial plane of a massive, spinning
  black hole, as observed by LISA},'' {\em Phys. Rev.} {\bf D62} (2000) 124021,
\href{http://www.arXiv.org/abs/gr-qc/0007074}{{\tt gr-qc/0007074}}.

\bibitem{Bardeen:1999px}
J.~M. Bardeen and G.~T. Horowitz, ``{The Extreme Kerr throat geometry: A Vacuum
  analog of AdS(2) x S**2},'' {\em Phys. Rev.} {\bf D60} (1999) 104030,
\href{http://www.arXiv.org/abs/hep-th/9905099}{{\tt hep-th/9905099}}.

\bibitem{Amsel:2009ev}
A.~J. Amsel, G.~T. Horowitz, D.~Marolf, and M.~M. Roberts, ``{No Dynamics in
  the Extremal Kerr Throat},'' {\em JHEP} {\bf 09} (2009) 044,
\href{http://www.arXiv.org/abs/0906.2376}{{\tt 0906.2376}}.

\bibitem{Dias:2009ex}
O.~J.~C. Dias, H.~S. Reall, and J.~E. Santos, ``{Kerr-CFT and gravitational
  perturbations},'' {\em JHEP} {\bf 08} (2009) 101,
\href{http://www.arXiv.org/abs/0906.2380}{{\tt 0906.2380}}.

\bibitem{Bredberg:2009pv}
I.~Bredberg, T.~Hartman, W.~Song, and A.~Strominger, ``{Black Hole
  Superradiance From Kerr/CFT},'' {\em JHEP} {\bf 04} (2010) 019,
\href{http://www.arXiv.org/abs/0907.3477}{{\tt 0907.3477}}.

\bibitem{Porfyriadis:2014fja}
A.~P. Porfyriadis and A.~Strominger, ``{Gravity waves from the Kerr/CFT
  correspondence},'' {\em Phys. Rev.} {\bf D90} (2014), no.~4, 044038,
\href{http://www.arXiv.org/abs/1401.3746}{{\tt 1401.3746}}.

\bibitem{Hadar:2014dpa}
S.~Hadar, A.~P. Porfyriadis, and A.~Strominger, ``{Gravity Waves from
  Extreme-Mass-Ratio Plunges into Kerr Black Holes},'' {\em Phys. Rev.} {\bf
  D90} (2014), no.~6, 064045,
\href{http://www.arXiv.org/abs/1403.2797}{{\tt 1403.2797}}.

\bibitem{Hadar:2015xpa}
S.~Hadar, A.~P. Porfyriadis, and A.~Strominger, ``{Fast plunges into Kerr black
  holes},'' {\em JHEP} {\bf 07} (2015) 078,
\href{http://www.arXiv.org/abs/1504.07650}{{\tt 1504.07650}}.

\bibitem{Gralla:2015rpa}
S.~E. Gralla, A.~P. Porfyriadis, and N.~Warburton, ``{Particle on the Innermost
  Stable Circular Orbit of a Rapidly Spinning Black Hole},'' {\em Phys. Rev.}
  {\bf D92} (2015), no.~6, 064029,
\href{http://www.arXiv.org/abs/1506.08496}{{\tt 1506.08496}}.

\bibitem{Hadar:2016vmk}
S.~Hadar and A.~P. Porfyriadis, ``{Whirling orbits around twirling black holes
  from conformal symmetry},'' {\em JHEP} {\bf 03} (2017) 014,
\href{http://www.arXiv.org/abs/1611.09834}{{\tt 1611.09834}}.

\bibitem{Hod:2017uof}
S.~Hod, ``{Marginally bound (critical) geodesics of rapidly rotating black
  holes},'' {\em Phys. Rev.} {\bf D88} (2013), no.~8, 087502,
\href{http://www.arXiv.org/abs/1707.05680}{{\tt 1707.05680}}.

\bibitem{AlZahrani:2010qb}
A.~M. Al~Zahrani, V.~P. Frolov, and A.~A. Shoom, ``{Particle Dynamics in Weakly
  Charged Extreme Kerr Throat},'' {\em Int. J. Mod. Phys.} {\bf D20} (2011)
  649--660,
\href{http://www.arXiv.org/abs/1010.1570}{{\tt 1010.1570}}.

\bibitem{Gralla:2017ufe}
S.~E. Gralla, A.~Lupsasca, and A.~Strominger, ``{Observational Signature of
  high-spin at the Event Horizon Telescope},'' {\em Mon. Not. Roy. Astron.
  Soc.} {\bf 475} (2018), no.~3, 3829--3853,
\href{http://www.arXiv.org/abs/1710.11112}{{\tt 1710.11112}}.

\bibitem{Gates:2018hub}
D.~Gates, D.~Kapec, A.~Lupsasca, Y.~Shi, and A.~Strominger, ``{Polarization
  Whorls from M87 at the Event Horizon Telescope},''
\href{http://www.arXiv.org/abs/1809.09092}{{\tt 1809.09092}}.

\bibitem{Gralla:2016qfw}
S.~E. Gralla, S.~A. Hughes, and N.~Warburton, ``{Inspiral into Gargantua},''
  {\em Class. Quant. Grav.} {\bf 33} (2016), no.~15, 155002,
\href{http://www.arXiv.org/abs/1603.01221}{{\tt 1603.01221}}.

\bibitem{10.2307/1969287}
S.~Bochner, ``Curvature and Betti numbers,'' {\em Annals of Mathematics} {\bf
  49} (1948), no.~2, 379--390.

\bibitem{1952AnMat..55..328Y}
K.~{Yano}, ``{Some remarks on tensor fields and curvature},'' {\em Annals of
  Mathematics} {\bf 55} (Dec., 1952) 328--347.

\bibitem{1973NYASA.224..125P}
R.~{Penrose}, ``{Naked Singularities},'' in {\em Sixth Texas Symposium on
  Relativistic Astrophysics}, D.~J. {Hegyi}, ed., vol.~224, p.~125.
\newblock Jan., 1973.

\bibitem{Floyd:1973aa}
R.~Floyd, {\em {The Dynamics of Kerr Fields}}.
\newblock PhD thesis, London University, London, UK, 1973.

\bibitem{Walker:1970un}
M.~Walker and R.~Penrose, ``{On quadratic first integrals of the geodesic
  equations for type [22] spacetimes},'' {\em Commun. Math. Phys.} {\bf 18}
  (1970)
265--274.

\bibitem{Carter:1968ks}
B.~Carter, ``{Hamilton-Jacobi and Schrodinger separable solutions of Einstein's
  equations},'' {\em Commun. Math. Phys.} {\bf 10} (1968), no.~4,
280--310.

\bibitem{Carter:1968rr}
B.~Carter, ``{Global structure of the Kerr family of gravitational fields},''
  {\em Phys. Rev.} {\bf 174} (1968)
1559--1571.

\bibitem{Compere:2019cqe}
G.~Comp\`ere, K.~Fransen, and C.~Jonas, ``{Transition from inspiral to plunge
  into a highly spinning black hole},'' accepted in Class.\ Quant.\ Grav., 
\href{http://www.arXiv.org/abs/1909.12848}{{\tt 1909.12848}}.

\bibitem{Compere:2012jk}
G.~Comp{\`e}re, ``{The Kerr/CFT correspondence and its extensions},'' {\em
  Living Rev. Rel.} {\bf 15} (2012) 11,
  \href{http://www.arXiv.org/abs/1203.3561}{{\tt 1203.3561}}.
[Living Rev. Rel.20,no.1,1(2017)].

\bibitem{Galajinsky:2010zy}
A.~Galajinsky, ``{Particle dynamics near extreme Kerr throat and
  supersymmetry},'' {\em JHEP} {\bf 11} (2010) 126,
\href{http://www.arXiv.org/abs/1009.2341}{{\tt 1009.2341}}.


\bibitem{Chen:2019hac}
B.~Chen, G.~Comp\`ere, Y.~Liu, J.~Long and X.~Zhang, ``{Spin and Quadrupole Couplings for High Spin Equatorial Intermediate Mass-ratio Coalescences},'' Class.\ Quant.\ Grav.\  \textbf{36} (2019) no.24, 245011, \href{http://www.arXiv.org/abs/1901.05370}{{\tt 1901.05370}}.


\bibitem{Burke:2019yek}
O.~Burke, J.~R. Gair, and J.~Sim\'on, ``{Transition from Inspiral to Plunge: A
  Complete Near-Extremal Waveform},''
\href{http://www.arXiv.org/abs/1909.12846}{{\tt 1909.12846}}.

\bibitem{Bardeen:1972fi}
J.~M. Bardeen, W.~H. Press, and S.~A. Teukolsky, ``{Rotating black holes:
  Locally nonrotating frames, energy extraction, and scalar synchrotron
  radiation},'' {\em Astrophys. J.} {\bf 178} (1972)
347.

\bibitem{Apte:2019txp}
A.~Apte and S.~A. Hughes, ``{Exciting black hole modes via misaligned
  coalescences: I. Inspiral, transition, and plunge trajectories using a
  generalized Ori-Thorne procedure},''
\href{http://www.arXiv.org/abs/1901.05901}{{\tt 1901.05901}}.

\bibitem{Yang:2012he}
H.~Yang, D.~A. Nichols, F.~Zhang, A.~Zimmerman, Z.~Zhang, and Y.~Chen,
  ``{Quasinormal-mode spectrum of Kerr black holes and its geometric
  interpretation},'' {\em Phys. Rev.} {\bf D86} (2012) 104006,
\href{http://www.arXiv.org/abs/1207.4253}{{\tt 1207.4253}}.

\bibitem{Hod:2012ax}
S.~Hod, ``{Spherical null geodesics of rotating Kerr black holes},'' {\em Phys.
  Lett.} {\bf B718} (2013) 1552--1556,
\href{http://www.arXiv.org/abs/1210.2486}{{\tt 1210.2486}}.

\end{thebibliography}
\end{document}